\documentclass[journal]{IEEEtran}
\ifCLASSINFOpdf
\else
\fi

\usepackage{color, colortbl}
\definecolor{Gray}{gray}{0.9}
\definecolor{LightCyan}{rgb}{0.88,1,1}
\definecolor{LightRed}{rgb}{1,0.88,0.88}
\definecolor{LightBlue}{rgb}{0.12,0.56,0.8}
\usepackage[table]{xcolor}

\usepackage[first=0,last=9]{lcg}

\usepackage{xfrac} 

\usepackage{amssymb}

\usepackage[english]{babel}
\usepackage{blindtext}

\usepackage{subcaption}
\usepackage{algorithm}
\usepackage{algorithmic}
\usepackage{amsmath}

\usepackage{multicol}
\usepackage{multirow}

\usepackage{colortbl}
\definecolor{White}{gray}{1}
\definecolor{Gray}{gray}{0.9}
\definecolor{black}{gray}{0.6}
\definecolor{Orange}{RGB}{255,153,100}
\definecolor{Bluee}{RGB}{225,250,255}
\definecolor{Blue}{RGB}{185,210,255}
\definecolor{Blu}{RGB}{145,170,255}
\definecolor{Bl}{RGB}{110,135,255}
\definecolor{Green}{RGB}{130,255,130}
\definecolor{Purple}{RGB}{183,130,255}
\definecolor{LightRed}{rgb}{1,0.3,.3}
\definecolor{LightGreen}{rgb}{0.26,1,0.6}
\definecolor{LightBlue}{rgb}{0.26,0.6,1}
\definecolor{ACMLightBlue}{cmyk}{0.49,0.01,0,0}
\newcolumntype{g}{>{\columncolor{Gray}}c}
\newcolumntype{d}{>{\columncolor{Bluee}}l}
\newcolumntype{b}{>{\columncolor{Blue}}c}
\newcolumntype{r}{>{\columncolor{Blu}}c}

\usepackage{balance}
\usepackage{hyperref}

\begin{document}
\title{Wanna Make Your TCP Scheme Great for Cellular Networks? Let Machines Do It for You!}
\author{Soheil Abbasloo,
        Chen-Yu Yen\IEEEauthorrefmark{2}\thanks{\IEEEauthorrefmark{2}Co-Primary author},
        H. Jonathan Chao,~\IEEEmembership{Fellow,~IEEE}
        \\
\IEEEauthorblockA{High Speed Networking Lab, New York University}
}


\maketitle

\begin{abstract}
Can we instead of designing yet another new TCP algorithm, design a TCP \textit{plug-in} that can enable machines to automatically boost the performance of the existing/future TCP designs in cellular networks? We answer this question by introducing DeepCC. DeepCC leverages advanced deep reinforcement learning (DRL) techniques to let machines automatically learn how to steer throughput-oriented TCP algorithms toward achieving applications' desired delays in a highly dynamic network such as the cellular network.
We used DeepCC plug-in to boost the performance of various old and new TCP schemes including TCP Cubic, Google's BBR, TCP Westwood, and TCP Illinois in cellular networks. Through both extensive trace-based evaluations and real-world experiments, we show that not only DeepCC can significantly improve the performance of TCP schemes, but also after accompanied by DeepCC, these schemes can outperform state-of-the-art TCP protocols including new clean-slate machine learning-based designs and the ones designed solely for cellular networks.
\end{abstract}


\IEEEpeerreviewmaketitle

\section{Introduction}
\IEEEPARstart{N}{early} after 30 years from one of the earliest versions of the TCP algorithm \cite{tahoa}, congestion control (CC) in packet-switched networks remains a very hot and active research topic. Every year, with the new waves of technologies and improvements in the design of the packet-switched networks, new TCP designs are proposed to do a better job of controlling the congestion in the network and satisfy the new delay/throughput demands of new emerging applications~\cite{bbr,vivace,copa,cubic,illi,aurora,indigo}. Cellular networks with their unique characteristics (highly variable channels, radio scheduling delays, deep per-user buffers, etc.) are among the most complicated network environments that experience a new wave of targeted TCP designs~\cite{sprout,verus,c2tcp,exll,tg}. Emerging 5G technology which holds the promise of improved latency, throughput, and reliability for the network is another example of the new improvements which adds fuel to the fire. 


\subsection{Motivations}
\subsubsection{Helping others instead of beating them; exploring new design space}
Interestingly, nearly all of the TCP schemes proposed during the last three decades have a common theme in their conclusions. Most of them conclude that ``\textit{... We designed a new TCP and showed that it beats the performance of other TCP schemes in the XYZ networks ...}''. But do we really require to put aside 3 decades of designs and replace current TCP schemes with completely new ones? Can't we change the main design strategy and instead of proposing yet another new TCP algorithm, come up with a \textit{framework} that can help existing TCP schemes and boost their performance considering new needs or new environments?


\subsubsection{Why a learning-based approach}
A possible approach for designing a plug-in for certain TCP schemes in cellular networks is that designers manually try to tune TCP schemes' parameters to adapt their logic to different cellular network scenarios. Although this approach is feasible, it won’t be a scalable approach. In other words, designers need to spend a lot of time on learning the logic of various TCP schemes, understanding their performance issues in different scenarios, and tune their proprietary parameters individually for various scenarios. That is a very challenging and time-consuming process and this needs to be done again for \textit{future} new TCP schemes. But what if we can let machines \textit{automatically} learn the behavior of TCP schemes in cellular networks and \textit{adapt} them to these highly dynamic environments? That can save a lot of time and resources. 

\subsubsection{Why DRL as the learning-based approach} 
Generally, learning algorithms can be categorized into  supervised, unsupervised, and reinforcement learning (RL). Supervised and unsupervised learning are usually one-shot and myopic (as in classification problems), while reinforcement learning is sequential and far-sighted (as in games)~\cite{deep_overview}.
Moreover, CC problem is a sequential decision-making problem that deals with how many packets should/can be sent to the network by different users through time. So, RL is a better fit in the context of CC. On the other hand, Deep neural networks (DNN) leverage powerful and non-linear function approximations that can provide rich representations of complex environments~\cite{mordatch2015interactive,duan2016benchmarking}. So, in the context of designing a TCP plug-in that should boost the performance of TCP schemes with their proprietary and sometimes complex state machines over cellular networks with their famous complicated network dynamics, DNNs greatly become handy for environment representation. 
These points and the recent advances in DRL and its successful examples of being applied to practical scenarios such as DeepMind's Atari~\cite{atari} and AlphaGo~\cite{go} motivated us to design a DRL-based plug-in which aims to boost the performance of the existing TCP schemes in cellular networks. 

\subsubsection{TCP and bufferbloat issue in cellular networks}
Different studies show that the TCP schemes, which are generally designed considering wired networks, perform poorly in cellular networks when the delay is considered as the performance metric~\cite{sprout, verus, c2tcp, exll}. This problem comes from the throughput-oriented nature of most of the current TCP schemes including the most popular one TCP Cubic. Large delays caused by the bufferbloat phenomenon will not make issues for the classic throughput-oriented applications such as web page download. However, emerging delay-sensitive applications such as real-time online gaming, virtual reality, augmented reality, vehicle to vehicle communications, etc. will suffer seriously from the bufferbloat phenomenon existed in cellular networks. Recently, these emerging applications and their new delay demands motivated the community toward a more delay-centric design~\cite{sprout, verus, c2tcp, exll, vivace, copa, ledbat}. Interestingly, the bufferbloat phenomenon and the delay issue of throughput-oriented TCP schemes in cellular networks reveal that these TCP schemes actually achieve very high throughput and link utilization therein~\cite{sprout, verus, c2tcp, exll}. This observation motivated us to address bufferbloat issue by controlling the delay of the throughput-oriented TCP schemes without compromising their throughput that much to have the best of both worlds (very low delay and high throughput).

\subsection{Challenges}
Improving a specific TCP scheme seems to have a straightforward procedure. First, analyze the scheme thoroughly. Second, find its problems, and then resolve them by proposing a solution. Although the procedure itself is clear, it is not clear how to really execute it even for a specific TCP scheme, let alone targeting a solution for different TCP schemes. So, one of the challenges is that
how we can come up with a general plug-in without digging into the details of or tying the solution to a certain TCP algorithm.

On the other hand, unique characteristics of cellular networks such as highly variable and unpredictable channels, radio downlink/uplink scheduling delays, self-inflicted queuing delays, impact of the history of users' traffic and their channel qualities on the scheduling of their packets at the base station (BTS), etc. make the task of applying DRL methods in cellular networks very demanding. For instance, as we show in section~\ref{sec_eval}, schemes that simply apply Vanilla DRL techniques perform poorly in cellular networks. 


Moreover, due to the real-time nature of the TCP algorithms and the very fast fluctuations of the cellular access links, achieving operational state gathering from the network, rapid exploration of huge action space, and execution of it, while having low overhead, become very challenging.



\subsection{Contributions and Sample of Results}
Our key contributions in this paper are:
\begin{enumerate}
\item By introducing DeepCC, we show that the default strategy of designing yet another new TCP is not necessarily the best strategy toward improving TCP.
\item To the best of our knowledge, DeepCC is the first learning-based plug-in for boosting the performance of the classic and modern TCP schemes in cellular networks (see section~\ref{sec_related_learning}).
\item We built, deployed, and successfully evaluated DeepCC through a Linux Kernel implementation and demonstrated how a modern tool such as DRL can be employed to help the task of CC in a very complex environment such as the cellular network. The added Kernel APIs and our modular framework are open to the public and can be exploited to design more TCP plug-ins.
\end{enumerate}

We showed through both real-world experiments and extensive trace-based evaluations using more than 25 LTE cellular traces that DeepCC can significantly improve the performance of different TCP schemes. For instance, when TCP Cubic~\cite{cubic} (the default TCP in Linux, Android, macOS, etc.) and BBR (a new TCP proposed by Google~\cite{bbr}) are enhanced using DeepCC, they can respectively achieve 300\% and 175\% lower queuing delay while they only compromise throughput about 6\%. We also showed that DeepCC not only can improve the performance of various TCP, but also after using DeepCC, classic TCP schemes such as TCP Illinois~\cite{illi} and TCP Westwood~\cite{west} can outperform the performance of state-of-the-art schemes either the clean-slate learning-based designs or the ones solely designed for cellular networks.


\section{Background \& Related Work}
\label{sec_related}

To have a sense about the position of our design among other existing works, here, we overview existing works and their background and leave more detailed comparisons to the evaluation section\footnote{For a brief overview of deep reinforcement learning, see Appendix~\ref{sec_back}}. During the last 3 decades, a plethora of CC schemes is proposed. Due to space limitations, we briefly discuss only the most related ones. To that end, after overviewing the main heuristic approach behind nearly all of the CC designs, we focus on three classes of CCs: 1) general CC schemes targeting general environments, 2) schemes targeted cellular networks with their unique challenges, and 3) schemes using learning-based techniques to overcome the congestion control. 

\subsection{Heuristic Approach}
Early results during the 1980s (e.g. \cite{jaf}) indicated that achieving the optimal point of operation in which all users get maximum throughput, minimum delay, and a fair share of the network resources in a fully distributed way is not feasible. These results motivated the heuristic approach toward the design of TCP which primarily was intended to be a fully end-to-end and distributed algorithm for controlling congestion in the network. 

\subsection{General Designs}
Among the early schemes, TCP Tahoe, TCP Reno~\cite{tahoa}, and TCP NewReno~\cite{newreno} can be named. These schemes introduced the heuristic AIMD-based algorithms based on the consideration of loss of packets as the key indicator of congestion. Later, different schemes (e.g. BIC~\cite{bic} and TCP Cubic~\cite{cubic}) attempted to replace the linear incremental functions of these proposals with better ones. For instance, Cubic (today's default TCP in most of the platforms) employed a cubic function. While these schemes hard-wire the loss of packets to certain control actions, another set of designs tried to find new ways of interpreting congestion in the network. 
TCP Vegas~\cite{vegas} was among the early schemes that introduced the use of delay as a congestion signal. 
Designs such as Compound TCP~\cite{compound} and TCP Illinois~\cite{illi} are among the schemes that followed this delay-based approach and combined it with the loss of packets as congestion indicators. Among some recent delay-based designs LEDBAT~\cite{ledbat} and Copa~\cite{copa} can be named. 

\subsection{CC for Cellular} The fact is general TCP designs take general assumptions about all network environments and use those general assumptions to perform predefined actions. However, this leads to the low performance of these schemes when those general assumptions are not met in a certain network environment. 
Therefore, some TCP designs attempted to laser-focus on a certain environment and use prior information about that environment during the design to boost the performance of TCP. Sprout~\cite{sprout}, Verus~\cite{verus}, ExLL~\cite{exll}, and C2TCP~\cite{c2tcp,c2tcp2} are among the recent end-to-end CC schemes that target cellular networks and consider their unique features during the design. Sprout uses a stochastic framework to predict the bandwidth of the cellular access link, while Verus calculates the congestion window (cwnd) using the delay profile of the network. ExLL attempts to infer the cellular bandwidth by looking into the pattern of data reception and control the cwnd on the receiver side, while C2TCP attempts to use techniques employed in AQM designs to bound the average delay of packets considering that a user's traffic is isolated from other users' traffic in cellular networks. In addition to these end-to-end designs, another category of schemes such as NATCP~\cite{natcp}, ABC~\cite{abc}, and TG~\cite{tg} leverage the use of direct feedback from the cellular network to do the task of CC. In other words, they abandon the fully end-to-end approach to improve the performance of TCP by using direct feedback from the network. However, these feedback-based approaches face deployment issues because they require changes in the network including the need for adding new devices and protocols to the network and extra coordination of different entities in the network.

\subsection{Learning-Based CCs}
\label{sec_related_learning}
Recently, a great deal of effort has been put into using learning-based designs to automatically perform the task of congestion control. The key idea behind this trend is to abandon hardwiring certain events to certain actions during the design phase (as it is done in classic TCP designs). Instead, schemes in this category argue that the space of possible events and possible actions corresponding to them is very huge, so it is better to let machines explore this huge space and learn the best actions by themselves. In other words, these designs try to replace current TCP with fully learning-based designs. For example, Remy~\cite{remy} attempts to prepare a mapping from all possible events to actions (including change in the cwnd) in a brute-force and offline manner, while PCC-Vivace~\cite{vivace} leverages online learning techniques to choose the best sending rates automatically. Indigo~\cite{indigo} uses imitation-learning while Aurora~\cite{aurora} leverages vanilla DRL techniques to determine the sending rates. However, DeepCC fundamentally differs from these schemes as it attempts to use learning-based techniques to help and boost the performance of the \textit{existing} TCP schemes instead of replacing them. Besides, later (in section~\ref{sec_eval}), we show that these fully learning-based schemes perform poorly in the context of highly dynamic networks such as cellular networks.

Also, TCP-RL~\cite{tcp-rl} attempts to use RL to tune the initial congestion window of TCP for a group of users. In particular, every 10 minutes, it changes the initial congestion window of users, who use the same service, based on their province, ISP, and IPs. Although changing initial congestion window (which basically only is used once at the start of transmission by TCP) in a 10-minute time scale may improve the average performance of a group of users using the same service through time, it really does not have that much of impact on the performance of users in highly variable networks such as cellular networks where users' cellular access link fluctuations happen at the order of milliseconds and there is no correlation between the changes of access link bandwidth of different users connected to even the same base station (e.g. a cellular user who is walking sees a completely different link compared to a cellular user who is in a stationary position). Interestingly, TCP-RL and its 10-minute decision-making loop highlight the great challenges in the design of DeepCC. DeepCC targets real-time decision-making loops at the scale of 10ms, i.e., $60000\times$ faster than what TCP-RL tries to achieve! In addition to a different scope of applicability, this fact puts tremendous pressure on the design aspects of DeepCC including its deployment-friendliness, operational state gathering from the network, rapid exploration of huge action space, and rapid execution of the actions while having low overhead.\footnote{Since there is no publicly available code for TCP-RL, we couldn't add it to our evaluations.}


\section{System Design Overview}
The fact that throughput-oriented TCP schemes achieve very high throughput but very large delays in cellular networks can be described as follows. Calculated cwnds of throughput-oriented TCP (which generally indicate the number of in-flight packets) are usually larger than the best values of cwnd at different times. The impact of large cwnd values is that the user can always expect to have enough packets at BTS to fully utilize the cellular access link when the capacity of the link increases (e.g. due to good quality of channel). However, having a large number of packets at BTS makes large self-inflicted queuing delays when the capacity of the cellular link drops (e.g. due to bad quality of channel) which leads to poor delay performance of them~\cite{sprout, c2tcp, verus}.

Hence, the throughput-orientedness of the TCP schemes can be controlled by controlling the \textit{maximum} values of the cwnd throughout time. More specifically, DeepCC controls the \textit{cap} of the cwnd value instead of the \textit{exact} values of the cwnd through time. This helps us consider the underlying TCP of the system as a black box and avoid controlling internal proprietary variables of different TCP schemes to have a general plug-in. This is possible thanks to the modularity of the current TCP stacks where the TCP layer has certain inputs/outputs independent of the choice of the TCP schemes. 

\begin{figure}[!t]
\centering
\includegraphics[width=.9\linewidth,height=2.3in]{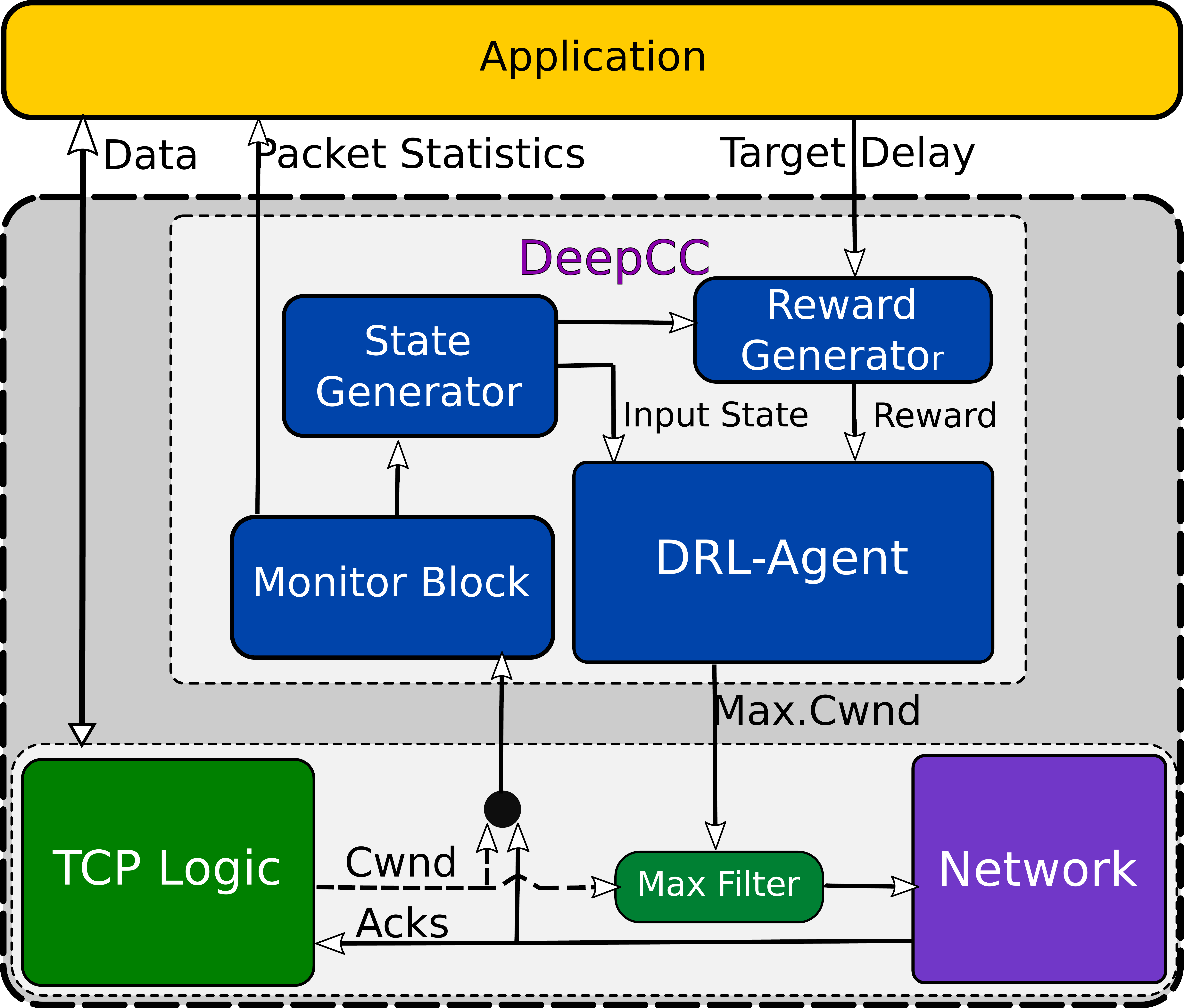}
\caption{DeepCC framework}
\label{fig_design}
\end{figure}
The big picture of the DeepCC plugin is shown in Fig.~\ref{fig_design}. DeepCC attempts to keep the average delay of packets below applications' desired Targets while keeping the throughput high. To that end, the application passes its desired Target delay as a socket-option parameter to DeepCC during the TCP socket creation. The Monitor block in DeepCC periodically collects cwnd of the system and required packet statistics from the Kernel. Every RTT, state generator block employs the information collected by Monitor block to generate a proper state vector declaring the state of the environment during that time period (detailed in section~\ref{sec_aux}). Every new state vector leads to the execution of new action by the DRL-agent. DRL-agent considers the generated state vector, the application's given Target, and the reward associated to the current state vector (which is generated by Reward block) to decide the proper maximum value of cwnd ($cwnd_{max}$) based on the previously learned behavior of the environment (detailed in section~\ref{sec_agent}).
At the end, the final refined cwnd considering this maximum value will be used to send packets into the network.

\noindent\textbf{A short Q\&A before going deep into the design:}\footnote{See section~\ref{sec_dis} for more}
\begin{enumerate}
\item Is DeepCC a plug-in for \textit{all} TCP schemes without any exception? No. The current version of DeepCC stands on top of the \textit{throughput-oriented} TCP to improve the delay performance of it. 
How to make DeepCC a plug-in for \textit{``everyone''} is a good motivation for future work.
\item Why is the throughput-oriented TCP chosen as the underlying TCP for DeepCC plug-in? Because today's most popular and widely used TCP schemes are throughput-oriented designs.
\end{enumerate}

\section{Components of DeepCC: Part I}
\label{sec_aux}

\subsection{Monitor Block}
Every RTT, the monitor block generates digested information to be used in the state generator block.
The statistics we use are the collectible signals which can be measured/monitored on the fly from the network. 
Even though we collect traces for training, where network bandwidth is revealed, we do not use it as state information because when the system is running online, bandwidth is unknown in advance. We consider the following statistics:
\begin{itemize}
\item $d$: The average packets' delay per RTT
\item $n$: The number of samples used for calculation of $d$
\item $p$: The average delivery rate (throughput\footnote{Throughout this paper we use delivery rate and throughput of the system interchangeably}) per RTT
\item $cwnd$: The current cwnd calculated by the underlying TCP.
\end{itemize}

In particular, Monitor block works as a shim layer and continuously generates the required packets' statistics by observing the incoming Ack packets. Monitor block periodically exploits generated packets' statistics and calculates $d$, $n$, and $p$ of each time period and passes them (combined with cwnd) to the state generator block. Having Monitor block implemented as a shim layer in the Kernel enables us to make the process of gathering required statistics independent of the underlying TCP schemes.

\subsection{State Generator Block}
The state generator is an information processing module that manages the desired Target delay information from the application ($Target$) and the network dynamics coming from the monitor block and generates the input state for DRL-agent. Each time the network monitor passes the packet statistics, the state generator updates the state vector.

Network dynamics in the wireless cellular environment are complicated and noisy. The learning process for DRL to extract the useful feature from the high dynamic range of state and map it to actions can be demanding. Another challenge is how the agent can effectively learn the policy of $cwnd_{max}$ based on the statistics in such a highly dynamic environment. The uncertainty of the environment makes the learning harder and makes the problem setting lose Markov property. The agent is easily confused to make a decision $a_t$, if the agent only relies on the observation at time $t$. To address these issues, we use two techniques: 1) the use of filter kernel and 2) the use of recurrent structure.

\textbf{Filter kernel}:\label{sec_kf} Inspired by the use of filter kernel in image processing area~\cite{fourier, imageproc, nixon_feature}, first, we apply a kernel to the state so that we can reduce the input space. We transform the delay $d$ with the $Target$ as a kernel:
\begin{equation}
\kappa (d) = \begin{cases}
0 & \text{if $d > Target $;}\\
1 & \text{else}\\
\end{cases}
\end{equation}

Now, we multiply $p$ and $n$ by the kernel: $\phi(p) = p\times\kappa$ and $\phi(n) = n\times\kappa$. Second, we encode the delay together with target: $\phi(d,Target) = \begin{bmatrix}
(1- \frac{d}{Target}) \times\kappa , \frac{d}{Target} \times(1-\kappa)
\end{bmatrix}.$\footnote{$\begin{bmatrix}x_1, ..., x_n\end{bmatrix}$ means concatenation of $x_1, ...,$ and $x_n$ . } The observed statistics at time $t$, after the above preprocessing, can be written as following vector:
\begin{equation}
o_t = \begin{bmatrix} \phi(p_t), \; \phi(n_t), \; \phi(d_t,Target), \; cwnd_t \end{bmatrix}
\label{features}
\end{equation}
Here, the intuition is that when system does not meet $Target$ (i.e., $d>Target$), the agent should focus more on delay signal ($d$) and consider performing the action which reduces the delay.

\textbf{Recurrent structure}: We also use a recurrent structure~\cite{LSTM, pearlmutter_rec} to address the lack of Markov property issue. Since at no time the agent has the direct knowledge of the current exact network bandwidth, the agent can only infer it like a latent variable from the observed statistics. In other words, because the agent has only access to a partial local observable state of the network, it can be confused if the decision only relies on $o_t$. 
Therefore, we represent the state at time $t$, $s_t$, by a finite history of the agent's interactions. In other words, $s_t$ becomes a vector (with length $m$) including the current observed statistics ($o_t$) concatenated to the history of observed statistics: $s_t = [o_t, o_{t-1}, ..., o_{t-m+1}]$.


\subsection{Reward Generator Block}
One of the important parts of any RL algorithm is the definition of the reward function which should reflect the main objective of the system.
DeepCC's objective is to keep the average delay of packets below the Target value (given by the application) while maximizing the throughput. So, DeepCC requires to have a reward function which reflects that objective correctly.
Intuitively, 
the RL agent is motivated to choose the actions which generate higher rewards and to avoid choosing the actions which lead to lower ones. When the environment is well-behaved (such as a game environment as in~\cite{go}) and does not show a very high random nature, the reward gained in a certain step can be considered as the result of the action chosen for that step. However, it is not necessarily the case for a dynamic and unpredictable environment such as a cellular network. For instance, when the reward function is simply defined as the throughput achieved by the system, it is not clear whether the currently gained very small reward is due to choosing a bad action by the agent or due to a very bad quality of channel in the wireless access link (which leads to low throughput). Therefore, the reward function needs to be carefully defined.

To that end, we use the following reward function where $w(n,d)$ is the moving average of the two recent values of $d$ ($w(n,d)=\frac{nd+n_{pre}d_{pre}}{n+n_{pre}}).$\footnote{$n_{pre}$ and $p_{pre}$ show the previous values of $n$ and $p$, respectively.}
\begin{equation}
\ r(n,d,p) = \begin{cases}
-\frac{w(n,d)}{Target}\times p\times n & \text{if $d>Target$;}\\
+\frac{w(n,d)}{Target}\times p\times n & \text{else}\\
\end{cases}
\label{eq_reward}
\end{equation}

If the agent observes a high delivery rate and a high number of Acks, while it observes a bad average delay ($d>Target$), most likely, its wrong choice of action was the source of achieving bad delay response. That is why in our reward function, the agent will be penalized/rewarded in proportion with the gained delivery rate, number of received Acks, and how far the average delay is from the Target delay. Therefore, using the reward function described in Eq.~\ref{eq_reward}, the agent will be motivated to first keep the average delay below $Target$ ($d<=Target$) (to receive less penalty). Then, it will be motivated to maximize its delivery rate and the number of received Acks (to collect more reward). 


\section{Components of DeepCC: Part II}
\label{sec_agent}
Now, we describe the core component of DRL block: DRL-agent. DRL-agent observes a set of network statistics (given by the state generator block) and collected reward (by the reward generator block) and outputs the $cwnd_{max}$ value.
The lower value of the $cwnd_{max}$ pushes the system toward being more delay-sensitive (less throughput-oriented), while the higher one pushes the system toward being more throughput-oriented (less delay-sensitive). The goal of DRL-agent is to learn the policy that makes underlying TCP gain more throughput while meeting the Target delay. The space of the action (i.e., the value of $cwnd_{max}$) at different times is immense. So, letting the DRL-agent find the best values of the action in the huge space in a timely manner is not feasible. To make the exploration phase feasible and efficient, we use the cwnd calculated by the underlying TCP as the base to calculate the $cwnd_{max}$. To that end, we define a parameter $\alpha$ which relates $cwnd_{max}$ value to the value of cwnd that the DRL-agent receives periodically from the state generator block using the following equation:
\begin{equation}
cwnd_{max}=2^{\alpha}\times cwnd
\label{cwnd}
\end{equation}
Now, instead of searching the entire space, DeepCC only chooses the $\alpha$ value restricted to $-1\le \alpha \le 1$. This greatly simplifies the exploration phase (and consequently lowers the convergence time), while $cwnd_{max}$ can still easily be increased exponentially when the DRL-agent chooses $\alpha=1$ consecutively (or decreased exponentially when $\alpha=-1$ is chosen consecutively).\footnote{Note that there is only one cwnd value per TCP socket. So, when the filter caps the value of cwnd to the desired maximum value, the cwnd of the socket is replaced by the capped value.} We will show in section~\ref{sec_eval} that this choice leads to very good system performance.

\subsection{The Learning Algorithm}
Here, we formulate the CC problem in a reinforcement learning setting  where the agent interacts sequentially with the environment with the goal of learning reward-maximizing behavior. At each time step $t$, agent observes state $s_t$ and chooses action $a_t$ according to a policy $\pi$ which maps states to actions. In return, the agent receives a reward $r_t = r(s_t,a_t)$ and a new state $s_{t+1}$. The return is defined as discounted sum of rewards $R_t = \sum_{t'=t}^T \gamma^{t'-t} r_{t'}$ and the agent's ultimate goal is to find the optimal policy $\pi_\theta$, with parameters $\theta$, which maximizes the expected return $J(\theta)=E[R_0]$.

DeepCC's agent is built on top of DDPG~\cite{ddpg} which is designed for learning policies in continuous action space. 
Our agent consists of a policy network ($\pi_\theta$) which determines a deterministic action $a_t = \pi_\theta(s_t)$ and a critic network ($Q_w$, with parameters $w$) that predicts the action-value function. The action-value function is defined as $Q_\pi(s_t,a_t) = E[R_t|s_t,a_t]$ and shows the expected return when taking action $a_t$ from state $s_t$, and subsequently acting following $\pi$. 
Considering $J(\theta)$ as the expected return, we have:
\begin{equation}
J(\theta) = E[Q_{\pi_\theta}(s,\pi_\theta(s))] \text{\ \ \ \ \ \ }  (s_0=s)
\end{equation}

In the learning algorithm, the parameters of the policy network are updated using policy gradient ascent. Leveraging deterministic policy gradient theorem~\cite{dpg}, the gradient of expected return can be computed as:
\begin{equation}
    \nabla_\theta J(\theta) = E [\nabla_a Q_{\pi_\theta}(s,a)|_{a=\pi_\theta(s)} \nabla_\theta \pi_\theta(s) ]
    \label{a_loss}
\end{equation}

Parameters of the critic network, that estimates the action-value of the policy $Q_w(s,a) \approx Q_{\pi_\theta}(s,a)$, are trained by minimizing the loss function ($L(w)$) which indicates a temporal difference error between a target function ($y_t$) and $Q_w$
\begin{equation}
    L(w) = E[(y_t-Q_w(s_t,a_t))^2]
    \label{c_loss}
\end{equation}

\noindent where $y_t = r_t + \gamma Q_{w'}(s_{t+1}, \pi_{\theta'}(s_{t+1}))$. $\pi_{\theta'}$ (target policy network) and $Q_{w'}$ (target critic network) are two separate deep neural networks with parameters $\theta'$ and $w'$, respectively.
$\theta'$ and $w'$ are exponentially weighted moving averages of $\theta$ and $w$. Employing the target networks is shown to make the training procedure more stable~\cite{DQN}.

The policy and critic networks are trained by iteratively updating their parameters using an experience replay. At each iteration, the tuple $(s_t,a_t,r_t,s_{t+1})$ is put in a finite-sized cache $D$ called replay buffer. The update of policy happens by applying the gradient with respect to $\theta$ in Eq.~\ref{a_loss} while the critic is updated by performing gradient descent step on Eq.~\ref{c_loss}. To calculate the gradient a minibatch of randomly chosen tuples from $D$ is employed. We summarize DRL-agent's learning algorithm in Algorithm~\ref{alg:ddpg}.

\begin{algorithm}[!t]
            \caption{DRL-Agent's Learning Algorithm} \label{alg:ddpg}
    \begin{algorithmic}
        \FOR {Learning is not finished}
        \FOR {$t=1, 2,..., T$}
        \STATE Select action $a_t = \pi_\theta(s_t)+ \mathcal{N}_t$ 
        \label{exploration}
        \STATE Execute $a_t$ and observe reward $r_t$ and new state $s_{t+1}$
        \STATE Store transition $(s_t,a_t,r_t,s_{t+1})$ in $D$
        \STATE Sample minibatch of $N$ transitions $(s_i,a_i,r_i,s_{i+1})$ from $D$
        \STATE Compute $y_i = r_i + \gamma Q_{w'}(s_{i+1},\pi_{\theta'}(s_{i+1} ))$
        \STATE Update the critic by minimizing the loss:
        \\ $L = \frac{1}{N} \sum_i (y_i - Q_w(s_i,a_i))^2$
        \label{alg-c}
        \STATE Update the policy using policy gradient:
        \\
        $ \nabla_{\theta}J \approx
         \frac{1}{N}\sum_i \nabla_a Q_w(s,a)|_{s=s_i,a=\pi_\theta(s_i)}\nabla_{\theta}\pi_\theta(s)|_{s=s_i}$
        \label{alg-a}
        \STATE Update the target networks:
        \\
        $
        \theta' \leftarrow \tau \theta+ (1-\tau)\theta', $
        \\
        $w' \leftarrow \tau w  + (1-\tau)w' 
        \label{alg-tg}
        $
        \ENDFOR
        \ENDFOR
    \end{algorithmic}
\label{ddpg}
\end{algorithm}

\subsection{Batch Normalization and Exploration}
\label{sec_explor}
\textbf{Batch Normalization:} 
To increase the training performance, we also use the Batch Normalization techniques.
Batch Normalization~\cite{batchnorm} is a technique that makes the optimization landscape smoother~\cite{NIPS2018_7515}. The smoothing effect on the optimization increases the training speed and makes our DRL block less sensitive to the variation of the hyperparameter setting. The Batch Normalization technique is implemented in our actor-network as an augmented layer after each hidden layer's output right before the non-linearity activation. 

\textbf{Exploration:} Adequate exploration in action space is crucial in our problem. The idea of exploration is to allow the agent to try different actions to improve the model. During exploration, although some actions that the agent chooses have a lower return at that moment, later after the agent may discover the trajectories with better policies, might lead to a higher return. The exploration techniques add noise to the actor as if the deterministic actor selecting an action in stochastic behavior. We experimented with different types of noise including variations of Ornstein-Uhlenbeck (OU) noise~\cite{OU} and variations of Gaussian noise. We observed that the uncorrelated additive Gaussian action space noise leads to better performance. Using noise ($\mathcal{N}$), the final generated action becomes $a_t= \mu_\theta(s_t) + \mathcal{N}$ (Algorithm~\ref{ddpg}, line~\ref{exploration}). Before using this equation, at the cold-start of training, we use an additional exploration method, in which we enforce the agent to walk over the action space (in a piece-wise manner) for a fixed amount of steps without experiencing noise.

\section{Implementation and Training}
Now, we examine DeepCC plug-in by using four different TCP schemes as the underlying TCP schemes. Without loss of generality, we chose TCP Cubic~\cite{cubic}, TCP Westwood~\cite{west}, TCP Illinois~\cite{illi}, and TCP BBR~\cite{bbr} as underlying schemes using DeepCC plug-in throughout this section. From now on, when a TCPx uses DeepCC plug-in, we add prefix character $D$ to the name of the TCPx scheme to refer to the new scheme (i.e., DCubic, DWest, DIlli, and DBBR). 
\label{sec_training}
\subsection{Implementation}
To have a general solution that can work with various TCP schemes, we have modified Linux Kernel 4.13 and added a plug-in that works independent of the TCP scheme of the system. It can be considered as a shim layer above the TCP layer. In particular, we modify the Kernel to calculate the required statistics, provide new socket options to collect the statistics needed from the Kernel, and provide new socket options to enforce a calculated action in the Kernel. For this prototype version, we use a user-space implementation of the DRL-agent using TensorFlow~\cite{tensor}, though putting the final learned model in the Kernel will improve the performance. 
The DRL-agent's actor and critic networks are implemented using two fully-connected hidden-layers with 1000 neurons on each layer.
For brevity, here, we skip the details of the implementation. The source code is available to the 
community at \url{https://github.com/Soheil-ab/DeepCC.v1.0.git}.

\subsection{Clean-Slate version}
Also, we use the same setting and DeepCC's framework and design a clean-slate learning TCP scheme that instead of capping the cwnd value of underlying TCP, it directly calculates the cwnd of the system. We use the same training setting that we use for training DeepCC to train Clean-Slate scheme and later compare the performance of the Clean-Slate version with DeepCC. 

\subsection{Training Phase}
We trained DeepCC over emulated LTE network environments using Mahimahi \cite{mahi} a trace-based emulator. For cellular environment, we employed cellular LTE traces 
including more than 100,000 variations of cellular link bandwidths, 
as our base environment. The statistics of the base cellular environment during training are shown in Table~\ref{table_learning_trace}. Later, during the evaluation phase (starting from section~\ref{sec_eval-general}), we don't use this cellular environment anymore and test the agent on other unseen environments to examine how good the agent can generalize the environment. Also, we used Target value of 50ms during the training phase (Since we do not use the absolute value of Target in our Reward function (we use the normalized value of delay), the learning phase is not that sensitive to the choice of the Target value. For instance, see Fig.~\ref{fig_aqm} and the discussion around it). The 50ms delay is a classic definition of resiliency requirement for the networks. For example, in optical networks,
devices are required to recover from failures by less than 50ms, because it was known that for video/audio signals, humans cannot detect less than 50ms signal mismatches~\cite{blsr_sonet,upsr_sonet,ring_sdh,protection_ansi,ring_eth}. 

\begin{figure}[!t]
    \centering
    \begin{minipage}[b]{0.48\linewidth}
        \centering
     \includegraphics[width=\linewidth,height=1.7in]{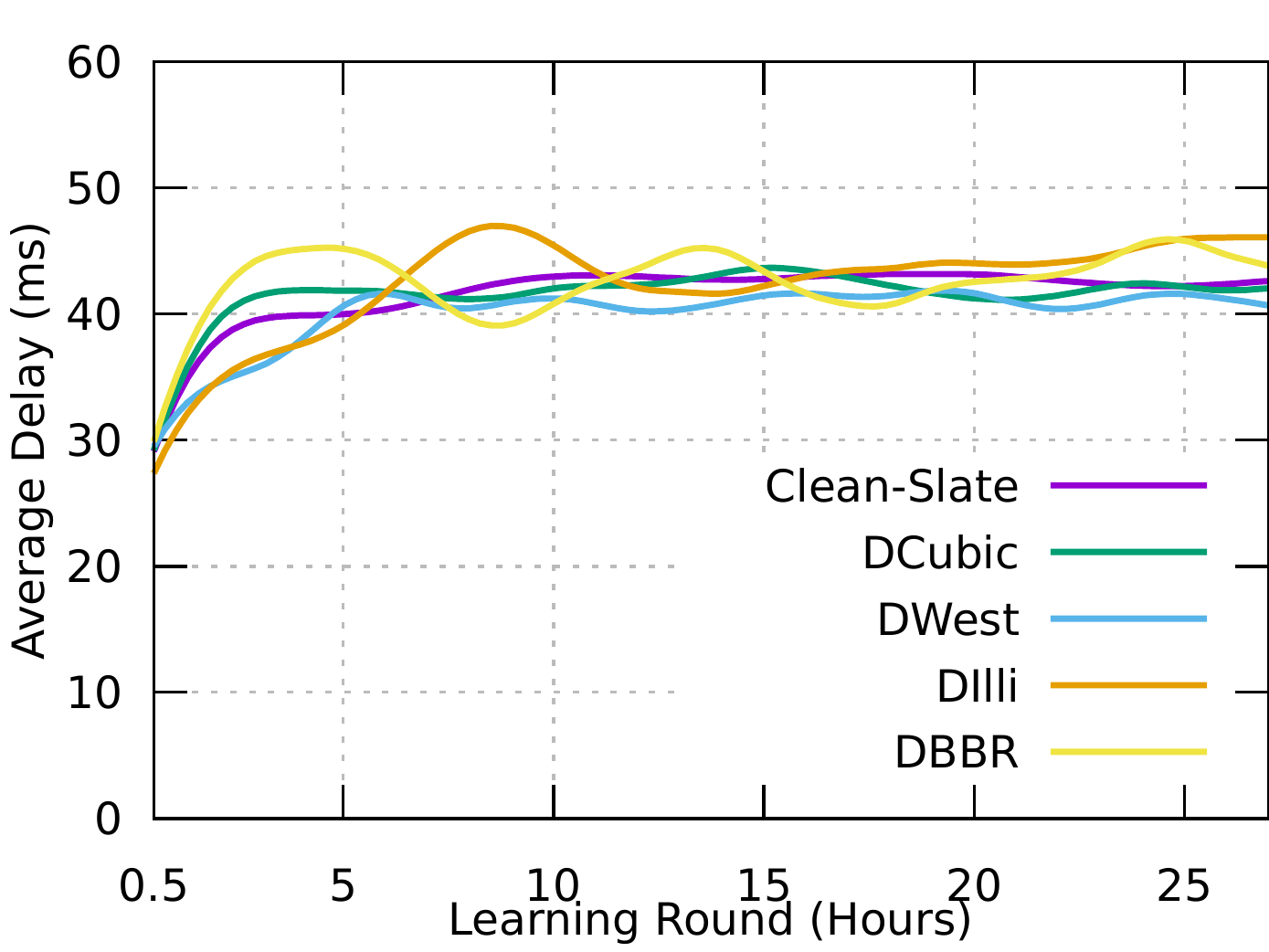}
    \end{minipage}        
    \begin{minipage}[b]{0.48\linewidth}
        \centering
     \includegraphics[width=\linewidth,height=1.7in]{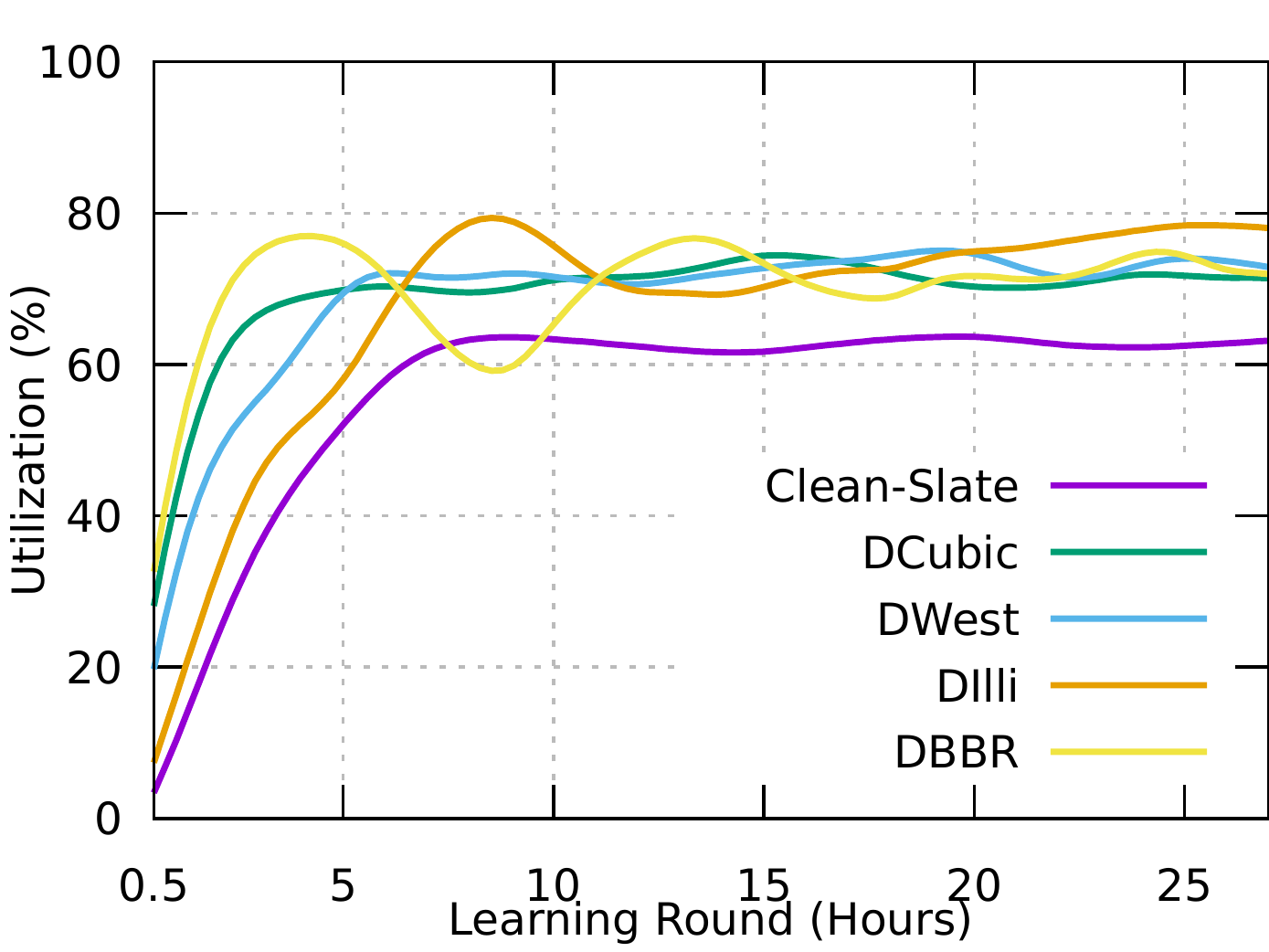}
    \end{minipage}        
     \caption{Average delay and utilization as a function of wall clock time for various schemes.}
     \label{fig_tr}
\end{figure}

To train the agent, we used Dell PowerEdge R730 servers with 48 cores. To observe the gradual improvements achieved during the learning, after every 30 minutes of training, we use the trained model, run an evaluation using the setting described in section~\ref{sec_eval}, and gather the average delay and utilization of the scheme over the entire trace used for the learning
(Totally, we trained the system over more than 2 million steps). Results are shown in Fig.~\ref{fig_tr}. While keeping average delay below the target value, DeepCC gradually learns to boost the utilization/throughput of the underlying TCP. For DeepCC, the behavior of underlying TCP is part of the dynamic nature of the environment that it observes. In other words, although the cellular environment is the same during the training of all TCP schemes, due to the different nature of these TCP schemes, DeepCC observes different environments when underlying TCP is different. That is why for different TCP schemes, DeepCC's learning curves are different. We use this trained model for all evaluations done throughout the rest of this paper. 

\begin{table}[!h] \renewcommand{\arraystretch}{0} \caption{Statistics of the cellular access link capacity used during training (bandwidth units are Mbps)} 
\small
\label{table_learning_trace}
\centering
\begin{tabular}{c| c | c |c| c}
Mean     & Std Deviation & Skewness     & Kurtosis     &    [Min,Max] \\ \hline
\rowcolor{Bluee}
12.7875    & 11.3804        & 1.6027        &6.5076        &    [0,90] \\
\end{tabular}
\end{table}

During the evaluation phases (starting from section~\ref{sec_eval-general}), agent experiences unseen LTE environments with different dynamics compared to the dynamics of the LTE environment used during the training. Besides, cellular traces used during the training are gathered from T-Mobile service provider. However, during the evaluation phases, in addition to new LTE traces gathered from T-Mobile service provider on different base stations, LTE traces from other service providers namely Verizon and AT\&T are also gathered and the agent is tested over all of them. Results reported in section~\ref{sec_scenarios} and section~\ref{sec_eval-real} show that with the described training setting, the trained model achieves good performance over unseen real LTE cellular networks and unseen twenty-seven LTE cellular traces.

\subsection{Clean-Slate vs DeepCC}
Interestingly, DeepCC enhanced schemes outperform the Clean-Slate scheme and achieve $10\%$ higher utilization. There is a subtle reason behind that. In Clean-Slate version, no changes occur in the cwnd until the end of each monitoring period (time of enforcing a new cwnd), because Clean-Slate scheme is the only entity controlling the cwnd. However, in DeepCC, during each monitoring period, underlying TCP can still actively change the values of cwnd to probe more bandwidth. This dynamism helps DeepCC achieve a more solid performance compared to Clean-Slate version and boost its throughput performance. Also, later in section~\ref{sec_eval-general}, we show that DeepCC performs better than Clean-Slate version in other scenarios.


\section{Evaluation: Trace-Based Emulations}
Here, we report the results of our trace-based evaluations comparing the performance of various TCP schemes in various cellular network conditions. 
\subsection{Setup}
\label{sec_eval-general}
\label{sec_eval}
\subsubsection{Emulator} To have more freedom to compare the performance of various schemes in a reproducible environment, we use Mahimahi~\cite{mahi} which is a trace-based network emulator. 
\begin{figure*}[!tp]
    \centering
    \begin{minipage}[b]{0.32\linewidth}
        \centering
        \includegraphics[width=0.95\linewidth,height=2in]{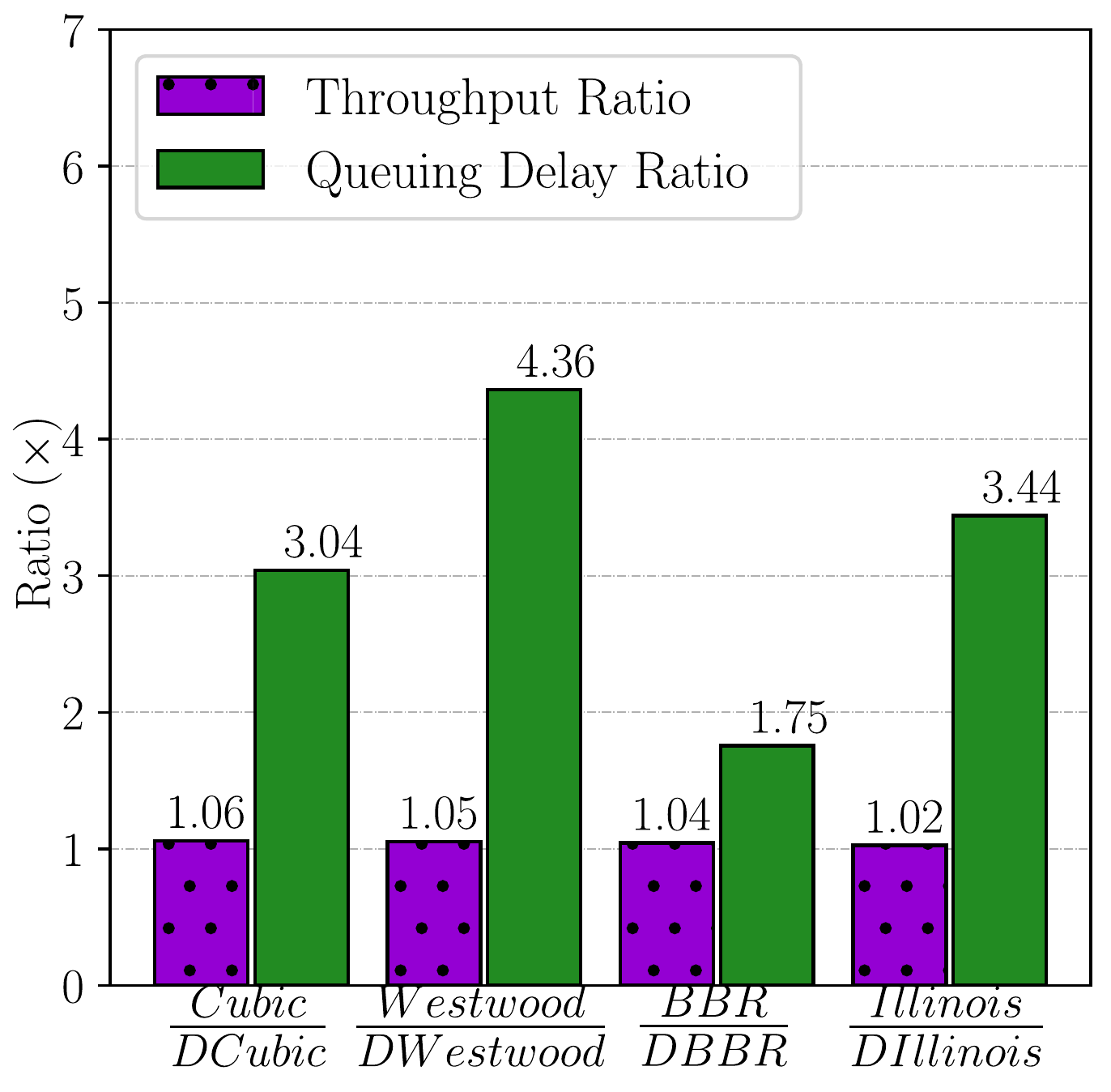}
        \subcaption{Stationary scenario} 
        \label{fig_static_q}
    \end{minipage}        
     \hfill
    \begin{minipage}[b]{0.32\linewidth}
        \centering
        \includegraphics[width=0.95\linewidth,height=2in]{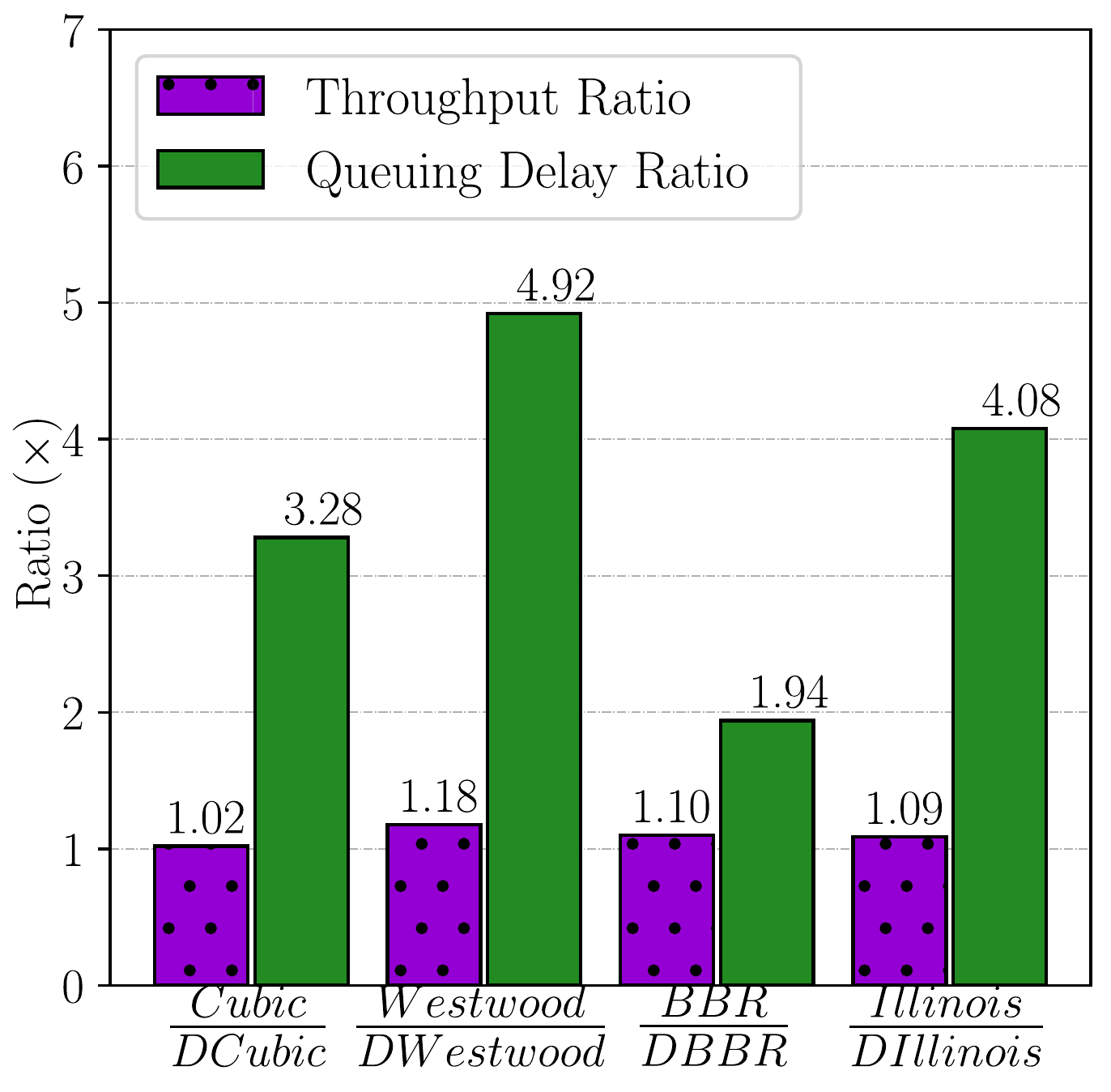}
        \subcaption{Moving scenario \#1} 
        \label{fig_mv1_q}
    \end{minipage}
    \hfill
    \begin{minipage}[b]{0.32\linewidth}
        \centering
        \includegraphics[width=0.95\linewidth,height=2in]{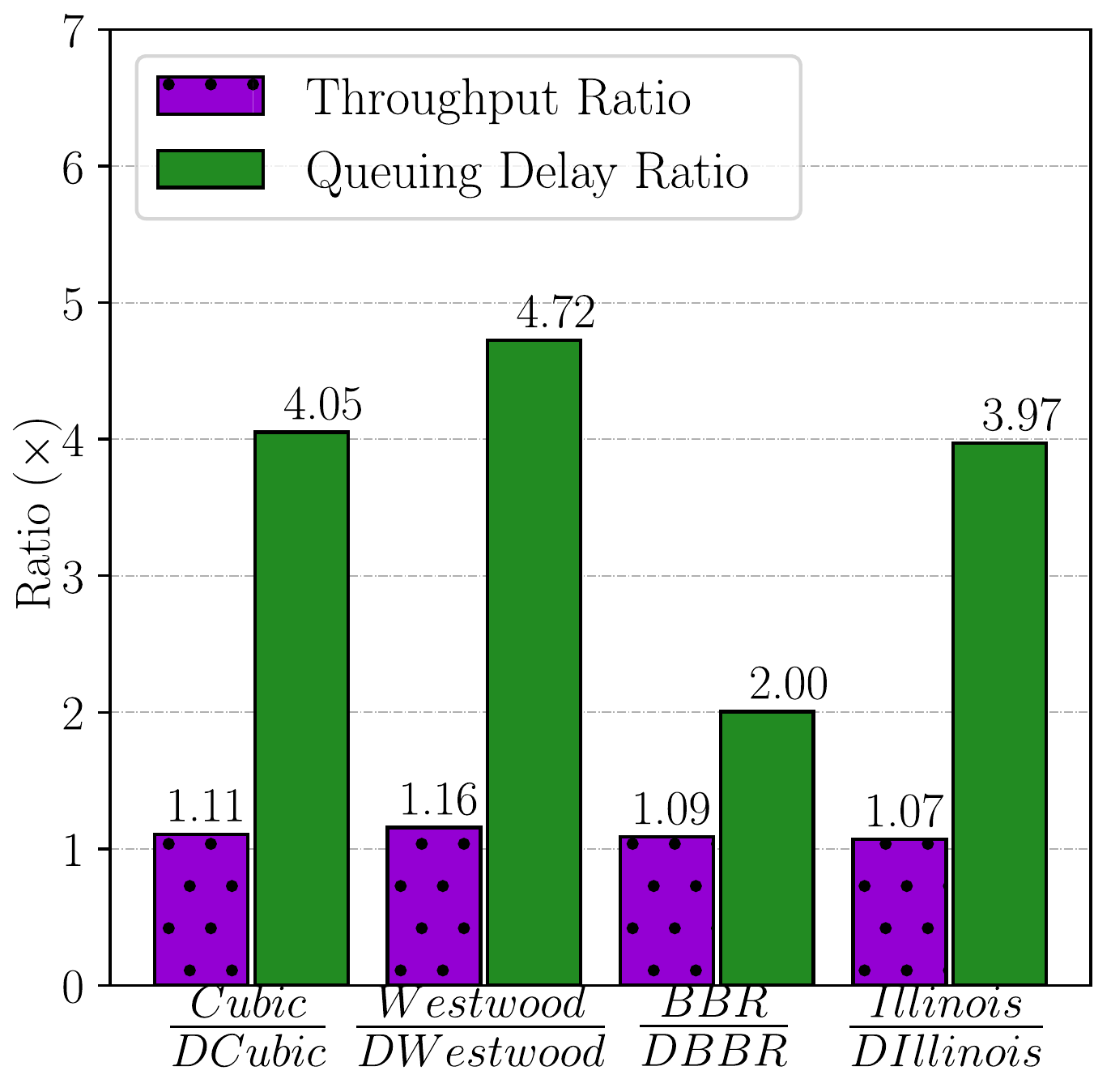}
        \subcaption{Moving scenario \#2} 
        \label{fig_mv2_q}
    \end{minipage}
    \caption{Ratio of Queuing delay and throughput of different schemes over their DeepCC enabled counterparts  ($\frac{TCP}{DeepTCP}$).} 
        
      \label{fig_del_util}
\end{figure*}


\subsubsection{Schemes Compared} To cover variety of designs, we compared DeepCC with 14 schemes each representing different design strategy. 
Among them, Sprout~\cite{sprout}, Verus~\cite{verus}, C2TCP~\cite{c2tcp}, Copa~\cite{copa}, and LEDBAT~\cite{ledbat} are delay-sensitive and delay-based, while Cubic~\cite{cubic} and Westwood~\cite{west}, BBR~\cite{bbr}, and Illinois~\cite{illi} are mainly throughput-oriented schemes. In addition, PCC-Vivace~\cite{vivace}, Remy~\cite{remy}, Indigo~\cite{indigo}, and Aurora~\cite{aurora} are learning-based approaches.\footnote{For Indigo and Aurora, we use the same model trained by their authors over variety of network scenarios. For Remy, similar to others (e.g. \cite{indigo,copa,aurora}) we use the model trained for a 100× range of link rates in prior work~\cite{copa}.} Moreover, we use our Clean-Slate scheme which has trained over the same setting that DeepCC is trained over.

\subsubsection{Setting} We set the minimum RTT of the network to 20ms and the buffer size to 150KB. Throughout the evaluation, unless it is mentioned, we choose the Target value of 50ms for DeepCC. Throughout the paper, we set $m=20$.


\subsubsection{Performance Metrics} We mainly use three performance metrics: averaged overall delay of packets
, average queuing delay, and averaged throughput/utilization throughout the run of an algorithm.\footnote{We have also used 95th percentile delay as another performance metric. We observed that all schemes perform relatively similar for both 95th percentile and average delay metrics. So, for brevity and considering DeepCC's objective of keeping the average delay below a target value, here, we skip reporting the 95th percentile results.} For utilization, we consider the average delivery rate of packets over the average capacity of the access link during the experiment. 

\subsection{Various Unseen Scenarios}
\label{sec_scenarios}
\begin{table}[!t] \renewcommand{\arraystretch}{0} 
\caption{Samples of statistics for the cellular access link traces from each of three scenarios (bandwidth units are Mbps)}
\label{table_trace_sample} 
\centering

\resizebox{\columnwidth}{!}{%
{\renewcommand{\arraystretch}{1.2}%

\begin{tabular}{c| c| c| c | c |c| c}
\#    &    Scen.     &    Mean     & Std dev.     & Skewness     & Kurtosis     &    [Min,Max] \\ \hline
\rowcolor{White}
(1)    &    Stationary&    16.0260    & 3.3751        & -0.4480        &6.1678        &    [0,34] \\
(2)    &    Stationary    &    21.5223    & 3.3262        & -0.3097        &5.8972        &    [0,50] \\ 
(3)    &    Stationary&    7.6064    & 2.8675        & -0.4376        &2.7488        &    [0,18] \\\hline
\rowcolor{Bluee}
(4)    &    Moving\#1&    11.8125    & 5.8968        & 0.1172        &2.2938        &    [0,28] \\
\rowcolor{Bluee}
(5)    &    Moving\#1&    7.2437    & 4.6935        & 0.1615        &1.8565        &    [0,28] \\
\rowcolor{Bluee}
(6)    &    Moving\#1&    3.9667    & 3.1363        & 1.2482        &4.4013        &    [0,18] \\ \hline
\rowcolor{Blue}
(7)    &    Moving\#2&    4.9487    & 5.3328        & 2.8663        &15.5169        &    [0,51] \\
\rowcolor{Blue}
(8)    &    Moving\#2&    18.0001    & 7.3393        & -0.6261        &2.3710        &    [0,48] \\
\rowcolor{Blue}
(9)    &    Moving\#2&    9.7758    & 9.5413        & 1.6356        &6.0892        &    [0,95] \\ \hline
\hline
\end{tabular}
}
}
\end{table}

Here, we examine the performance of DeepCC in 3 different general scenarios: 
\begin{enumerate}
\item Stationary scenario: When the cellular user is in static position and does not move 
\item Moving scenario \#1: When the cellular user is walking in crowded places
\item Moving scenario \#2: When the cellular user is riding a bus, a taxi, or driving a car
\end{enumerate}
To that end, We use a combination of 23 LTE cellular traces that we have collected in NYC (using Saturatr~\cite{sprout} tool) and 5 other LTE traces gathered by prior work~\cite{sprout,c2tcp} from 3 cellular network providers in the US. In particular, for the stationary scenario, we collected five new traces in residential buildings and in Times Square (a crowded place with different moving obstacles in the environment) and used two additional static traces collected in NYC by prior work~\cite{c2tcp}. For moving scenario \#1, we collected and used twelve different new traces in different crowded areas of NYC. For moving scenario \#2, we collected six new traces while the cellular user was riding taxi/bus and used three additional traces collected in Boston by prior work~\cite{sprout} while user was driving. An important note here is that these traces are gathered from three different cellular LTE providers and from different base stations. So, they can represent a good set of LTE cellular environments with different network dynamics for examining performance of DeepCC over unseen LTE environments. 
To quantify the differences among traces gathered, we report statistics of nine of these traces in Table~\ref{table_trace_sample}. Also, to have a qualitative sense about these traces, we also report a few minutes of these nine traces in Appendix~\ref{sec_trace}.

\subsection{Performance Results: With \& Without DeepCC}
\begin{figure*}[!t]
    \centering
    \begin{minipage}[b]{0.32\linewidth}
        \centering
        \includegraphics[width=\linewidth,height=2in]{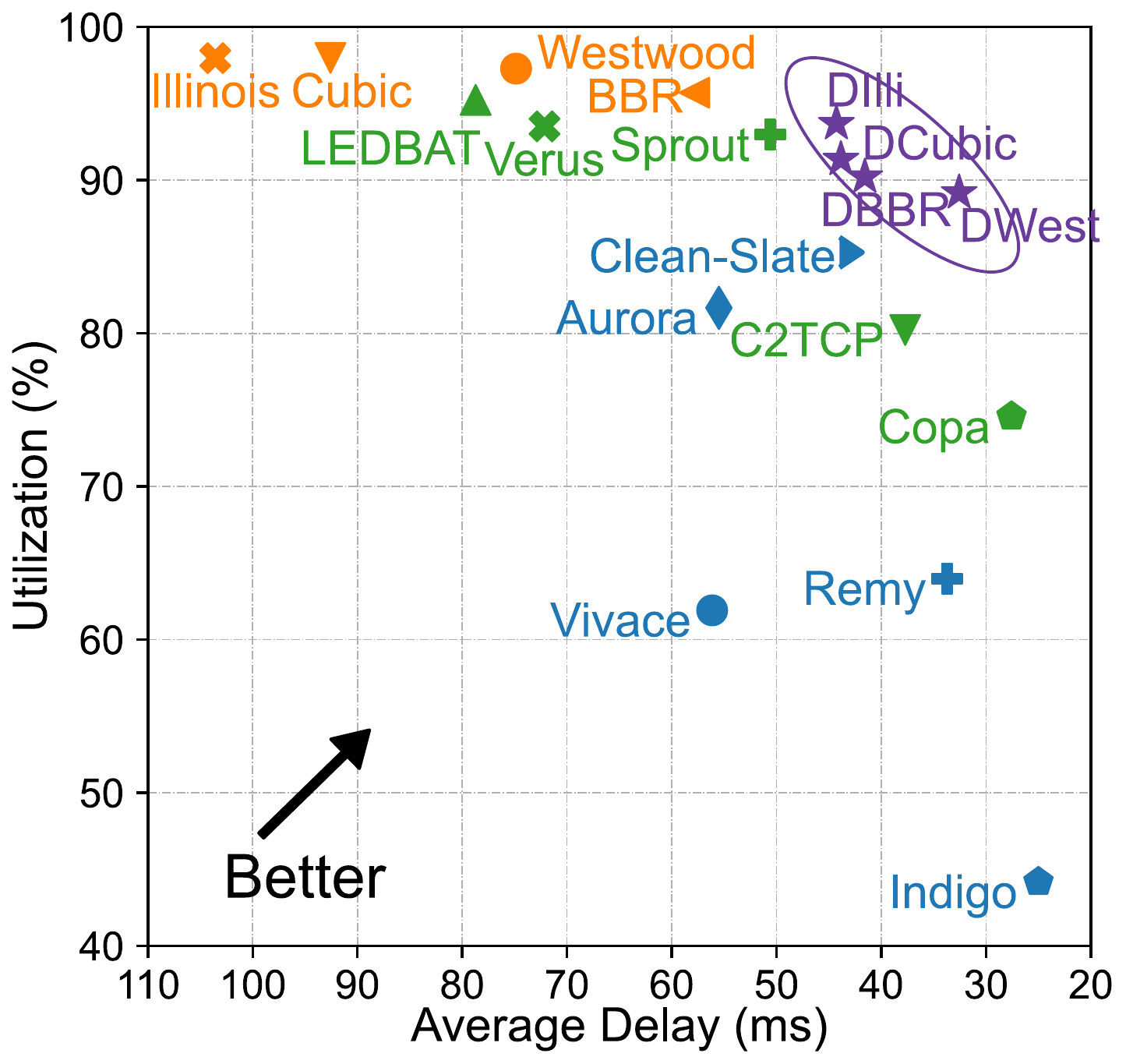}
        \subcaption{Stationary scenario} 
        \label{fig_static}
    \end{minipage}        
     \hfill
    \begin{minipage}[b]{0.32\linewidth}
        \centering
        \includegraphics[width=\linewidth,height=2in]{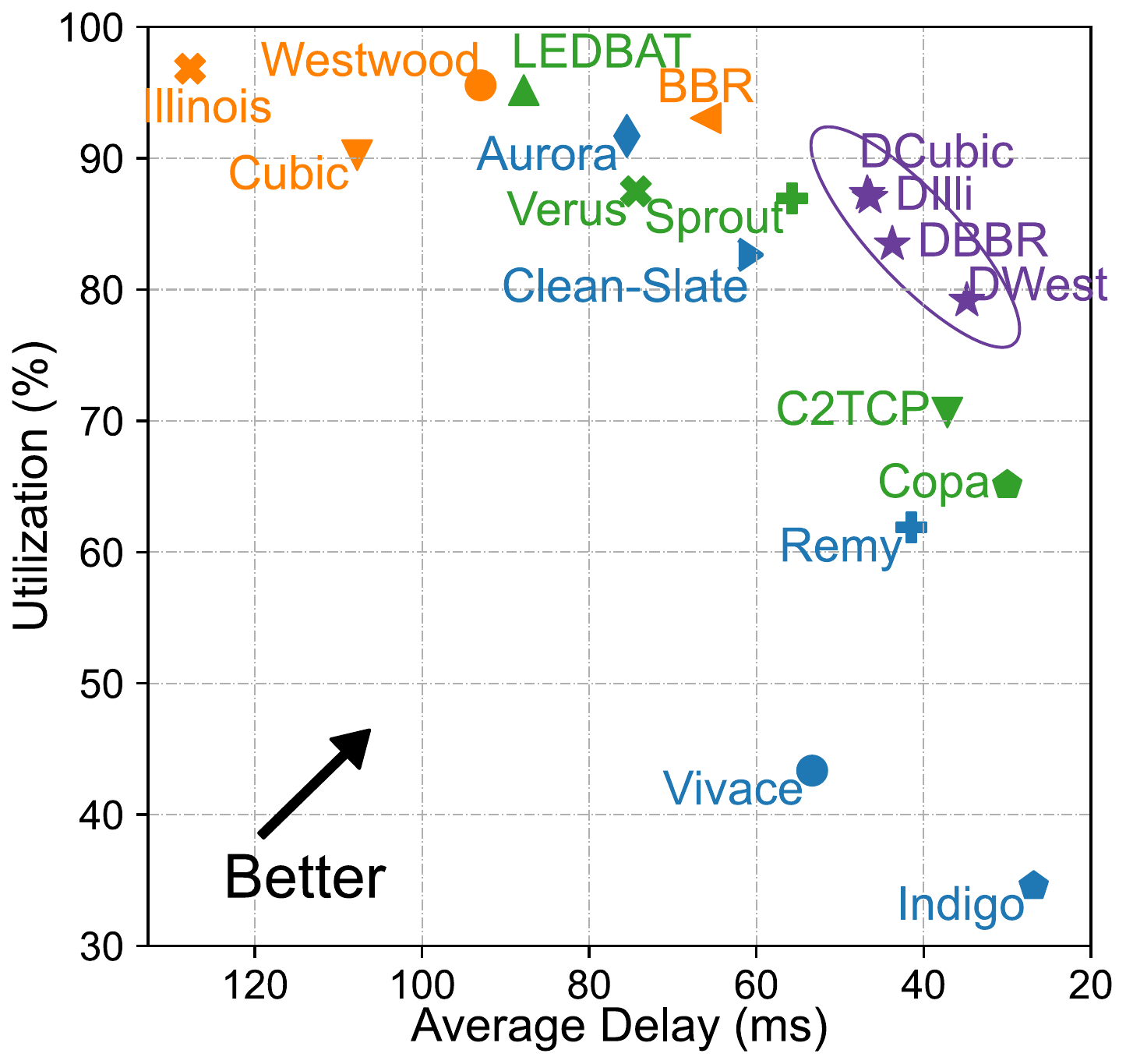}
        \subcaption{Moving scenario \#1} 
        \label{fig_mv1}
    \end{minipage}
    \hfill
    \begin{minipage}[b]{0.32\linewidth}
        \centering
        \includegraphics[width=\linewidth,height=2in]{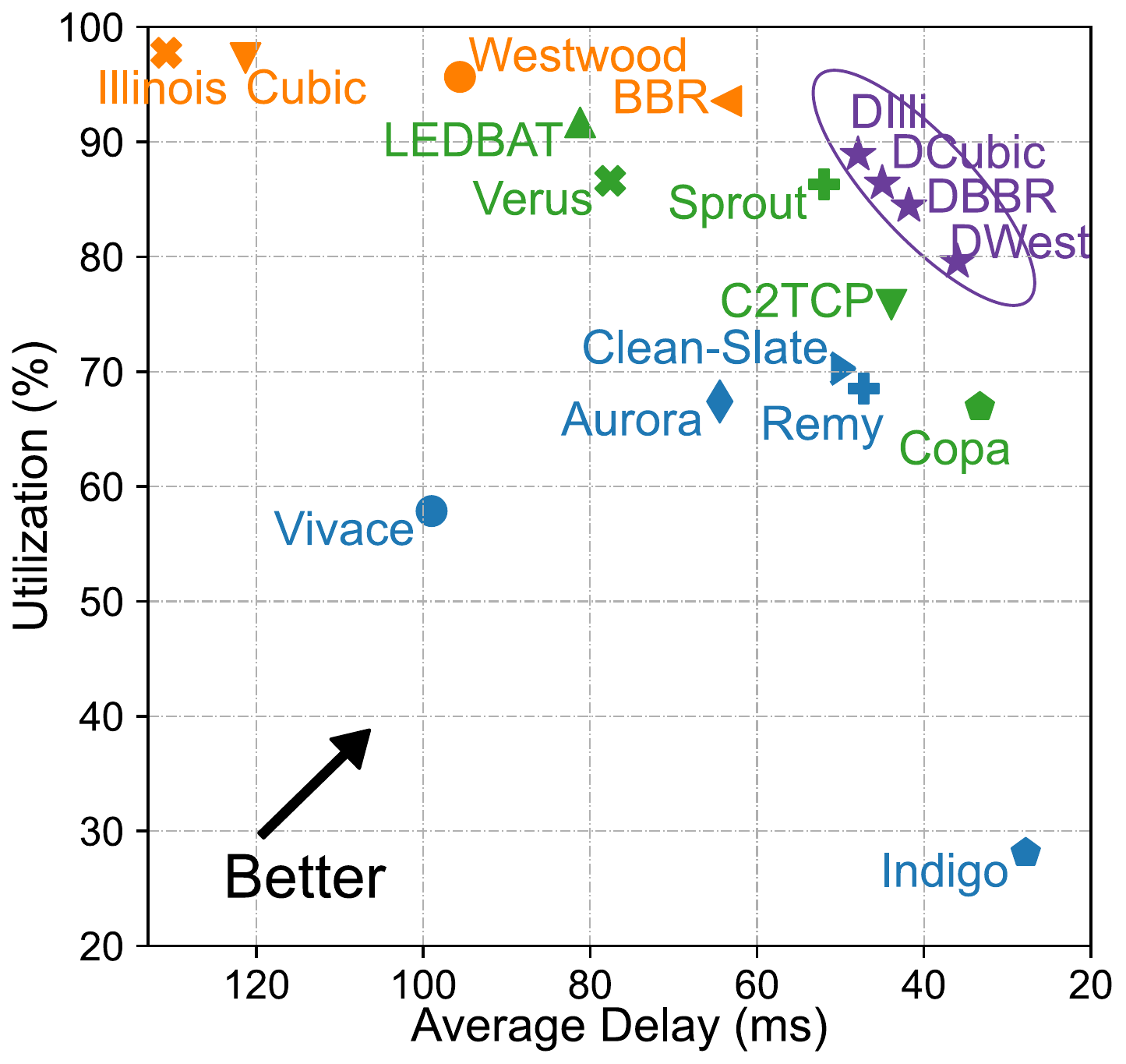}
        \subcaption{Moving scenario \#2} 
        \label{fig_mv2}
    \end{minipage}        
        \caption{Average end-to-end delay and utilization results of various schemes in different scenarios}
      \label{fig_static-mv}
\end{figure*}

To highlight the significant improvements achieved by using DeepCC plug-in for different underlying TCP schemes, we start by focusing on the averaged queuing delay 
and averaged throughput that schemes can achieve with and without using DeepCC plug-in over different scenarios. To that end, we average performance of a scheme over all environments in a scenario and report the queuing delay and throughput of a certain TCP scheme over the queuing delay and throughput of that scheme when it is accompanied by DeepCC. In other words, this ratio indicates how much performance improvements DeepCC can bring to the table. Results are shown in Fig.~\ref{fig_del_util}. 

Fig.~\ref{fig_del_util} shows clear delay improvements gained by using DeepCC plug-in for all schemes in all scenarios. For instance, DeepCC reduces queuing delay of Illinois about $4\times$ while it's throughput is only 9\% lower than Illinois in the moving scenario\#1 (Fig.~\ref{fig_mv1_q}). As another example, Dcubic causes $4\times$ lower queuing delay while it only reduces Cubic's throughput 11\% in the Moving scenario\#2.


\underline{\textit{Remark 1}}: The key takeaway from Fig.~\ref{fig_del_util} is that DeepCC significantly enhances the delay performances of all four TCP schemes while it only compromises a few percentages of their throughput. 

\subsection{Performance Results: DeepCC vs. State-of-the-Art}
Now, we compare DeepCC's performance with all other state-of-the-art schemes. Fig.~\ref{fig_static}, Fig.~\ref{fig_mv1}, and Fig.~\ref{fig_mv2} show the averaged end-to-end delay and utilization of all schemes over all cellular traces used in the stationary scenario, moving scenario \#1, and moving scenario \#2, respectively. 
We depict different classes of CC designs with a different color in Fig.~\ref{fig_static-mv}. In particular, we use orange for mainly throughput-oriented schemes, green for delay-oriented designs, blue for learning-based schemes, and violet for DeepCC enabled schemes.

\underline{\textit{Remark 2}}: DeepCC helps TCP schemes to become performance frontiers and operate close to the right top region of the graphs in Fig.~\ref{fig_static-mv}. 
As Table~\ref{table_trace_sample} illustrates, cellular LTE traces used here are very different from each other (and from the cellular environment during the training). So, although the agent trained over LTE environment with certain statistics, results reported in Fig.~\ref{fig_static-mv} illustrate that DRL-agent's trained model is not overfitted during the training and it can achieve a good generalization of the environment and perform well for different unseen LTE environments. For more discussion on generalization, see section~\ref{sec_general}. 

\underline{\textit{Remark 3}}: DeepCC outperforms its Clean-Slate version up to $2\times$ in terms of delay (Fig.\ref{fig_mv1}) and up to $20\%$ in terms of utilization/throughput (Fig.\ref{fig_mv2}). Generally, as we showed in Fig.~\ref{fig_tr}, during the training phase, Clean-Slate scheme learned a model that achieves lower performance compared to DeepCC counterparts. So, its performance was expected to remain lower than DeepCC in other unseen scenarios. Besides, in Clean-Slate version, no change occurs in the cwnd until the end of each monitoring period, because Clean-Slate scheme is the only entity controlling the cwnd. However, in DeepCC, during each monitoring period, underlying TCP actively can change the values of cwnd. This dynamism brings a more solid performance for DeepCC compared to Clean-Slate version. 

\underline{\textit{Remark 4}}: Throughput-oriented schemes such as Cubic, Westwood, and Illinois intend to fill up the buffers. As expected, this leads to their good utilization, while it causes bufferbloat/delay problems for them. 

\begin{figure*}[!t]
    \begin{minipage}[b]{.64\linewidth}
    \centering
    \begin{minipage}[b]{.48\linewidth}
        \centering
        \includegraphics[width=\linewidth,height=1.6in]{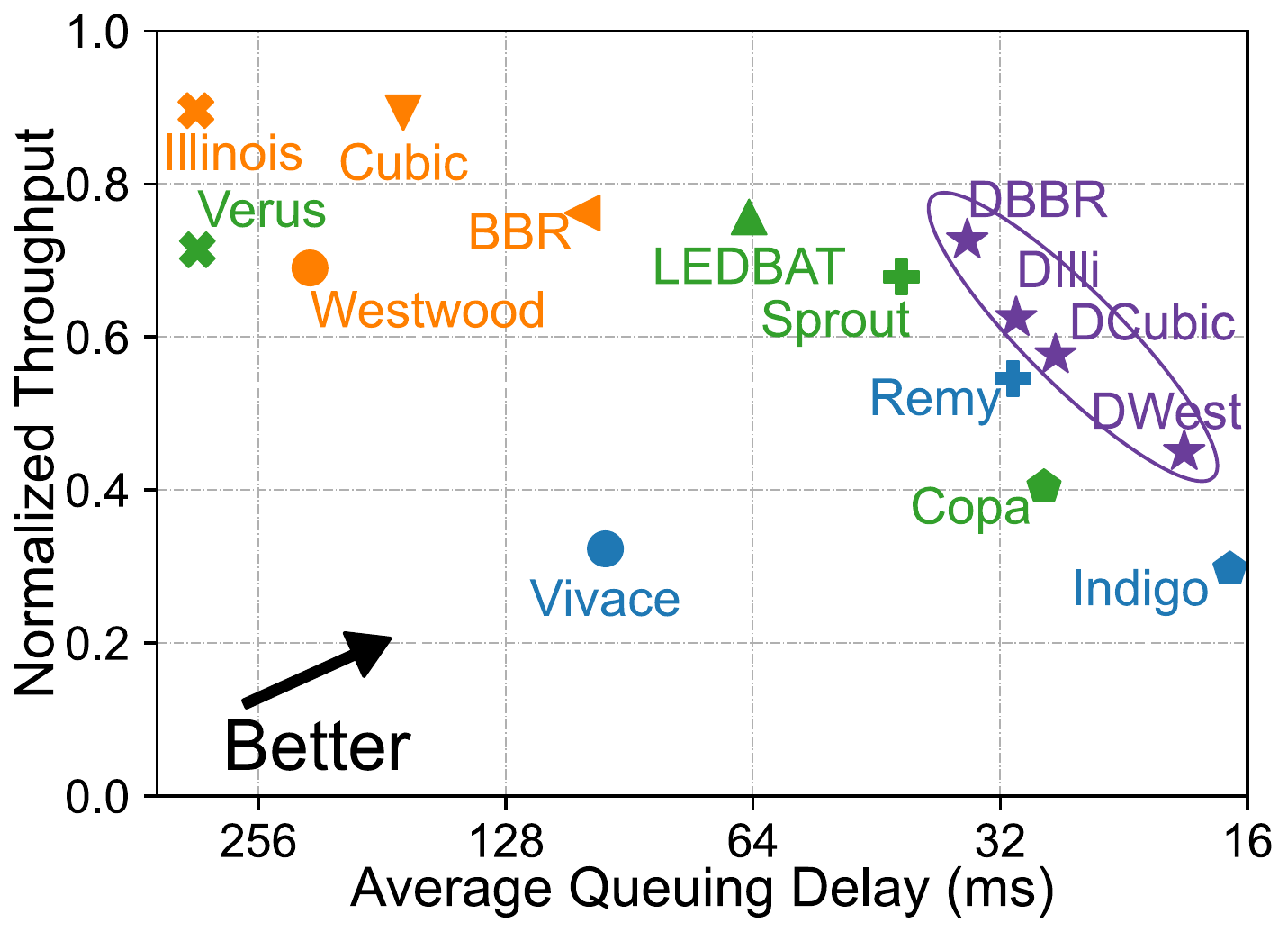}
    \end{minipage}        
    \hfill
    \begin{minipage}[b]{.48\linewidth}
        \centering
        \includegraphics[width=\linewidth,height=1.6in]{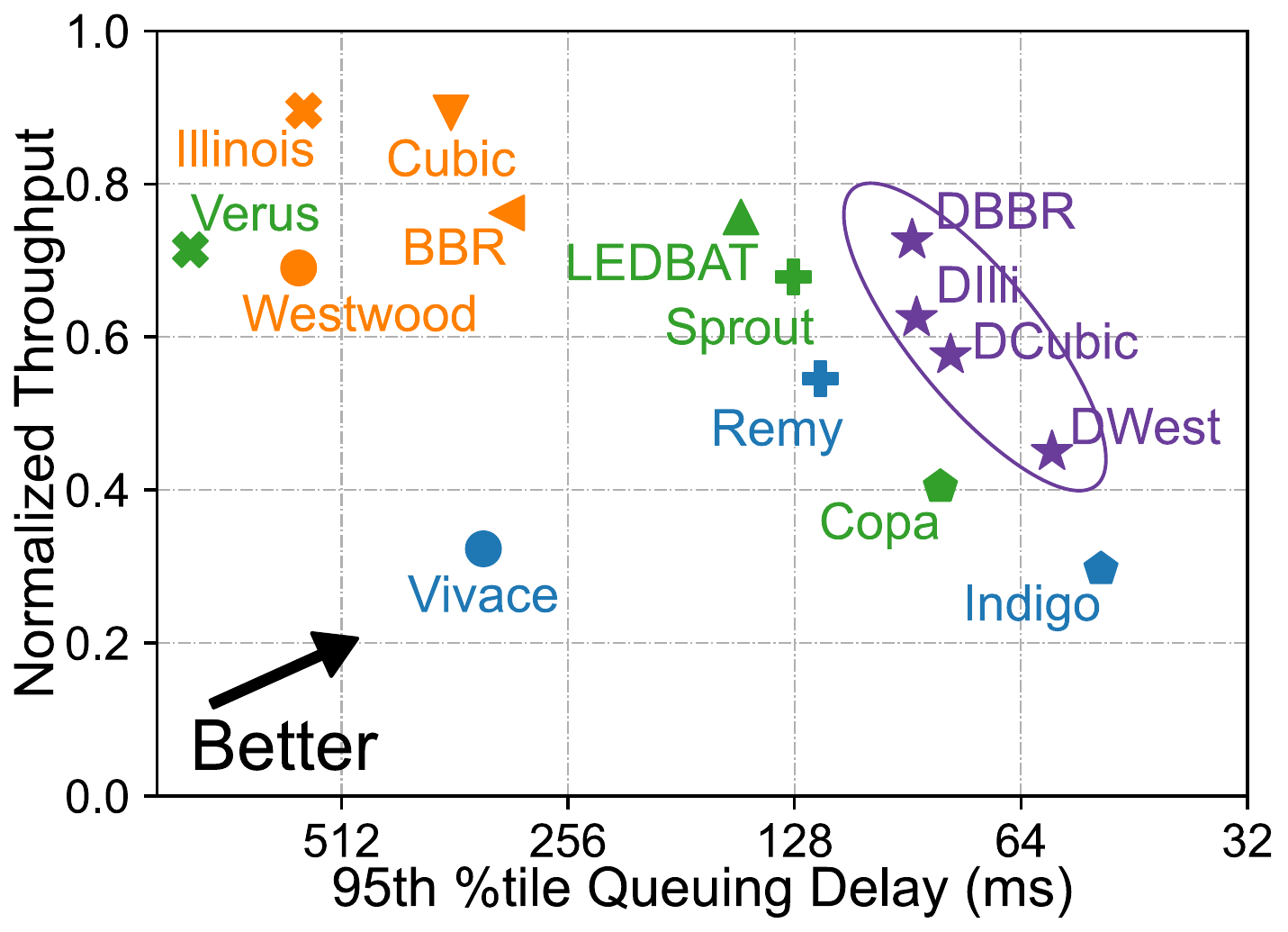}
    \end{minipage}        
    \caption{Performance results of various schemes over all in-field tests}
    \label{fig_real}
    \end{minipage}    
    \hfill
    \begin{minipage}[b]{.32\linewidth}
        \centering
        \includegraphics[width=\linewidth,height=1.5in]{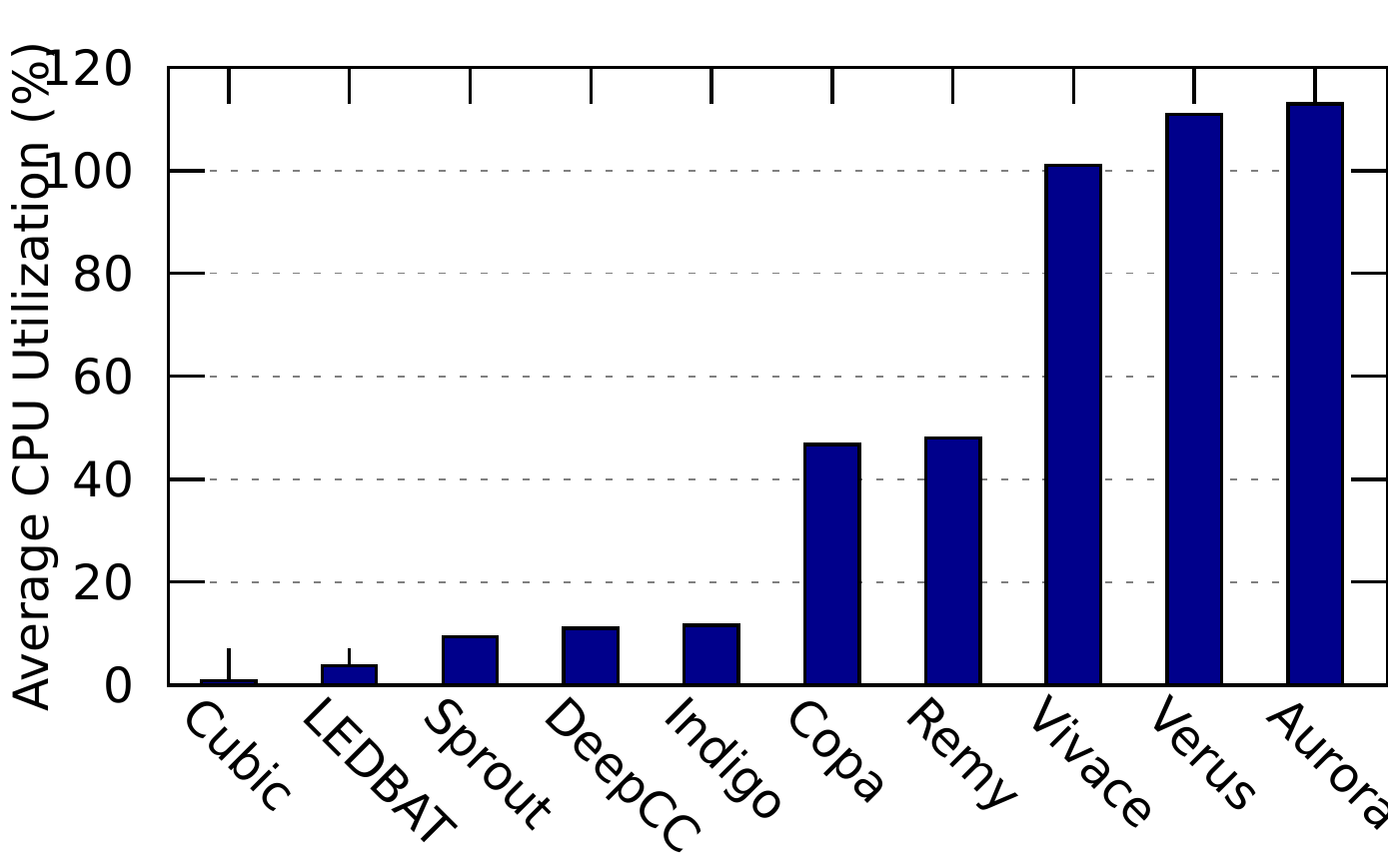}
        \caption{The average CPU utilization of various TCP schemes}
        \label{fig_cpu}
    \end{minipage}
\end{figure*}


\underline{\textit{Remark 5}}: Sprout and Verus (which are designed for cellular networks) use the delay of packets in a white-box approach to predict future packet arrivals or make a delay profile of the network respectively. BBR also uses a white-box approach. Modeling network in a certain way helps these schemes perform better in terms of delay compared to Cubic, Westwood, and Illinois. However, the unpredictable nature of the cellular environment leads to the inaccuracy of their network models. That's why at the end of the day, the white-box approach cannot keep up with the dynamics of the cellular network and delay performance of these schemes remains behind the top 6 schemes (e.g. in the stationary scenario, they achieve $\sim1.6-2.3\times$ more delay compared to DWest). 

\underline{\textit{Remark 6}}: Learning-based approaches (e.g. Aurora, Indigo, and Vivace) cannot do a good job in cellular networks. The key reasons are the fast link fluctuations and intrinsic uncertainty property of the cellular environments. These properties cause wrong interpretation of the network for these schemes leading to their poor performance. For instance, we observed that Vivace (an online learning-based approach) experiences convergence problems and most of the time it settles down on wrong sending rates in the cellular networks. Lack of mechanisms to deal with high uncertainty in highly dynamic environments makes these schemes not suitable for cellular networks. LEDBAT (which is originally designed for wired network) suffers from the same issue and cannot keep up with the dynamics of the network. Among other schemes (excluding DeepCC assisted ones!), Copa and C2TCP (which is designed for cellular networks) achieve a better trade-off between delay and throughput. 

\section{Evaluation: Real-World Experiments}
\label{sec_eval-real}
To evaluate the performance of DeepCC in the real-world where other flows coexist in the network, cellular packet schedulers are employed to schedule the traffic, and various unknown traffic policer and AQMs are used on Internet paths, we made a testbed. In particular, we employed three GENI~\cite{geni} servers (where our patched Kernels are installed on) located in New York, Indiana, and Virginia and six cellular clients (with vanilla Kernels) located in different dynamic and crowded places in New York City and New Jersey including Times Square, Herald Square, Washington Square, NYU's Bobst library, and Newport Center. Using them, we performed our tests over two different cellular LTE network providers (namely TMobile and AT\&T). We repeated each test two times. Considering all combinations of server-client pairs, the minimum RTT (mRTT) in our testbed spans from 30-70ms. For reporting an aggregated view of the performance across all experiments, we use the average normalized throughput, average queuing delay, and 95th percentile queuing delay. For each experiment, the throughput of each scheme is normalized to the flow achieving the highest throughput in that experiment.\footnote{For DeepCC, we set the Target to have 50ms delay budget for the average queuing delay.} 
Each experiment lasts 15 seconds. Half of them have one flow and the other half have three flows which start at 0, 5, and 10 seconds (after the start of the experiment). 
The overall results are shown in Fig.~\ref{fig_real}.\footnote{Since both DeepCC and C2TCP require different patched Kernels, we could not include C2TCP in our in-field tests where flows are sent back-to-back. We didn't have this problem during our traced-based tests, because we could switch from one Kernel to another one without time constraint.}

\underline{\textit{Key Takeaway}}: The relative performance of the schemes in Fig.~\ref{fig_real} is very close to the results reported in the trace-based evaluations done in section~\ref{sec_scenarios}. The bottom line is that DeepCC not only improves the performance of the classic and modern TCP schemes but also makes them the performance frontiers outperforming state-of-the-art schemes. Also, results from the real-world evaluations indicate that training over an emulated environment can provide us with models that can achieve good performances in real-world scenarios that the agent has not seen before. 

\section{Evaluation: Deep Dive}
\subsection{Overhead}
Having an extra module as a plug-in might arise the concern of its overhead on the system. As mentioned earlier, the current version of DeepCC is only a prototype/proof-of-concept version that is not fully implemented in the kernel (DRL-Agent block is implemented at user-space). This greatly impacts the overall overhead of the current version. However, here, we would like to show that using a plug-in, especially when that plug-in is based on a modern and (seemingly) complicated tool such as DRL, will not necessarily lead to high overhead on the system when it is compared to the overhead of the current state-of-the-art TCP schemes (which are also implemented at user-space). To illustrate that, we use various state-of-the-art TCP schemes and send traffic from a server to a client over an arbitrary LTE trace for about 8 minutes and measure the average CPU utilization of these schemes on the sender side. Moreover, to have a sample of a fully optimized and implemented scheme in the kernel as the base, we use Cubic Scheme (by measuring iperf's CPU utilization). As results reported in Fig.~\ref{fig_cpu} show, DeepCC has way lower overhead compared to most of the state-of-the-art schemes, though its overhead (mainly due to its user-space DRL-Agent block) is still required to be reduced. We leave the optimization of the DRL-agent and its kernel implementation to our future work.

\subsection{Under The Hood}
\begin{figure}[!h]
    \centering
    \includegraphics[width=\linewidth,height=3.6in]{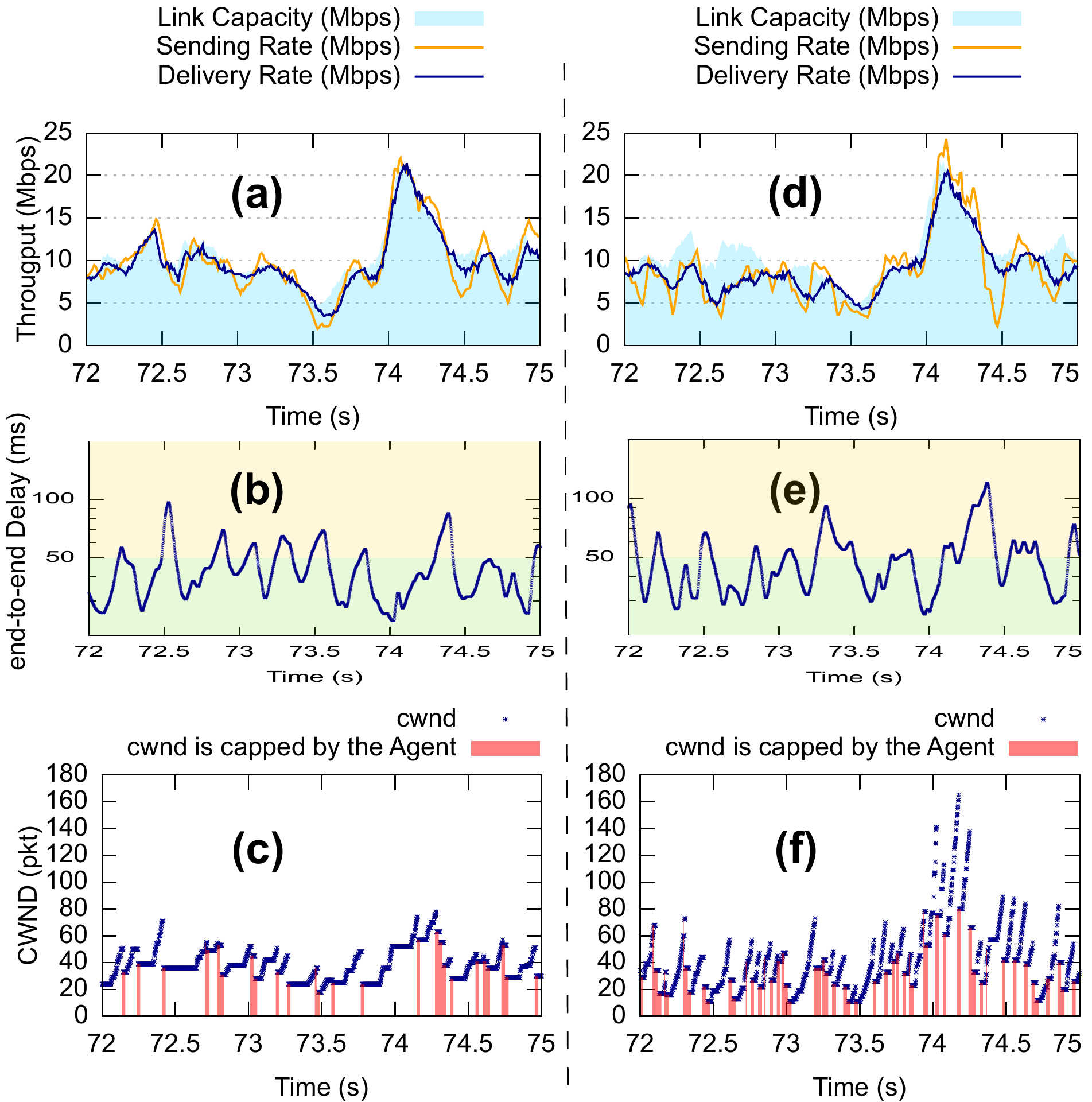}
    \caption{Dynamics of DeepCC enhanced versions of Cubic (left) and Illinois (right) for a cellular trace}
    \label{fig_uhood}    
\end{figure}

It is already a well-known fact that we still lack a theoretical understanding of many methods that are currently used in learning-based approaches, particularly in deep learning~\cite{alchamy,lecun_alchamy}. In other words, why learning-based methods work very well in practice is still an open question seeking solid theoretical answers. That being said, here, we attempt to intuitively explain how DeepCC performs. To that end, we pick an arbitrary cellular trace and send traffic (following the setting in section~\ref{sec_scenarios}) over that using DCubic and Dillinois schemes and record the variations of cwnd, $cwnd_{max}$, delay, and throughput. To have a better picture, we zoom in to an arbitrary three seconds of the evaluation (Fig.~\ref{fig_uhood}). Generally, there are two key things that DeepCC learns: 1) when (how frequently) to cap the cwnd? and 2) by how much? Note that at every monitoring time period, DeepCC observes 100 ($=5\times 20$) different input statistics. Following all of them to find the exact answers to the mentioned two questions is impractical here. 

\begin{figure*}[!h]
    \centering
    \begin{minipage}[b]{0.32\linewidth}
        \centering
        \includegraphics[width=\linewidth,height=1.8in]{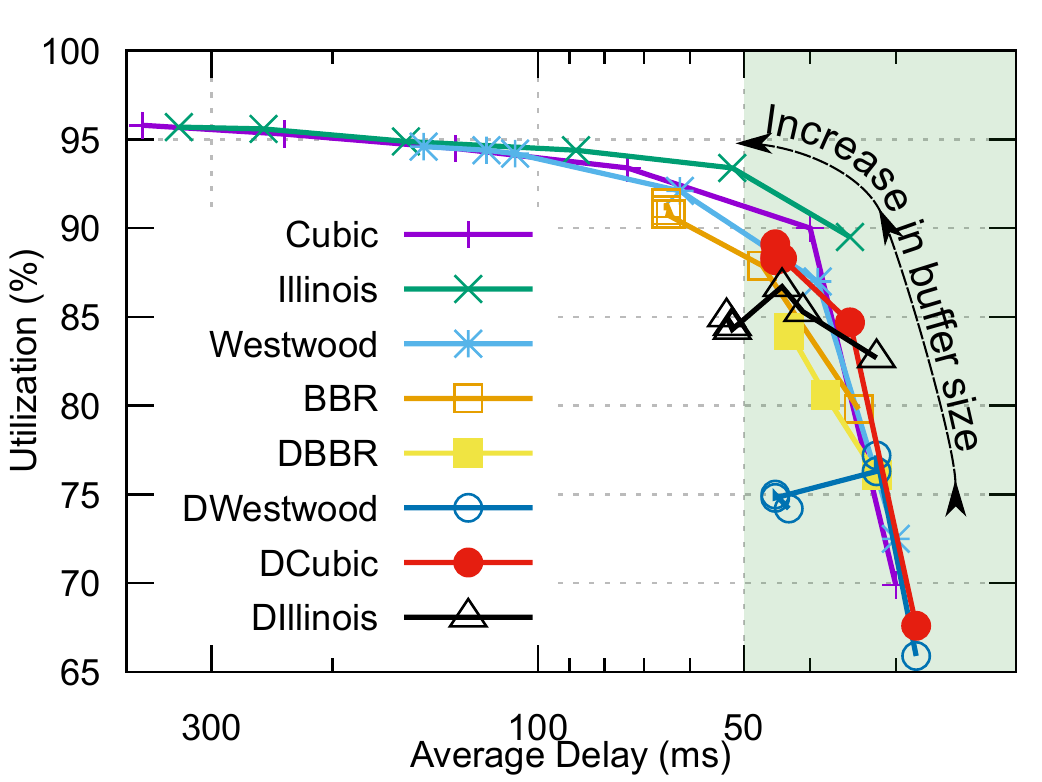}
        \caption{Impact of buffer size on Performance}
        \label{fig_q}
    \end{minipage}        
    \hfill
    \begin{minipage}[b]{0.32\linewidth}
        \centering
        \includegraphics[width=\linewidth,height=1.8in]{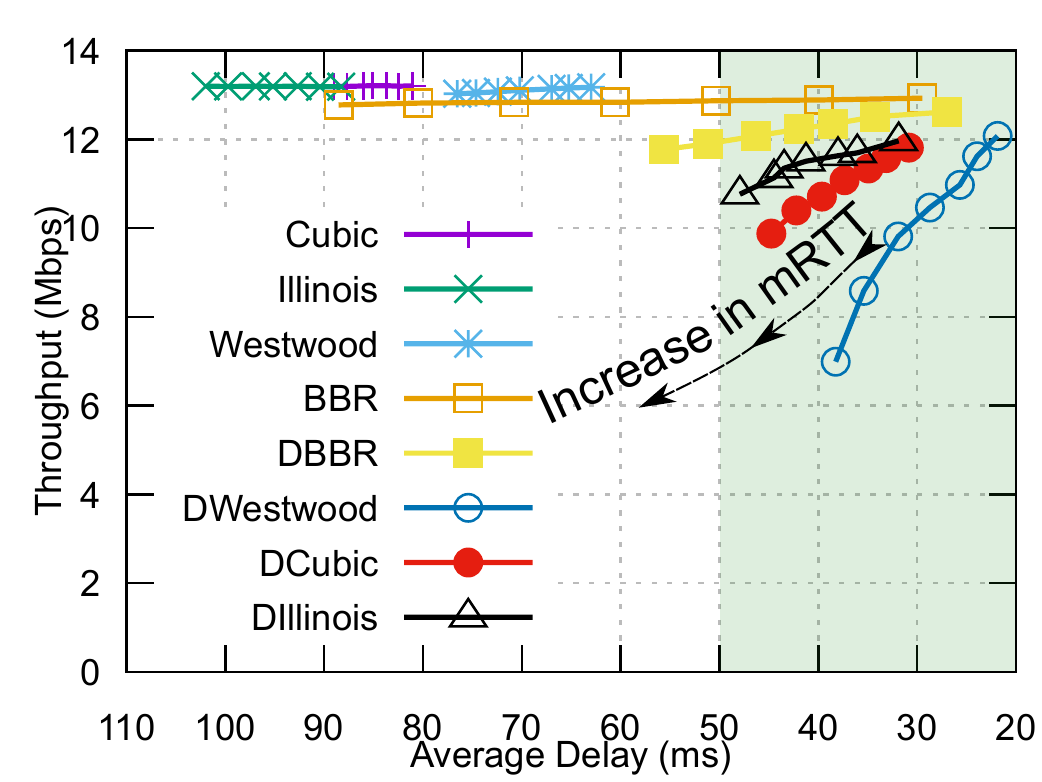}
        \caption{Impact of mRTT on Performance}
        \label{fig_mrtt}
    \end{minipage}
    \hfill
    \begin{minipage}[b]{.32\linewidth}
        \centering
        \includegraphics[width=\linewidth,height=1.8in]{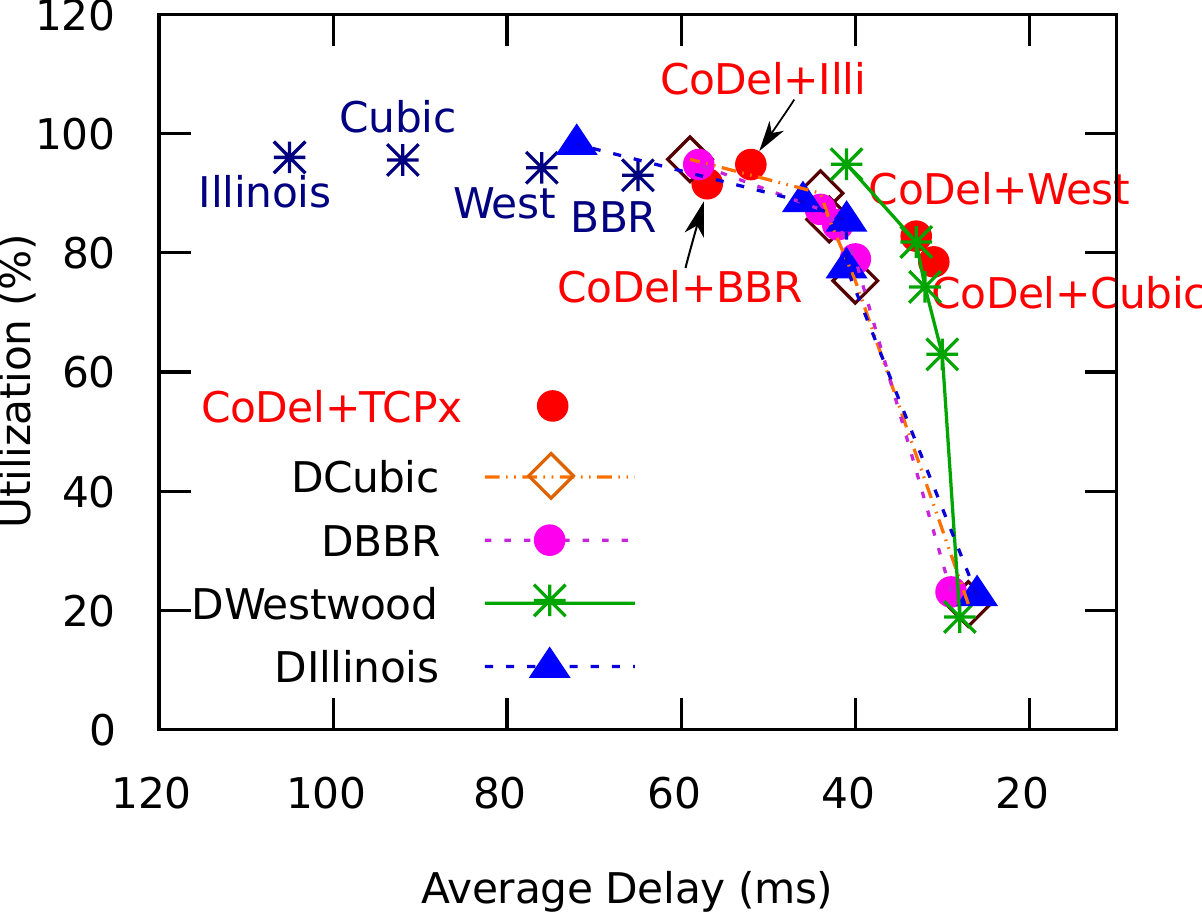}
        \caption{Schemes with/w.o DeepCC with diff. Targets}
        \label{fig_aqm}
    \end{minipage}
\end{figure*}

How frequently does DeepCC cap the cwnd? It depends on the underlying TCP's behavior. For instance, DeepCC allows Cubic to control the growth of cwnd for most of the times (blue dots in Fig.~\ref{fig_uhood}.c), while occasionally it caps the cwnd to lower values than the ones calculated by Cubic (red shaded regions in Fig.~\ref{fig_uhood}.c) to keep the average delay below the Target value (50ms in this case). For example, during $[73.5s,74.25s]$, when available link capacity is increased, DeepCC only caps cwnd 3times at 73.6s, 73.8, and 74.2s (Fig.~\ref{fig_uhood}.c). The first 2 times partly happen because DeepCC does not observe any improvement in the gained throughput (note that around 73.6s and 73.8s, link capacity (blue shaded region in Fig.~\ref{fig_uhood}.a) temporarily remains constant), while the 3rd time partly happens because available capacity suddenly drops at 74.2s. Also, note that DeepCC chooses a different amount of reduction in cwnd in these 3 events, due to the different dynamics of its input signals.

On the other hand, when underlying TCP is Illinois, DeepCC caps cwnd of the system more often, because generally as Fig.~\ref{fig_uhood}.f illustrates, cwnd growth of Illinois is more aggressive compared to Cubic's scenario. So, DeepCC has learned to control it more frequently to effectively manage its bufferbloat issue. Also, Fig.~\ref{fig_uhood} depicts that in some cases, similar to $[72.6s,73s]$ duration in Fig.~\ref{fig_uhood}.d, DeepCC can behave conservatively to control the underlying TCP's delay and compromise a bit of throughput. 

\subsection{Impact of Buffer Size on Performance}
DeepCC tries to resolve the bufferbloat issue of the underlying TCP scheme and to keep the average delay of packets around the desired value no matter how much buffer exists in the network. Therefore, it is expected to have very low sensitivity to the size of the buffer in the network. This property of DeepCC helps the loss-based TCP schemes (which normally fill up the buffer until they see a packet loss) achieve controlled self-inflicted delays in the cellular networks independent of the buffer sizes in the network. To show that, we vary the buffer size from 30KB to 1MB and explore the performance of Cubic, Westwood, Illinois, BBR, and their DeepCC counterparts across various buffer sizes. Fig.~\ref{fig_q} shows the results. As expected, DeepCC can effectively control the delay of underlying schemes below target (green region), while without DeepCC, increasing buffer size leads to high delays of the schemes under the tests.

\subsection{Impact of mRTT on Performance}
In this paper, we trained DeepCC in a cellular environment with a fixed mRTT ($20ms$). Here, we show that the learned model can still perform well when environments' mRTT differs compared to the training setting, though training over multiple mRTT can improve the learned model. To that end, we arbitrarily pick one of our traces and change the mRTT from 4ms to 30ms, while fixing the Target to 50ms and record the performance of different schemes. Results are shown in Fig.~\ref{fig_mrtt}. Without using DeepCC, TCP schemes either show low sensitivity to the changes of mRTT, because they have already reached very high delays (e.g. Cubic, Illinois, and Westwood), or show high sensitivity to the values of mRTT that can end up to having high delays (e.g. BBR). In contrast, When DeepCC is enabled, these schemes show low sensitivity to mRTT and DeepCC can effectively keep their performance in the desired operation range (green region). Note that with a fixed Target, for large mRTT values, there is a lower budget for queuing delay. So, achieving that lower queuing delay requires compromising more throughput.

\subsection{A Flexible End-to-End Scheme vs. an In-Network AQM Scheme}
\label{sec_target}
Here, we compare the performance of our fully end-to-end scheme with a delay-centric in-network AQM scheme CoDel~\cite{codel}. We use various TCP schemes at sender combined with CoDel as AQM scheme in the network (CoDel+TCPx), While for other schemes we use normal FIFO queues. Moreover, here, we examine the flexibility of DeepCC for providing various Target delays by varying Target delay of the system. Although we only used one Target value during the training of our model, we show that the trained model performs well for the other Targets as well. More training using more Target values will increase the accuracy and performance of the model. However, we deliberately decided to use only one Target to see how well DRL-agent can generalize the environment. So, without loss of generality, we use one of our traces and change the Target value from 25ms to 100ms (with mRTT=20ms). The average delay and utilization results are shown in Fig.~\ref{fig_aqm}. CoDel can improve the performance of TCP schemes very well. However, CoDel similar to other AQM designs such as BoDe~\cite{bode} and PIE~\cite{pie} requires changes in the network devices which leads to having extra CAPEX costs for the cellular network providers. On the other hand, DeepCC as a fully end-to-end and deployment-ready approach, which does not require any changes in the network, can improve classic schemes such as Westwood to even outperform the performance of new TCP schemes (e.g. BBR and Cubic) combined with a modern in-network AQM schemes such as CoDel. 
\subsection{Impact of Filter Kernel on Performance}
\begin{figure}[!b]
    \begin{minipage}[b]{\linewidth}
        \centering
        \begin{minipage}[b]{0.49\linewidth}
            \centering
            \includegraphics[width=\linewidth,height=1.4in]{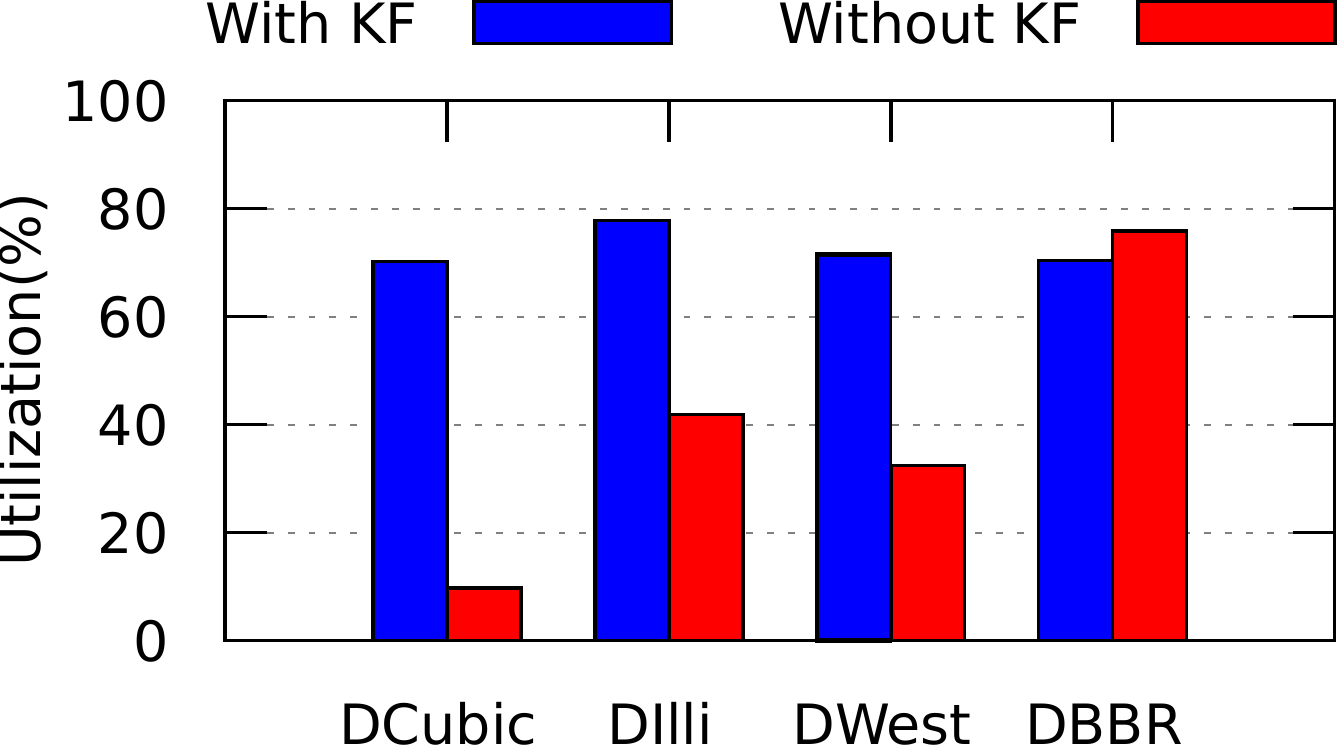}
        \end{minipage}        
        \hfill
        \begin{minipage}[b]{0.49\linewidth}
            \centering
            \includegraphics[width=\linewidth,height=1.4in]{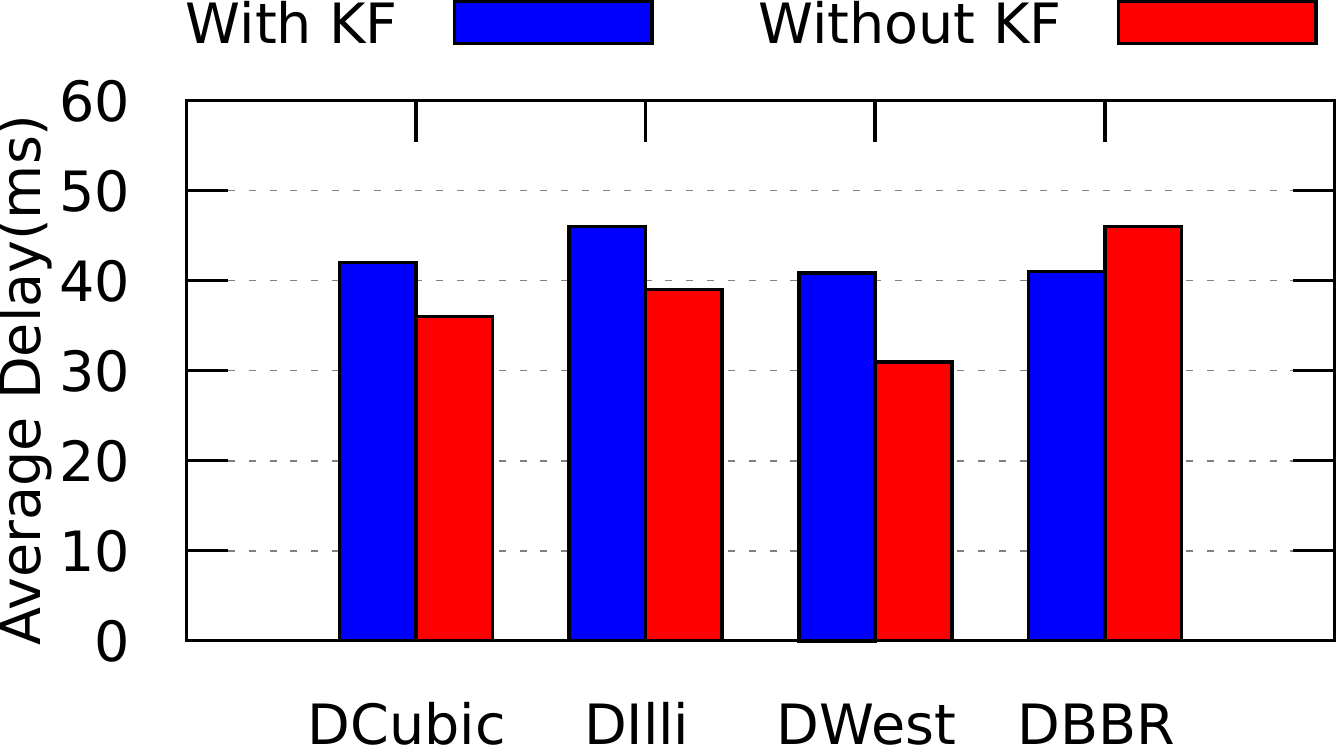}
        \end{minipage}
        \caption{The util. (left) and the delay (right) of schemes after training with/w.o using filter kernels}
        \label{fig_fk}
    \end{minipage}
\end{figure}

In our context, there are two fully dynamic entities: 1) cellular network itself and 2) behavior of the underlying TCP scheme which make the learning phase normally a very long process. In such a dynamic environment the use of approaches such as filter kernel will guide the system toward a faster and more efficient training phase.
To investigate the impact of our filter kernel (detailed in section~\ref{sec_kf}) on the performance of DeepCC, we compare DeepCC with another version of it which simply uses the raw state inputs (without filtering them). We use the new version to train four TCP schemes (for the same amount of time used for training schemes in section~\ref{sec_training}) following the instructions provided in section \ref{sec_training}. After the training phase, we compare their performance results with the results of their DeepCC counterparts (using the filter kernel) over an arbitrary cellular trace. As expected, results shown in Fig.~\ref{fig_fk} confirm that using filter kernel in DeepCC plug-in can lead to significant performance benefits. 

\subsection{Fairness}
To investigate the impact of DeepCC on the fairness property of the underlying scheme, we use 4 servers to send 4 separate DeepCC enhanced flows to the client (servers are connected to the client through a switch). 
We use Mahimahi at the client to control the client's bottleneck link properties. Particularly, we set the client's bandwidth to 24Mbps, the unidirectional link delay to 5ms, and the buffer size to the BDP (bandwidth-delay product).
Note that in cellular networks, flows can be put into separate bearers identifying separate classes with different delay requirements to avoid a bandwidth-hungry application taking entire bandwidth from a delay-sensitive application, if both are destined to the same user~\cite{lte-book}. Therefore, here, we consider that flows are in the same class and have the same Target delay. 
We use Jain index metric~\cite{jain} to quantify the fairness and compare it among various schemes with and without DeepCC. Jain index, $\mathcal{J}$, for $n$ competing flows with rates $(r_1, r_2, \dots, r_n)$ is defined by $\mathcal{J} (r_1, r_2, \dots, r_n) =\frac{\overline{\mathbf{r}}^2}{\overline{\mathbf{r}^2}}= \frac{( \sum_{i=1}^n r_i )^2}{n \times \sum_{i=1}^n {r_i}^2}$.
As this equation reveals, Jain Index is a number between $1/n$ (worst index) and 1
(best index). 
To do the evaluation, we fix number of competing flows and the scheme under the test. Then we let all flows start at the same time and give them 60s to settle down on the bottleneck link. At the end, we calculate the Jain indices. The Jain indices of various schemes with and without DeepCC are summarized in Table~\ref{table_jindex}. 
\begin{table}[!b] \renewcommand{\arraystretch}{1} \caption{Jain index of different schemes with and without using DeepCC plug-in} 
\label{table_jindex} \centering
\begin{tabular}{c|c|c|c}
Scheme             & 2 flows    &   3 flows & 4 flows \\ \hline
\rowcolor{ACMLightBlue}
DIllinois            & 0.985 & 0.976 &0.977  \\ 
Illinois            & 0.738 & 0.868    &0.761  \\ \hline
\rowcolor{ACMLightBlue}
DWestwood            & 0.982 & 0.980 &0.974  \\ 
Westwood            & 0.987 & 0.986    &0.978  \\ \hline
\rowcolor{ACMLightBlue}
DCubic                & 0.969 & 0.953 &0.936  \\ 
Cubic                & 0.973 & 0.941    &0.918  \\ \hline
\rowcolor{ACMLightBlue}
DBBR                & 0.967 & 0.979 &0.981  \\ 
BBR                    & 0.855 & 0.953    &0.833  \\ \hline
\hline
\end{tabular}
\end{table}

As Table~\ref{table_jindex} illustrates, 
the DeepCC enhanced schemes' fairness property is at least as good as the underlying TCP schemes' fairness. Discussions around Fig.~\ref{fig_uhood} can shed light on the possible reasons. When DeepCC observes that a certain scheme has less aggressive cwnd growth, it allows the underlying TCP scheme to control the calculation of cwnd for most of the time, while occasionally it overrules the value of cwnd. Therefore, if an underlying TCP scheme is already a fair protocol, most likely after using DeepCC, it still remains a fair protocol (e.g. Cubic and Westwood in Table~\ref{table_jindex}). However, if the underlying TCP is aggressive (so, potentially shows lower fairness (e.g. Illinois in Table~\ref{table_jindex})), DeepCC acts more and caps the cwnd values more frequently (as shown in Fig.~\ref{fig_uhood}.f). So, this can provide room for other flows to potentially have more chances to compete with that flow. In these cases, using DeepCC can lead to tangible fairness index improvements. 

\subsection{Non-Cellular Bottleneck}
\label{sec_non-cel}
\begin{figure}[!t]
    \centering
    \begin{minipage}[b]{0.49\linewidth}
        \centering
        \includegraphics[width=0.95\linewidth,height=1.1in]{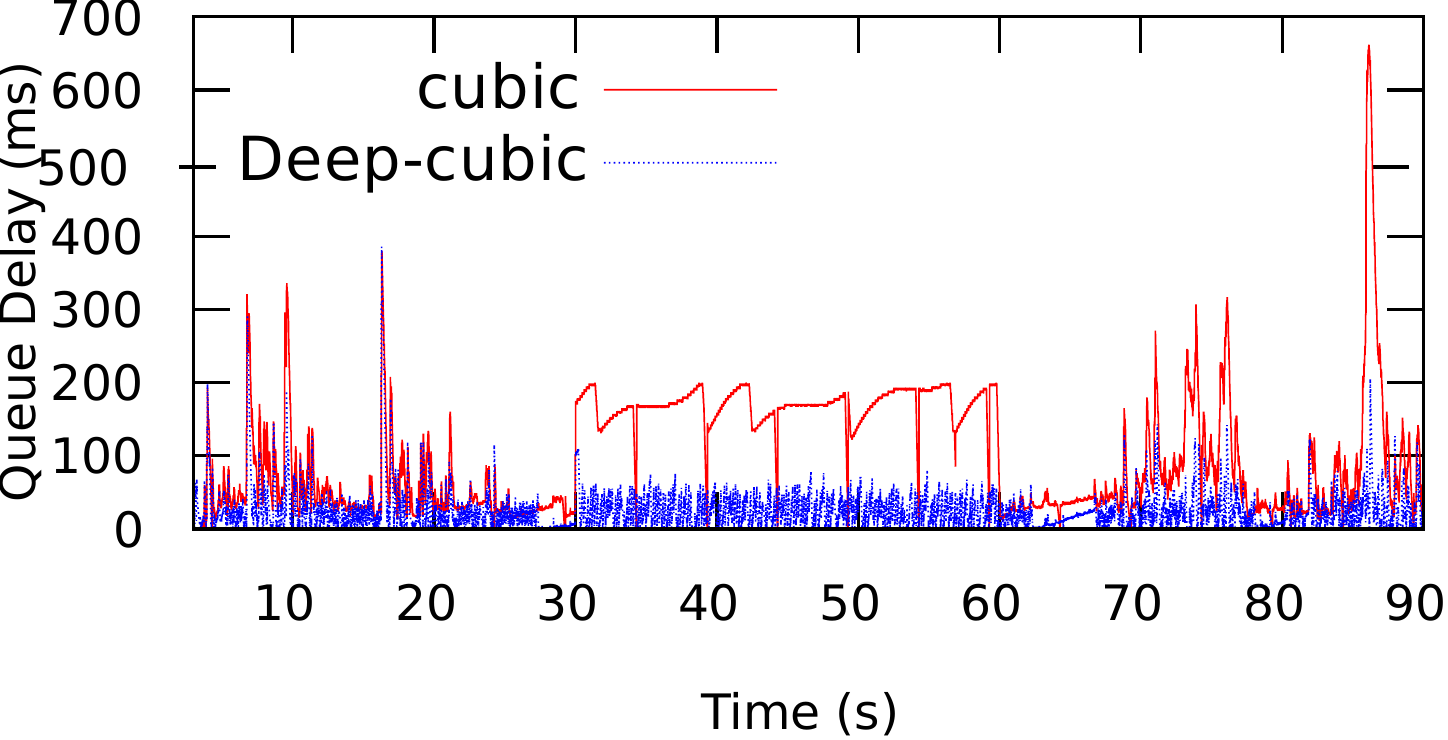}
    \end{minipage}        
    \begin{minipage}[b]{0.49\linewidth}
        \centering
        \includegraphics[width=0.95\linewidth,height=1.1in]{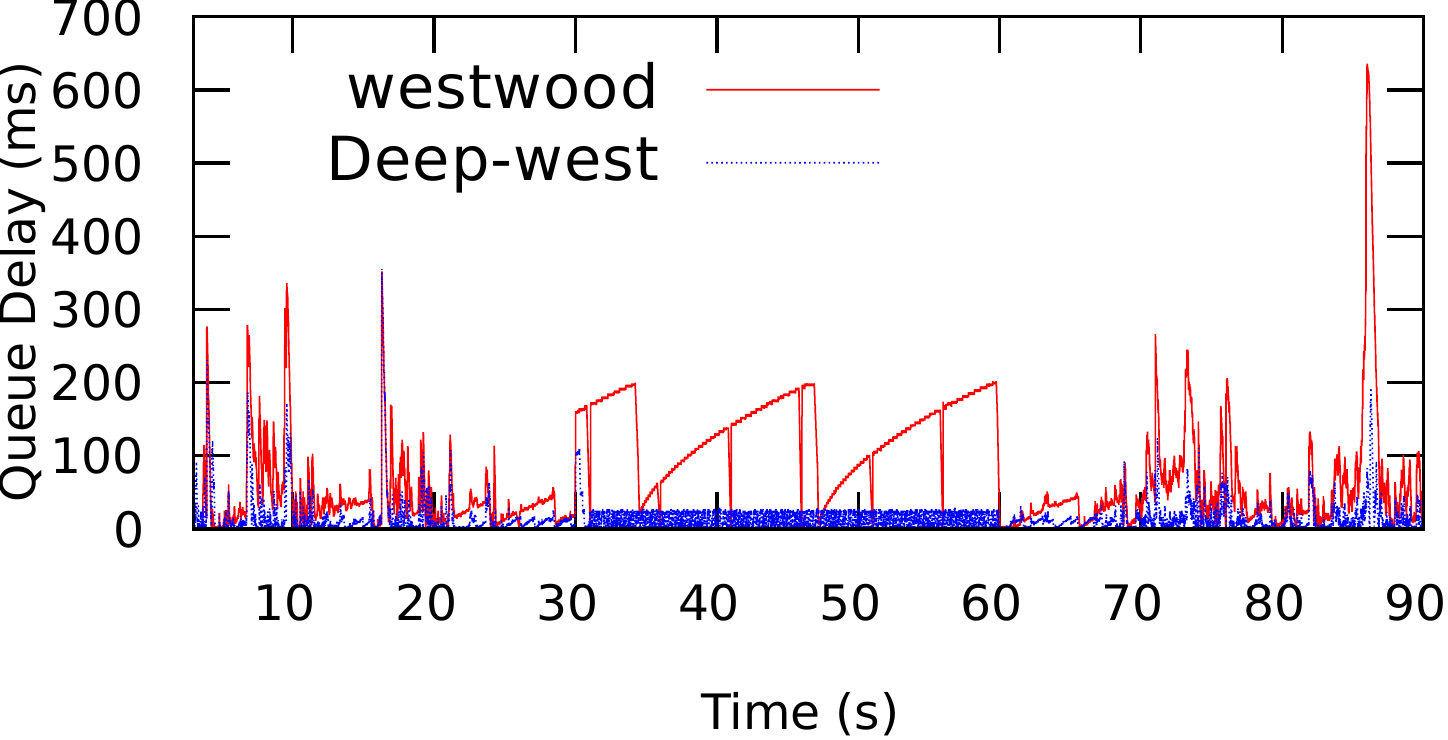}
    \end{minipage}        
    \begin{minipage}[b]{0.49\linewidth}
        \centering
        \includegraphics[width=0.95\linewidth,height=1.1in]{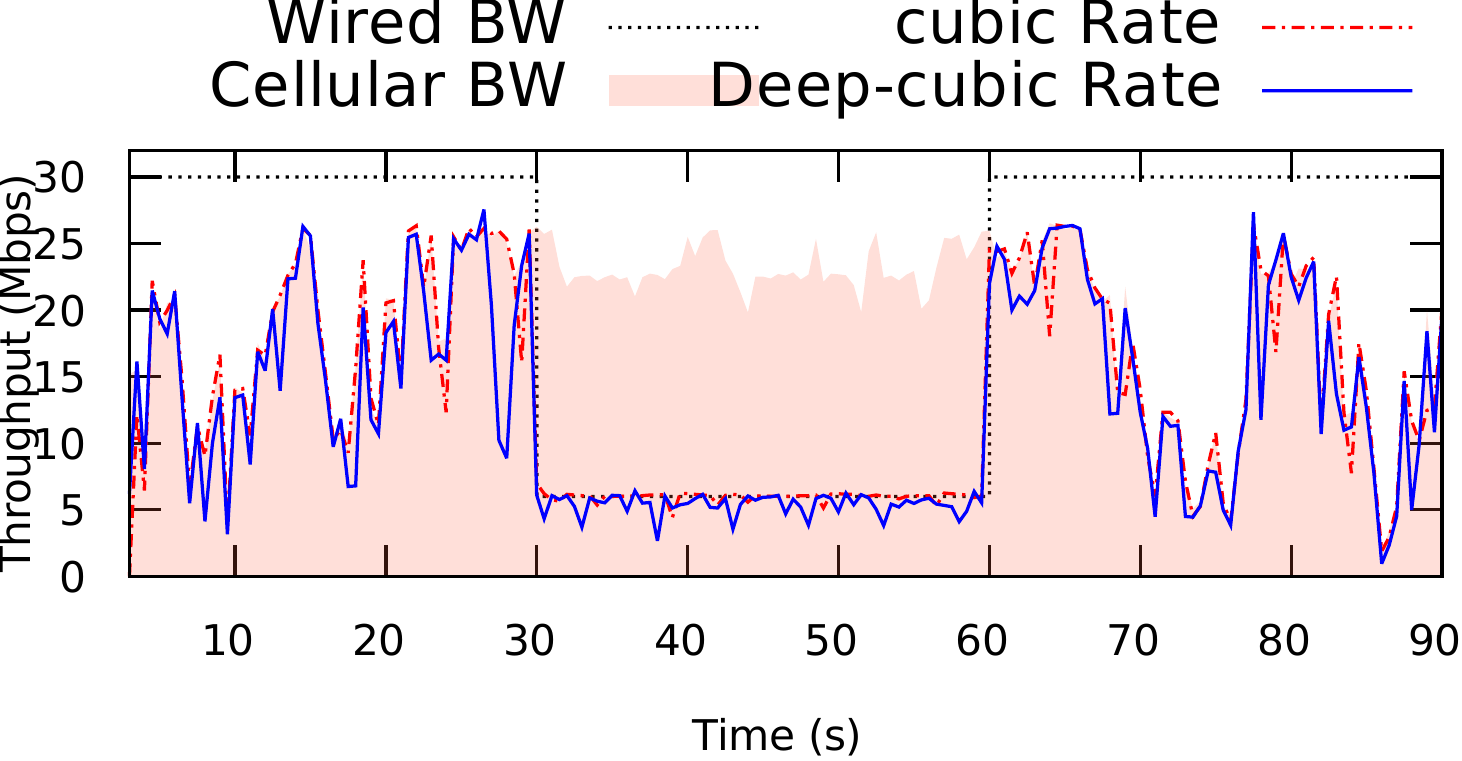}
    \end{minipage}        
    \begin{minipage}[b]{0.49\linewidth}
        \centering
        \includegraphics[width=0.95\linewidth,height=1.1in]{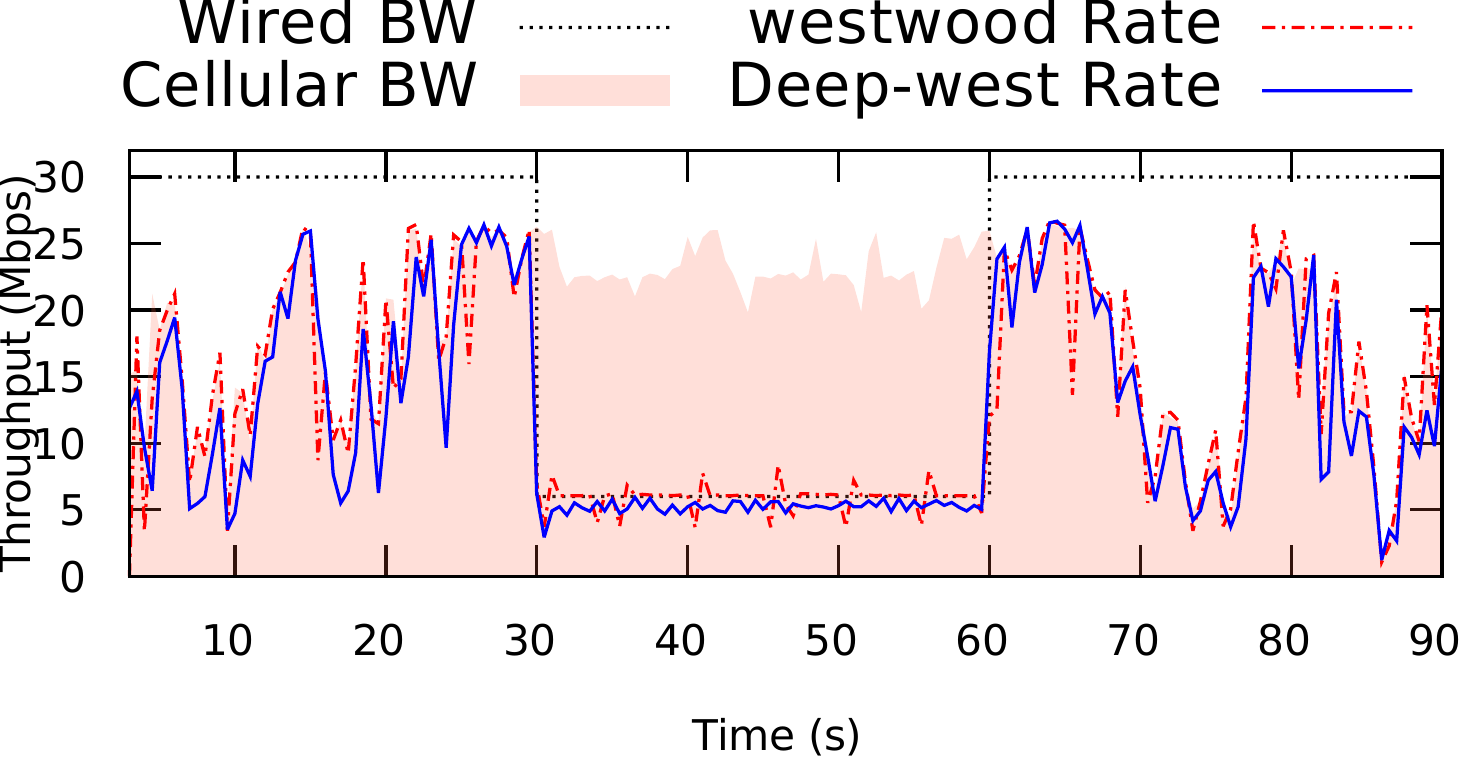}
    \end{minipage}        
    \caption{The performance when bottleneck link switches between cellular and non-cellular links}
    \label{fig_noncel}
\end{figure}

In recent trends and cellular architectures data is pushed very close to the end-users (e.g. MEC~\cite{mec}, Mobile CDN, etc.), so the assumption that cellular access links are the bottleneck links in the network is valid. That being said, in this section, we investigate the performance of DeepCC in scenarios where the cellular link is not the bottleneck link. In the first scenario, we throttle the bandwidth of a wired link located on the path of the traffic from a server to a cellular client to make a non-cellular bottleneck link. In particular, we throttle the wired bandwidth from 30Mbps (which is higher than the maximum cellular link trace's bandwidth used during the test) to 6Mbps (which is less than the used cellular access link's bandwidth during a particular period (check Fig.~\ref{fig_noncel})). After 30 seconds, we roll back the bandwidth of the wired link to 30Mbps. The delay and throughput of schemes with and without using DeepCC plug-in are shown in Fig.~\ref{fig_noncel} (Due to space limitation and the fact that results for BBR and Illinois are similar to the results for Cubic and Westwood, here, we only report the graphs for Cubic and Westwood). As Fig.~\ref{fig_noncel} illustrates, DeepCC performs very well even when bottleneck link changes among cellular and non-cellular links. 

\section{Discussion}
\label{sec_dis}

\subsection{Generalization and Transfer Concern}
\label{sec_general}
Both traced-based evaluations and real-world experiments in section~\ref{sec_eval} indicate that DeepCC learned the policy that performs well in different cellular LTE scenarios. Although performance results are promising, understanding, formulation, measurement, and improving generalization in the context of RL (and deep learning) and whether a learned model can be successfully transferred are still active research topics in machine learning community~\cite{G5,G6,G7,G2,G3,G4,G1}.

\subsection{Does DeepCC guarantee the delay performance?} After 3 decades of active research, none of the proposed CC designs can provide performance guarantees over uncertain network conditions of the Internet. That is partly due to the nondecentralized nature of Power (see section ~\ref{sec_related}). Also, to have a guarantee on delay performance, it is already proven that admission control is a must have feature in the network~\cite{sharpedge}. In other words, without having admission control by network operators and by simply using an end-to-end TCP protocol, guaranteeing a delay for the end-users is not feasible. Besides, Internet and its underlying packet-switch architecture are evolved on the basis of having a \textit{best-effort} nature. That said, the essence of having the objective of achieving/having performance guarantees for an end-to-end CC scheme which at the end of the day runs on a best-effort medium (i.e., Internet) is debatable.

\subsection{Can DeepCC Be Used in Other Networks?}
Although our framework is general and can be used in other networks, the objective that we chose for the current version of DeepCC suites cellular networks and not a general network. The objective of meeting applications' desired target delays is feasible in today's cellular networks due to their distinguishing key characteristics. For instance, combination of having different bearers for different QoS requirements and isolation of different clients' traffic from each other (through using per-client separate queues at BTS) make it possible to avoid the competition among multiple flows with contradicting objectives such as achieving ultra-low latency for one flow and achieving very high throughput for another one. Also, emerging solutions such as 5G's network slicing~\cite{5g_slicing} can greatly help this isolation. Another unique property of cellular networks is the existence of wireless scheduler at BTS to directly control and resolve the fairness issue among flows and make inter-protocol fairness a less relevant concern therein. So, it can be seen that without these properties, no TCP scheme can achieve ultra-low latency for a flow when this flow coexists with a throughput-hungry flow (which is not necessarily using the same TCP scheme). 




\subsection{Setting Target Delay}
Different approaches can be used to let applications set the target delay. 
For instance, instead of giving total freedom to applications, they may just choose which ``class'' of applications they belong to. Then, based on the predefined class properties, target delays can be set at the underlying layers. 
An important point is that as we showed in Fig.~\ref{fig_design}, DeepCC provides continuous packet statistics to the applications. So, applications can even adjust the target dynamically based on the available statistics. For example, if mRTT of the network (part of available statistics) is larger than their desired target, they can change target accordingly during the life of a session. Also, they can set a relative value for the target considering mRTT of the network. For instance, applications can specify $Target=2\times mRTT$.

\subsection{What If an Application Only Cares About Throughput?}
Part of the flexibility of having a TCP plug-in such as DeepCC is that it can be disabled/enabled. We have made new socket options to enable/disable DeepCC per connection. As we showed in section~\ref{sec_eval}, when throughput is considered as the only performance metric, current throughput-oriented TCP schemes already perform very well in cellular networks. So, a throughput-oriented application such as web page download can simply disable DeepCC plugin and roll-back to default TCP, if it prefers.



\section{Final Note}
Instead of proposing yet another new TCP scheme, we presented a new design direction and demonstrated that machines can automatically learn to significantly boost the performance of existing TCP designs in highly dynamic networks. 
We hope that DeepCC's framework and its performance benefits motivate the community toward the design of more \textit{plug-in}-based approaches benefiting current/future protocols to alleviate congestion, particularly in a highly dynamic environment such as the cellular network and for emerging delay-centric applications and services.

\bibliographystyle{IEEEtran}

\begin{thebibliography}{10}
\providecommand{\url}[1]{#1}
\csname url@samestyle\endcsname
\providecommand{\newblock}{\relax}
\providecommand{\bibinfo}[2]{#2}
\providecommand{\BIBentrySTDinterwordspacing}{\spaceskip=0pt\relax}
\providecommand{\BIBentryALTinterwordstretchfactor}{4}
\providecommand{\BIBentryALTinterwordspacing}{\spaceskip=\fontdimen2\font plus
\BIBentryALTinterwordstretchfactor\fontdimen3\font minus
  \fontdimen4\font\relax}
\providecommand{\BIBforeignlanguage}[2]{{%
\expandafter\ifx\csname l@#1\endcsname\relax
\typeout{** WARNING: IEEEtran.bst: No hyphenation pattern has been}%
\typeout{** loaded for the language `#1'. Using the pattern for}%
\typeout{** the default language instead.}%
\else
\language=\csname l@#1\endcsname
\fi
#2}}
\providecommand{\BIBdecl}{\relax}
\BIBdecl

\bibitem{tahoa}
V.~Jacobson, ``Congestion avoidance and control,'' in \emph{ACM SIGCOMM CCR},
  vol.~18, no.~4.\hskip 1em plus 0.5em minus 0.4em\relax ACM, 1988, pp.
  314--329.

\bibitem{bbr}
N.~Cardwell, Y.~Cheng, C.~S. Gunn, S.~H. Yeganeh, and V.~Jacobson, ``Bbr:
  Congestion-based congestion control,'' \emph{Queue}, vol.~14, no.~5, p.~50,
  2016.

\bibitem{vivace}
M.~Dong, T.~Meng, D.~Zarchy, E.~Arslan, Y.~Gilad, B.~Godfrey, and M.~Schapira,
  ``$\{$PCC$\}$ vivace: Online-learning congestion control,'' in \emph{15th
  $\{$USENIX$\}$ Symposium on Networked Systems Design and Implementation
  ($\{$NSDI$\}$ 18)}, 2018, pp. 343--356.

\bibitem{copa}
V.~Arun and H.~Balakrishnan, ``Copa: Practical delay-based congestion control
  for the internet,'' in \emph{15th $\{$USENIX$\}$ Symposium on Networked
  Systems Design and Implementation ($\{$NSDI$\}$ 18)}, 2018, pp. 329--342.

\bibitem{cubic}
S.~Ha, I.~Rhee, and L.~Xu, ``Cubic: a new tcp-friendly high-speed tcp
  variant,'' \emph{ACM SIGOPS Operating Systems Review}, vol.~42, no.~5, pp.
  64--74, 2008.

\bibitem{illi}
S.~Liu, T.~Ba{\c{s}}ar, and R.~Srikant, ``Tcp-illinois: A loss-and delay-based
  congestion control algorithm for high-speed networks,'' \emph{Performance
  Evaluation}, vol.~65, no. 6-7, pp. 417--440, 2008.

\bibitem{aurora}
N.~Jay, N.~Rotman, B.~Godfrey, M.~Schapira, and A.~Tamar, ``A deep
  reinforcement learning perspective on internet congestion control,'' in
  \emph{International Conference on Machine Learning}, 2019, pp. 3050--3059.

\bibitem{indigo}
F.~Y. Yan, J.~Ma, G.~D. Hill, D.~Raghavan, R.~S. Wahby, P.~Levis, and
  K.~Winstein, ``Pantheon: the training ground for internet congestion-control
  research,'' in \emph{2018 $\{$USENIX$\}$ Annual Technical Conference
  ($\{$USENIX$\}$$\{$ATC$\}$ 18)}, 2018, pp. 731--743.

\bibitem{sprout}
K.~Winstein, A.~Sivaraman, H.~Balakrishnan \emph{et~al.}, ``Stochastic
  forecasts achieve high throughput and low delay over cellular networks.'' in
  \emph{NSDI}, 2013, pp. 459--471.

\bibitem{verus}
Y.~Zaki, T.~P{\"o}tsch, J.~Chen, L.~Subramanian, and C.~G{\"o}rg, ``Adaptive
  congestion control for unpredictable cellular networks,'' in \emph{ACM
  SIGCOMM CCR}, vol.~45, no.~4.\hskip 1em plus 0.5em minus 0.4em\relax ACM,
  2015, pp. 509--522.

\bibitem{c2tcp}
S.~Abbasloo, Y.~Xu, and H.~J. Chao, ``C2tcp: A flexible cellular tcp to meet
  stringent delay requirements,'' \emph{IEEE Journal on Selected Areas in
  Communications}, vol.~37, no.~4, pp. 918--932, 2019.

\bibitem{exll}
S.~Park, J.~Lee, J.~Kim, J.~Lee, S.~Ha, and K.~Lee, ``Exll: an extremely
  low-latency congestion control for mobile cellular networks,'' in
  \emph{Proceedings of the 14th International Conference on emerging Networking
  EXperiments and Technologies}.\hskip 1em plus 0.5em minus 0.4em\relax ACM,
  2018, pp. 307--319.

\bibitem{tg}
P.~Szil{\'a}gyi and A.~Terzis, ``Mobile content delivery optimization based on
  throughput guidance,'' in \emph{ICCRG meeting IETF93 (work in progress)},
  2015.

\bibitem{deep_overview}
Y.~Li, ``Deep reinforcement learning: An overview,'' \emph{arXiv preprint
  arXiv:1701.07274}, 2017.

\bibitem{mordatch2015interactive}
I.~Mordatch, K.~Lowrey, G.~Andrew, Z.~Popovic, and E.~V. Todorov, ``Interactive
  control of diverse complex characters with neural networks,'' in
  \emph{Advances in Neural Information Processing Systems}, 2015, pp.
  3132--3140.

\bibitem{duan2016benchmarking}
Y.~Duan, X.~Chen, R.~Houthooft, J.~Schulman, and P.~Abbeel, ``Benchmarking deep
  reinforcement learning for continuous control,'' in \emph{International
  Conference on Machine Learning}, 2016, pp. 1329--1338.

\bibitem{atari}
V.~Mnih, K.~Kavukcuoglu, D.~Silver, A.~Graves, I.~Antonoglou, D.~Wierstra, and
  M.~Riedmiller, ``Playing atari with deep reinforcement learning,''
  \emph{arXiv preprint arXiv:1312.5602}, 2013.

\bibitem{go}
D.~Silver, A.~Huang, C.~J. Maddison, A.~Guez, L.~Sifre, G.~Van Den~Driessche,
  J.~Schrittwieser, I.~Antonoglou, V.~Panneershelvam, M.~Lanctot \emph{et~al.},
  ``Mastering the game of go with deep neural networks and tree search,''
  \emph{nature}, vol. 529, no. 7587, p. 484, 2016.

\bibitem{ledbat}
D.~Rossi, C.~Testa, S.~Valenti, and L.~Muscariello, ``Ledbat: The new
  bittorrent congestion control protocol.'' in \emph{ICCCN}, 2010, pp. 1--6.

\bibitem{west}
C.~Casetti, M.~Gerla, S.~Mascolo, M.~Y. Sanadidi, and R.~Wang, ``Tcp westwood:
  end-to-end congestion control for wired/wireless networks,'' \emph{Wireless
  Networks}, vol.~8, no.~5, pp. 467--479, 2002.

\bibitem{jaf}
J.~Jaffe, ``Flow control power is nondecentralizable,'' \emph{IEEE Transactions
  on Communications}, vol.~29, no.~9, pp. 1301--1306, 1981.

\bibitem{newreno}
T.~Henderson, S.~Floyd, A.~Gurtov, and Y.~Nishida, ``The newreno modification
  to tcp's fast recovery algorithm,'' Tech. Rep., 2012.

\bibitem{bic}
L.~Xu, K.~Harfoush, and I.~Rhee, ``Binary increase congestion control (bic) for
  fast long-distance networks,'' in \emph{Proceedings-IEEE INFOCOM},
  vol.~4.\hskip 1em plus 0.5em minus 0.4em\relax IEEE, 2004, pp. 2514--2524.

\bibitem{vegas}
L.~S. Brakmo, S.~W. O'Malley, and L.~L. Peterson, \emph{TCP Vegas: New
  techniques for congestion detection and avoidance}.\hskip 1em plus 0.5em
  minus 0.4em\relax ACM, 1994, vol.~24, no.~4.

\bibitem{compound}
K.~Tan, J.~Song, Q.~Zhang, and M.~Sridharan, ``A compound tcp approach for
  high-speed and long distance networks,'' in \emph{Proceedings-IEEE INFOCOM},
  2006.

\bibitem{c2tcp2}
S.~Abbasloo, T.~Li, Y.~Xu, and H.~J. Chao, ``Cellular controlled delay tcp
  (c2tcp),'' in \emph{IFIP Networking Conference, 2018}, 2018.

\bibitem{natcp}
S.~Abbasloo, Y.~Xu, H.~J. Chao, H.~Shi, U.~C. Kozat, and Y.~Ye, ``Toward
  optimal performance with network assisted tcp at mobile edge,'' \emph{2nd
  {USENIX} Workshop on Hot Topics in Edge Computing (HotEdge 19)}, 2019.

\bibitem{abc}
P.~Goyal, M.~Alizadeh, and H.~Balakrishnan, ``Rethinking congestion control for
  cellular networks,'' in \emph{Proceedings of the 16th ACM Workshop on Hot
  Topics in Networks}.\hskip 1em plus 0.5em minus 0.4em\relax ACM, 2017, pp.
  29--35.

\bibitem{remy}
\BIBentryALTinterwordspacing
K.~Winstein and H.~Balakrishnan, ``Tcp ex machina: Computer-generated
  congestion control,'' in \emph{Proceedings of the ACM SIGCOMM 2013 Conference
  on SIGCOMM}.\hskip 1em plus 0.5em minus 0.4em\relax ACM, 2013. [Online].
  Available: \url{http://doi.acm.org/10.1145/2486001.2486020}
\BIBentrySTDinterwordspacing

\bibitem{tcp-rl}
X.~Nie, Y.~Zhao, Z.~Li, G.~Chen, K.~Sui, J.~Zhang, Z.~Ye, and D.~Pei, ``Dynamic
  tcp initial windows and congestion control schemes through reinforcement
  learning,'' \emph{IEEE Journal on Selected Areas in Communications}, vol.~37,
  no.~6, pp. 1231--1247, 2019.

\bibitem{fourier}
R.~N. Bracewell and R.~N. Bracewell, \emph{The Fourier transform and its
  applications}.\hskip 1em plus 0.5em minus 0.4em\relax McGraw-Hill New York,
  1986, vol. 31999.

\bibitem{imageproc}
J.~R. Jensen and K.~Lulla, ``Introductory digital image processing: a remote
  sensing perspective,'' 1987.

\bibitem{nixon_feature}
M.~Nixon and A.~S. Aguado, \emph{Feature extraction and image processing for
  computer vision}.\hskip 1em plus 0.5em minus 0.4em\relax Academic Press,
  2012.

\bibitem{LSTM}
\BIBentryALTinterwordspacing
S.~Hochreiter and J.~Schmidhuber, ``Long short-term memory,'' \emph{Neural
  Comput.}, vol.~9, no.~8, pp. 1735--1780, Nov. 1997. [Online]. Available:
  \url{http://dx.doi.org/10.1162/neco.1997.9.8.1735}
\BIBentrySTDinterwordspacing

\bibitem{pearlmutter_rec}
B.~A. Pearlmutter, ``Gradient calculations for dynamic recurrent neural
  networks: A survey,'' \emph{IEEE Transactions on Neural networks}, vol.~6,
  no.~5, pp. 1212--1228, 1995.

\bibitem{ddpg}
T.~P. Lillicrap, J.~J. Hunt, A.~Pritzel, N.~Heess, T.~Erez, Y.~Tassa,
  D.~Silver, and D.~Wierstra, ``Continuous control with deep reinforcement
  learning,'' \emph{CoRR}, vol. abs/1509.02971, 2015.

\bibitem{dpg}
D.~Silver, G.~Lever, N.~Heess, T.~Degris, D.~Wierstra, and M.~Riedmiller,
  ``Deterministic policy gradient algorithms,'' in \emph{Proceedings of the
  31st International Conference on Machine Learning}, ser. Proceedings of
  Machine Learning Research, E.~P. Xing and T.~Jebara, Eds., vol.~32,
  no.~1.\hskip 1em plus 0.5em minus 0.4em\relax Bejing, China: PMLR, 22--24 Jun
  2014, pp. 387--395.

\bibitem{DQN}
V.~Mnih, K.~Kavukcuoglu, D.~Silver, A.~A. Rusu, J.~Veness, M.~G. Bellemare,
  A.~Graves, M.~Riedmiller, A.~K. Fidjeland, G.~Ostrovski \emph{et~al.},
  ``Human-level control through deep reinforcement learning,'' \emph{Nature},
  vol. 518, no. 7540, p. 529, 2015.

\bibitem{batchnorm}
S.~Ioffe and C.~Szegedy, ``Batch normalization: Accelerating deep network
  training by reducing internal covariate shift,'' \emph{arXiv preprint
  arXiv:1502.03167}, 2015.

\bibitem{NIPS2018_7515}
S.~Santurkar, D.~Tsipras, A.~Ilyas, and A.~Madry, ``How does batch
  normalization help optimization?'' in \emph{Advances in Neural Information
  Processing Systems 31}, S.~Bengio, H.~Wallach, H.~Larochelle, K.~Grauman,
  N.~Cesa-Bianchi, and R.~Garnett, Eds.\hskip 1em plus 0.5em minus 0.4em\relax
  Curran Associates, Inc., 2018, pp. 2483--2493.

\bibitem{OU}
G.~E. Uhlenbeck and L.~S. Ornstein, ``On the theory of the brownian motion,''
  \emph{Phys. Rev.}, vol.~36, pp. 823--841, Sep 1930.

\bibitem{tensor}
\BIBentryALTinterwordspacing
M.~Abadi \emph{et~al.}, ``{TensorFlow}: Large-scale machine learning on
  heterogeneous systems,'' 2015, software available from tensorflow.org.
  [Online]. Available: \url{https://www.tensorflow.org/}
\BIBentrySTDinterwordspacing

\bibitem{mahi}
R.~Netravali, A.~Sivaraman, K.~Winstein, S.~Das, A.~Goyal, and H.~Balakrishnan,
  ``Mahimahi: A lightweight toolkit for reproducible web measurement,'' 2014.

\bibitem{blsr_sonet}
Telcordia, ``Gr-1230-core: Sonet bidirectional line switched ring equipment
  generic criteria,'' 1998.

\bibitem{upsr_sonet}
------, ``Gr-1400-core: Sonet dual-fed unidirectional path switched ring (upsr)
  equipment generic criteria,'' 1999.

\bibitem{ring_sdh}
ITU-T, ``Itu-t g.841: Types and characteristics of sdh network protection
  architectures,'' 1998.

\bibitem{protection_ansi}
ANSI, ``Ansi t1.105.02: Synchronous optical network (sonet)-automatic
  protection,'' 2000.

\bibitem{ring_eth}
ITU-T, ``Itu-t g.8032/y.1344: Ethernet ring protection switching,'' 2015.

\bibitem{geni}
\BIBentryALTinterwordspacing
M.~Berman \emph{et~al.}, ``Geni: A federated testbed for innovative network
  experiments,'' \emph{Computer Networks}, vol.~61, pp. 5 -- 23, 2014, special
  issue on Future Internet Testbeds â Part I. [Online]. Available:
  \url{http://www.sciencedirect.com/science/article/pii/S1389128613004507}
\BIBentrySTDinterwordspacing

\bibitem{alchamy}
\BIBentryALTinterwordspacing
A.~Rahimi and B.~Recht, ``Back when we were kids,'' 2017, "Test of Time" award
  talk at NIPS. [Online]. Available:
  \url{https://www.youtube.com/watch?v=Qi1Yry33TQE}
\BIBentrySTDinterwordspacing

\bibitem{lecun_alchamy}
\BIBentryALTinterwordspacing
Y.~Lecun, ``My take on ali rahimi's "test of time" award talk at nips.'' 2017.
  [Online]. Available:
  \url{https://www.facebook.com/yann.lecun/posts/10154938130592143}
\BIBentrySTDinterwordspacing

\bibitem{codel}
K.~Nichols and V.~Jacobson, ``Controlling queue delay,'' \emph{Communications
  of the ACM}, vol.~55, no.~7, pp. 42--50, 2012.

\bibitem{bode}
S.~Abbasloo and H.~J. Chao, ``Bounding queue delay in cellular networks to
  support ultra-low latency applications,'' \emph{arXiv preprint
  arXiv:1908.00953}, 2019.

\bibitem{pie}
R.~Pan, P.~Natarajan, C.~Piglione, M.~S. Prabhu, V.~Subramanian, F.~Baker, and
  B.~VerSteeg, ``Pie: A lightweight control scheme to address the bufferbloat
  problem,'' in \emph{High Performance Switching and Routing (HPSR), 2013 IEEE
  14th International Conference on}.\hskip 1em plus 0.5em minus 0.4em\relax
  IEEE, 2013, pp. 148--155.

\bibitem{lte-book}
S.~Sesia, M.~Baker, and I.~Toufik, \emph{LTE-the UMTS long term evolution: from
  theory to practice}.\hskip 1em plus 0.5em minus 0.4em\relax John Wiley \&
  Sons, 2011.

\bibitem{jain}
R.~K. Jain, D.-M.~W. Chiu, and W.~R. Hawe, ``A quantitative measure of fairness
  and discrimination,'' 1984.

\bibitem{mec}
\BIBentryALTinterwordspacing
``Mobile edge computing introductory technical white paper,'' etsi.org, Tech.
  Rep., 2014. [Online]. Available:
  \url{https://portal.etsi.org/Portals/0/TBpages/MEC/Docs/Mobile-edge\_Computing\_-\_Introductory\_Technical\_White\_Paper\_V1\%2018-09-14.pdf}
\BIBentrySTDinterwordspacing

\bibitem{G5}
\BIBentryALTinterwordspacing
A.~Barreto \emph{et~al.}, ``Successor features for transfer in reinforcement
  learning,'' in \emph{Advances in Neural Information Processing Systems
  30}.\hskip 1em plus 0.5em minus 0.4em\relax Curran Associates, Inc., 2017,
  pp. 4055--4065. [Online]. Available:
  \url{http://papers.nips.cc/paper/6994-successor-features-for-transfer-in-reinforcement-learning.pdf}
\BIBentrySTDinterwordspacing

\bibitem{G6}
A.~Nagabandi, I.~Clavera, S.~Liu, R.~S. Fearing, P.~Abbeel, S.~Levine, and
  C.~Finn, ``Learning to adapt in dynamic, real-world environments through
  meta-reinforcement learning,'' \emph{arXiv preprint arXiv:1803.11347}, 2018.

\bibitem{G7}
I.~Higgins \emph{et~al.}, ``Darla: Improving zero-shot transfer in
  reinforcement learning,'' in \emph{Proceedings of the 34th International
  Conference on Machine Learning-Volume 70}.\hskip 1em plus 0.5em minus
  0.4em\relax JMLR. org, 2017, pp. 1480--1490.

\bibitem{G2}
\BIBentryALTinterwordspacing
K.~Cobbe, O.~Klimov, C.~Hesse, T.~Kim, and J.~Schulman, ``Quantifying
  generalization in reinforcement learning,'' in \emph{Proceedings of the 36th
  International Conference on Machine Learning}, ser. Proceedings of Machine
  Learning Research, K.~Chaudhuri and R.~Salakhutdinov, Eds., vol.~97.\hskip
  1em plus 0.5em minus 0.4em\relax Long Beach, California, USA: PMLR, 09--15
  Jun 2019, pp. 1282--1289. [Online]. Available:
  \url{http://proceedings.mlr.press/v97/cobbe19a.html}
\BIBentrySTDinterwordspacing

\bibitem{G3}
\BIBentryALTinterwordspacing
H.~Wang, S.~Zheng, C.~Xiong, and R.~Socher, ``On the generalization gap in
  reparameterizable reinforcement learning,'' \emph{CoRR}, vol. abs/1905.12654,
  2019. [Online]. Available: \url{http://arxiv.org/abs/1905.12654}
\BIBentrySTDinterwordspacing

\bibitem{G4}
\BIBentryALTinterwordspacing
Z.~Allen{-}Zhu, Y.~Li, and Y.~Liang, ``Learning and generalization in
  overparameterized neural networks, going beyond two layers,'' \emph{CoRR},
  vol. abs/1811.04918, 2018. [Online]. Available:
  \url{http://arxiv.org/abs/1811.04918}
\BIBentrySTDinterwordspacing

\bibitem{G1}
C.~Finn, T.~Yu, J.~Fu, P.~Abbeel, and S.~Levine, ``Generalizing skills with
  semi-supervised reinforcement learning,'' \emph{arXiv preprint
  arXiv:1612.00429}, 2016.

\bibitem{sharpedge}
\BIBentryALTinterwordspacing
S.~Abbasloo and H.~J. Chao, ``Sharpedge: An asynchronous and core-agnostic
  solution to guarantee bounded-delays,'' \emph{CCF Transactions on Networking,
  2020}, 2019. [Online]. Available: \url{https://arxiv.org/abs/2001.00112}
\BIBentrySTDinterwordspacing

\bibitem{5g_slicing}
K.~Sparks \emph{et~al.} (2018) 5g network slicing whitepaper.
  \url{https://transition.fcc.gov/bureaus/oet/tac/tacdocs/reports/2018/5G-Network-Slicing-Whitepaper-Finalv80.pdf}.

\bibitem{sutton}
\BIBentryALTinterwordspacing
R.~Sutton and A.~Barto, \emph{Reinforcement Learning: An Introduction}, ser.
  Adaptive Computation and Machine Learning series.\hskip 1em plus 0.5em minus
  0.4em\relax MIT Press, 2018. [Online]. Available:
  \url{https://books.google.com/books?id=uWV0DwAAQBAJ}
\BIBentrySTDinterwordspacing

\bibitem{Bellman}
R.~Bellman, \emph{Dynamic Programming}, 1st~ed.\hskip 1em plus 0.5em minus
  0.4em\relax Princeton, NJ, USA: Princeton University Press, 1957.

\end{thebibliography}

\appendices
\balance
\begin{figure*}[!t]
    \centering
    \begin{minipage}[b]{0.32\textwidth}
        \centering
        \includegraphics[width=0.95\linewidth,height=1in]{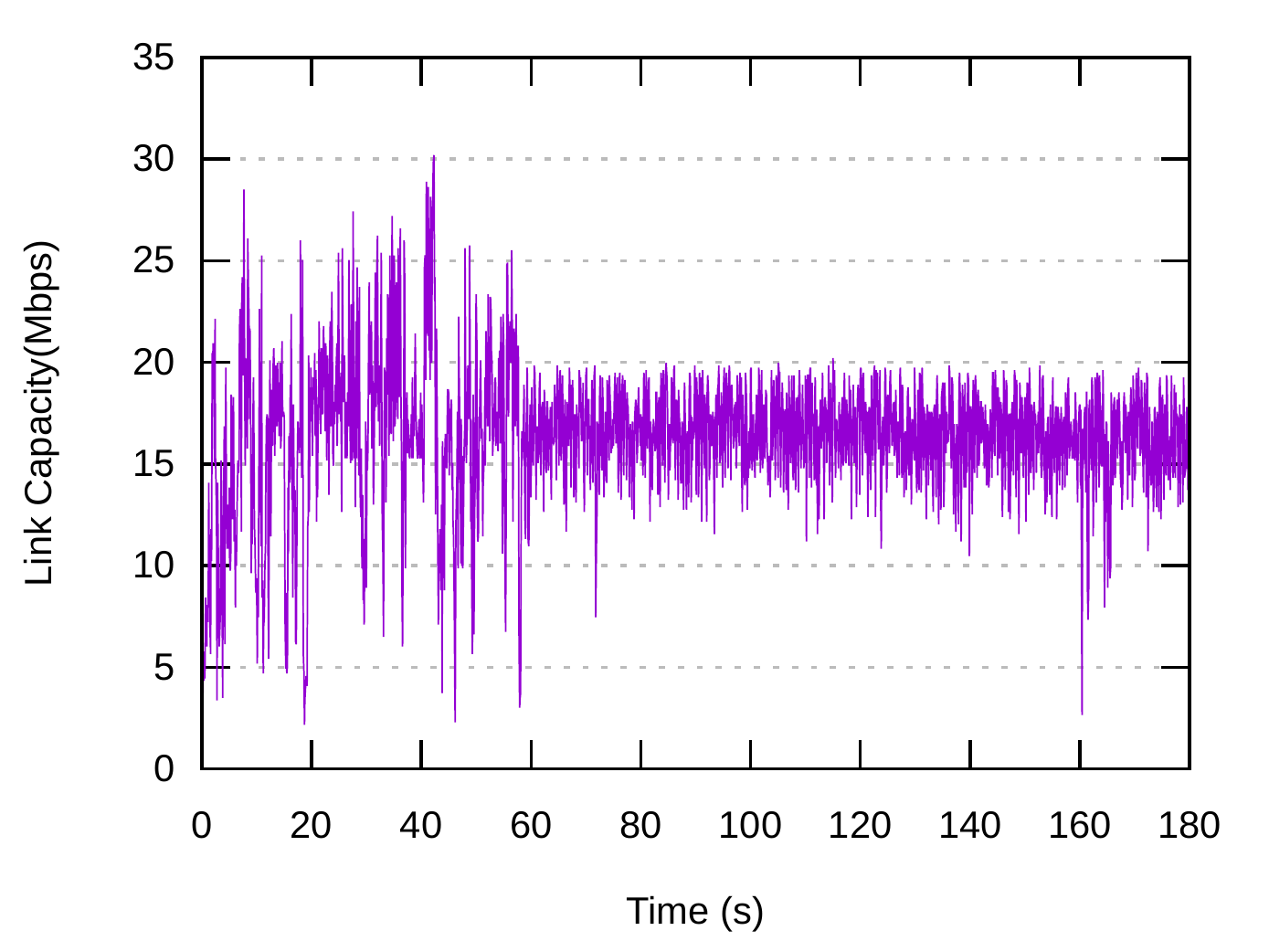}
        \subcaption{Trace \# (1) }
    \end{minipage}        
    \begin{minipage}[b]{0.32\textwidth}
        \centering
        \includegraphics[width=0.95\linewidth,height=1in]{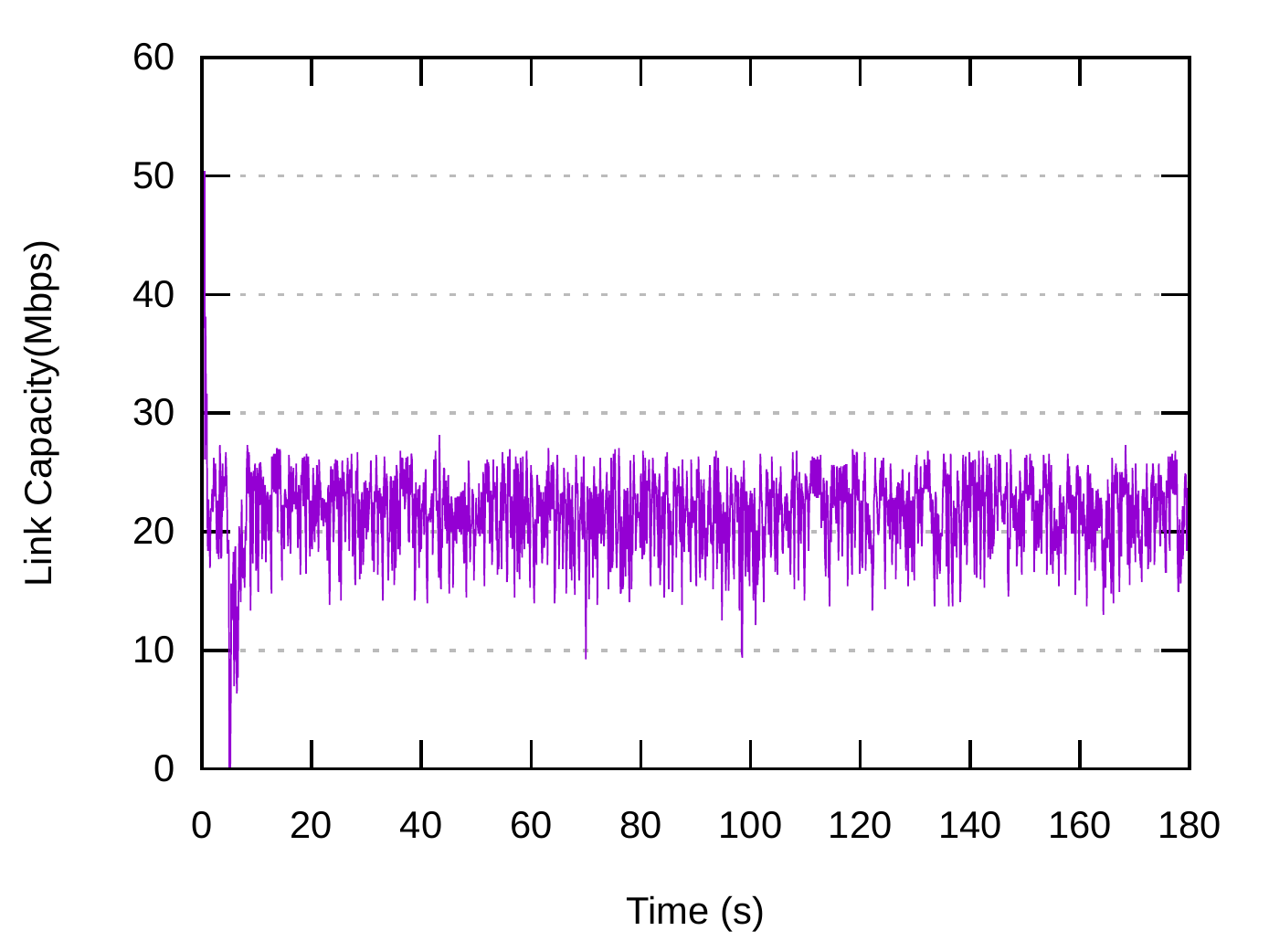}
        \subcaption{Trace \# (2) }
    \end{minipage}        
    \begin{minipage}[b]{0.32\textwidth}
        \centering
        \includegraphics[width=0.95\linewidth,height=1in]{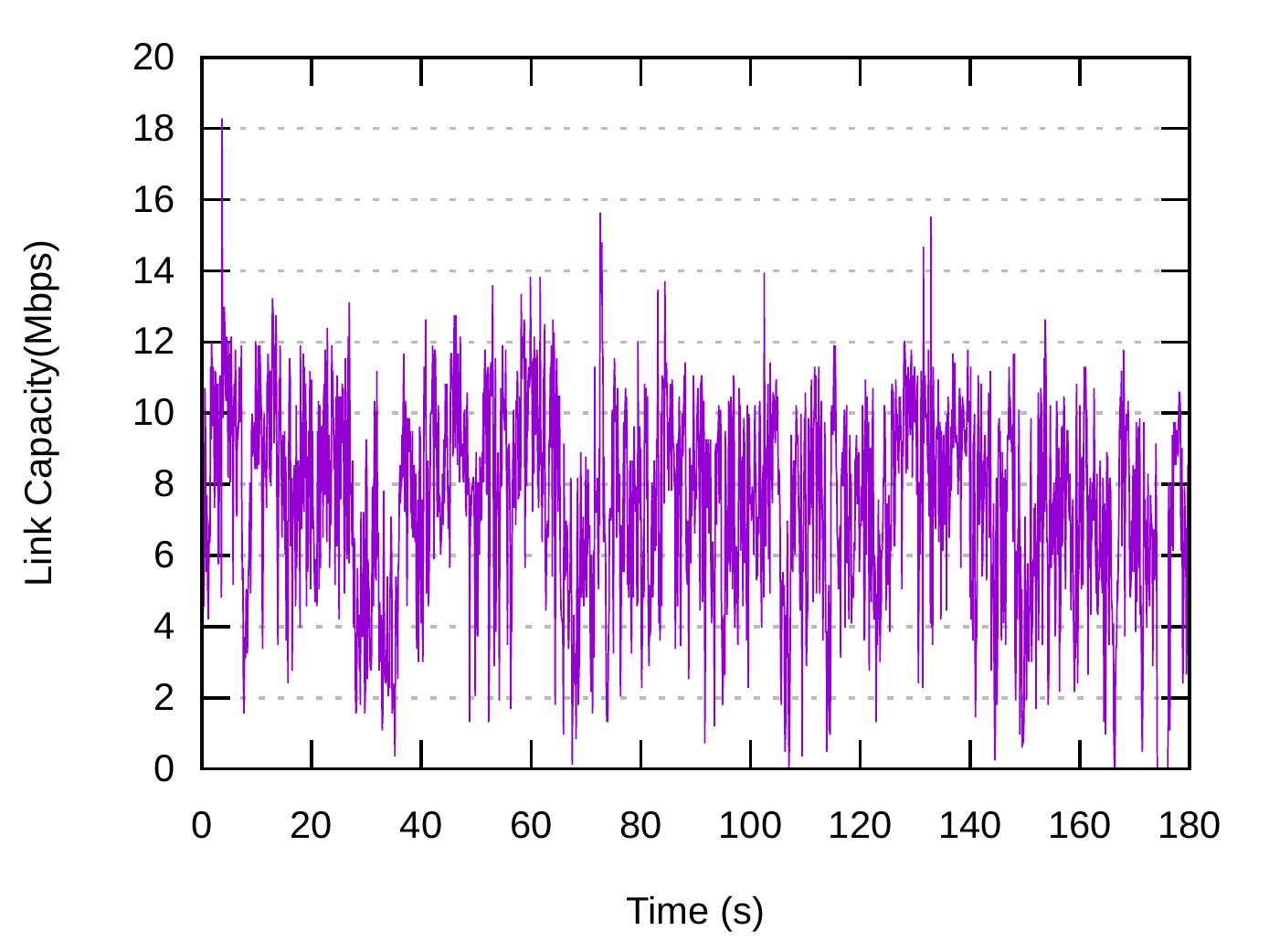}
        \subcaption{Trace \# (3) }
    \end{minipage}        
     \begin{minipage}[b]{0.32\textwidth}
        \centering
        \includegraphics[width=0.95\linewidth,height=1in]{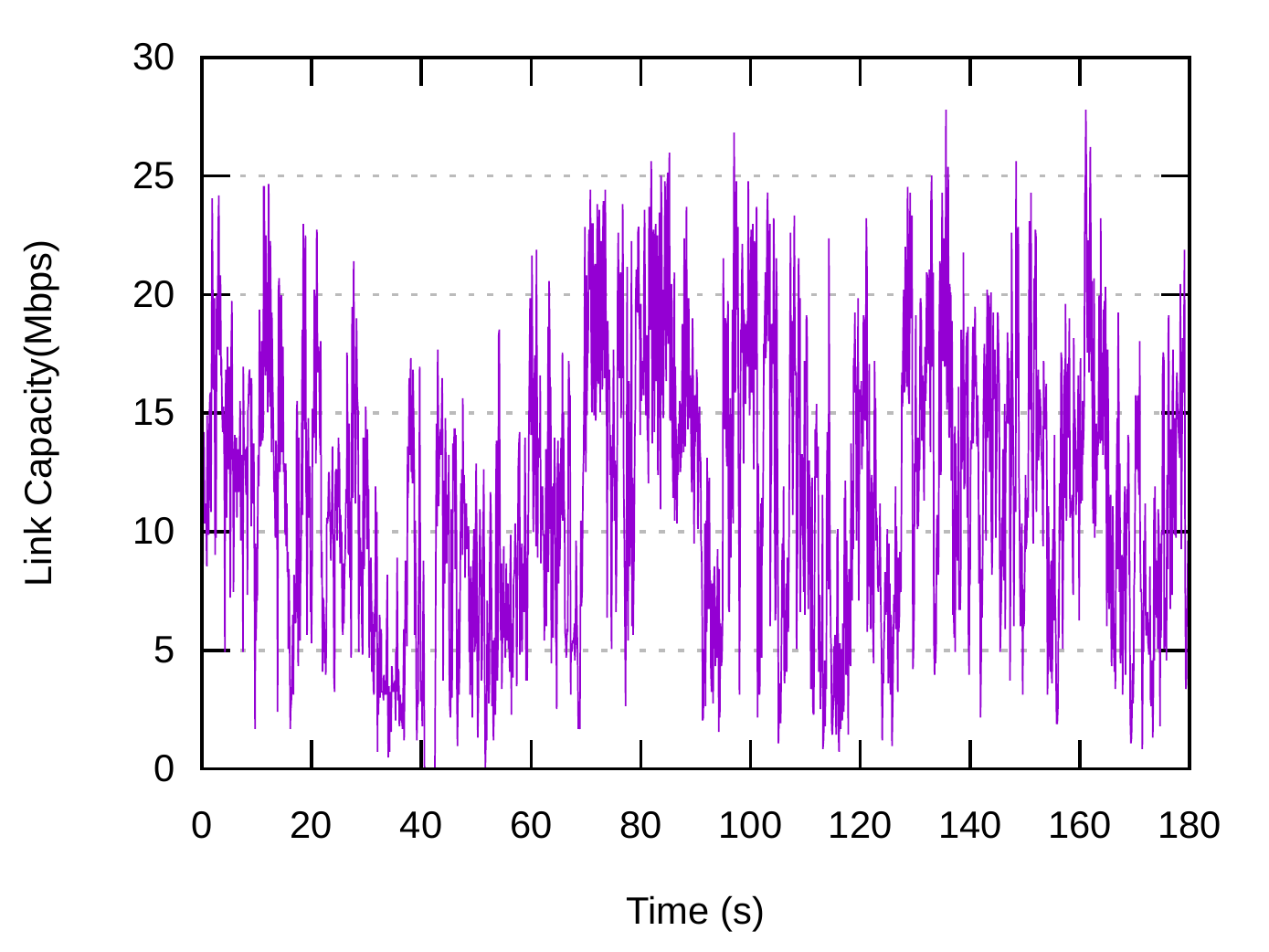}
        \subcaption{Trace \# (4) }
    \end{minipage}        
     \begin{minipage}[b]{0.32\textwidth}
        \centering
        \includegraphics[width=0.95\linewidth,height=1in]{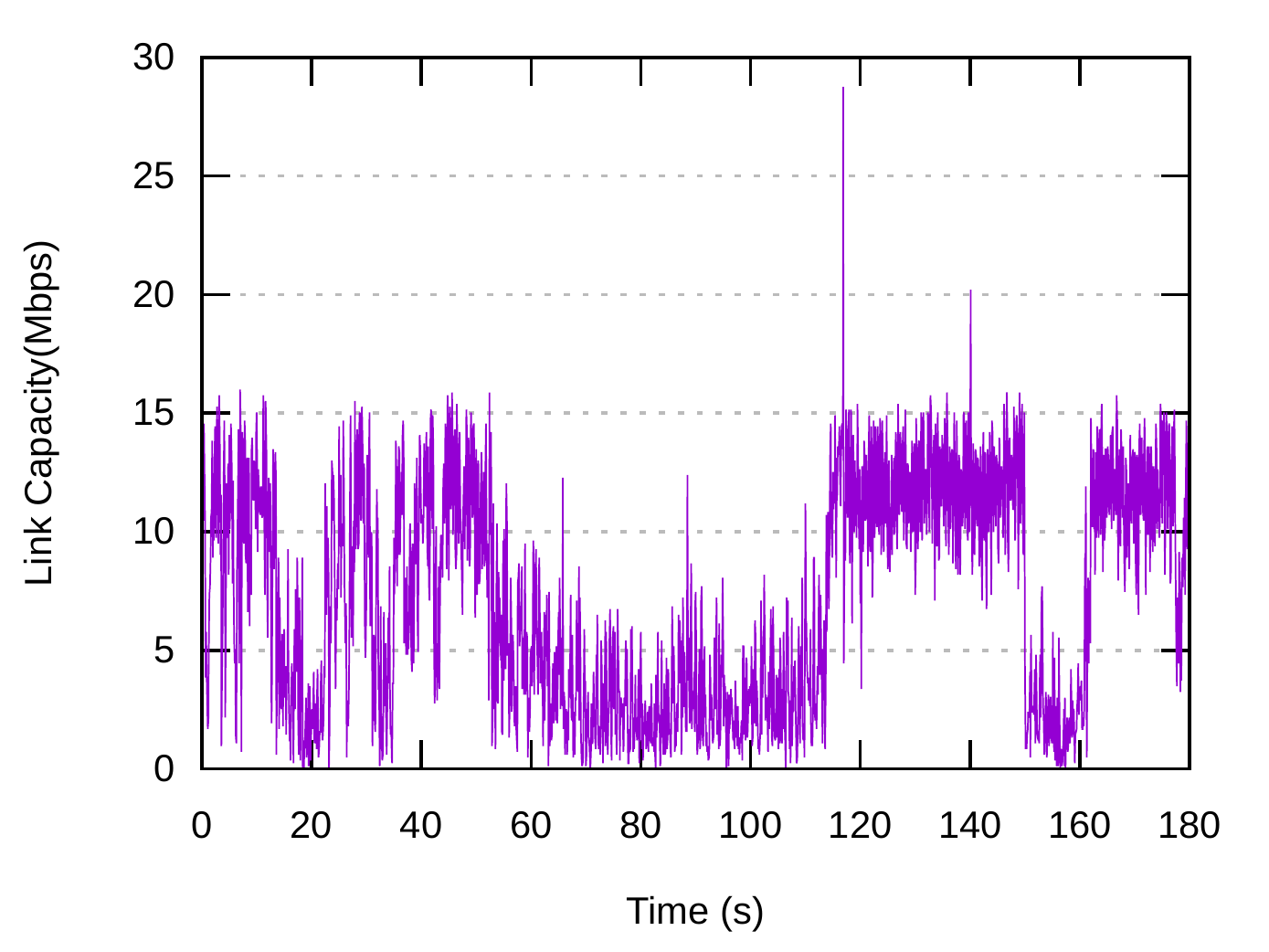}
        \subcaption{Trace \# (5) }
    \end{minipage}        
     \begin{minipage}[b]{0.32\textwidth}
        \centering
        \includegraphics[width=0.95\linewidth,height=1in]{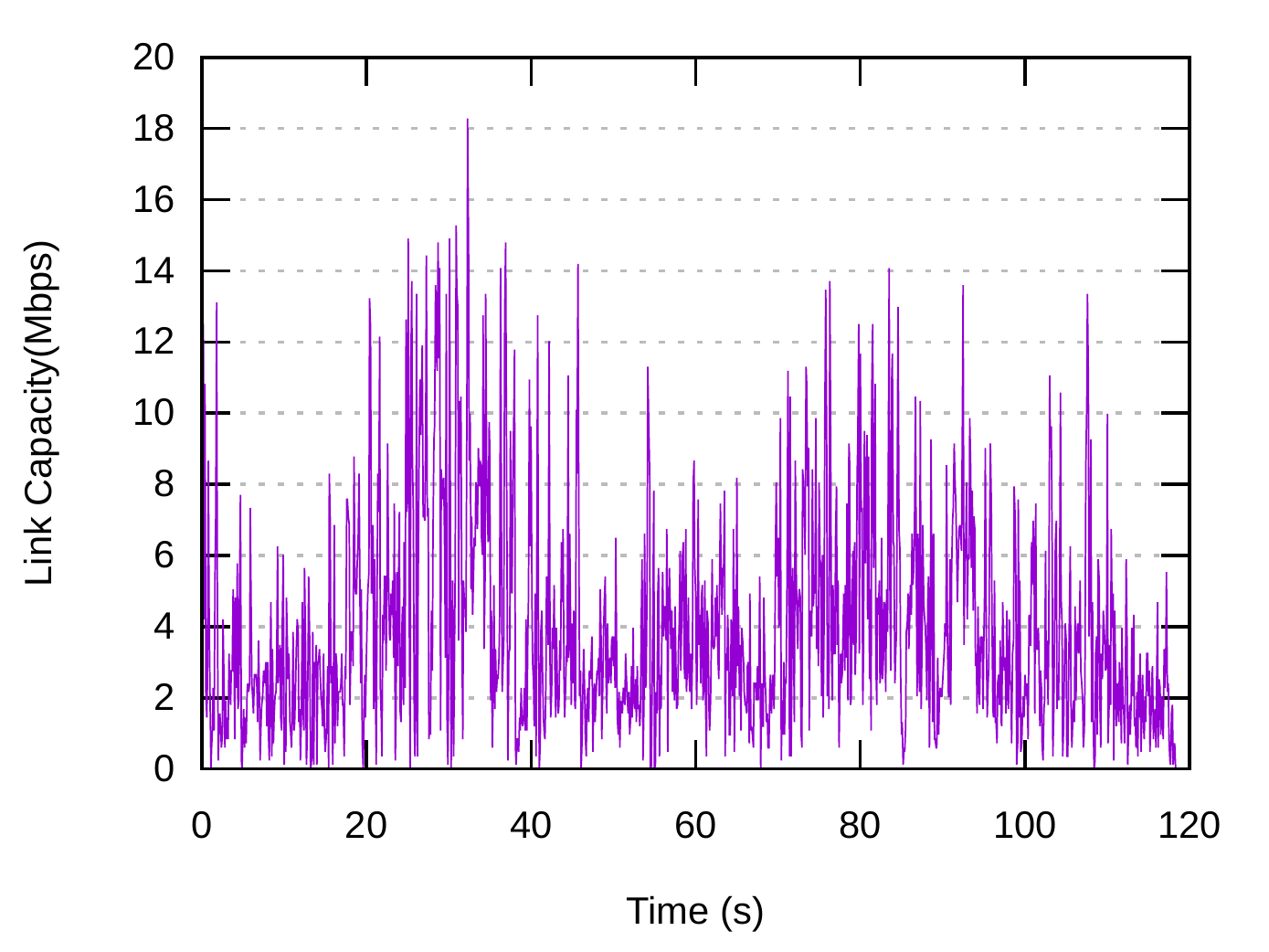}
        \subcaption{Trace \# (6) }
    \end{minipage}        
     \begin{minipage}[b]{0.32\textwidth}
        \centering
        \includegraphics[width=0.95\linewidth,height=1in]{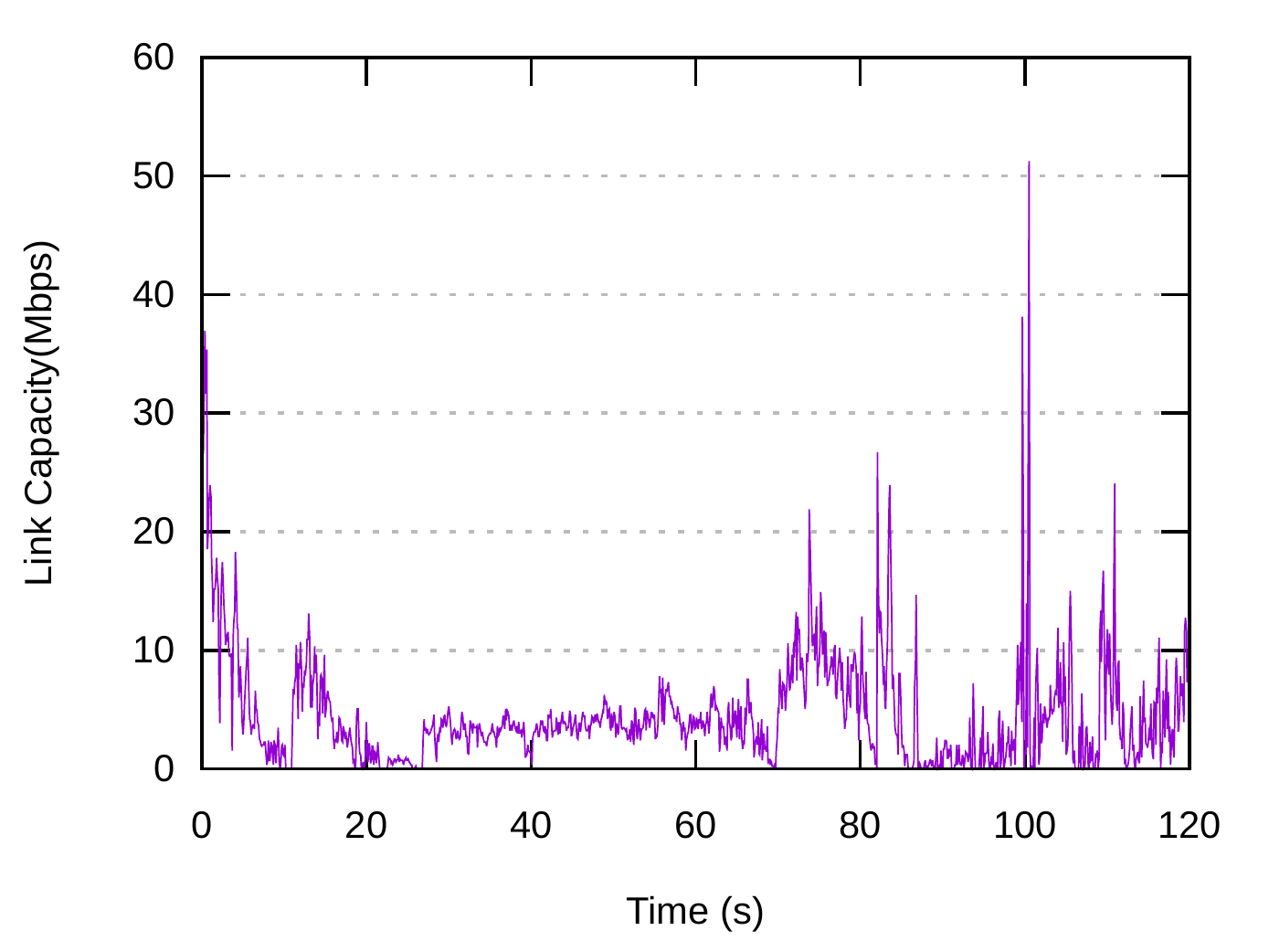}
        \subcaption{Trace \# (7) }
    \end{minipage}        
     \begin{minipage}[b]{0.32\textwidth}
        \centering
        \includegraphics[width=0.95\linewidth,height=1in]{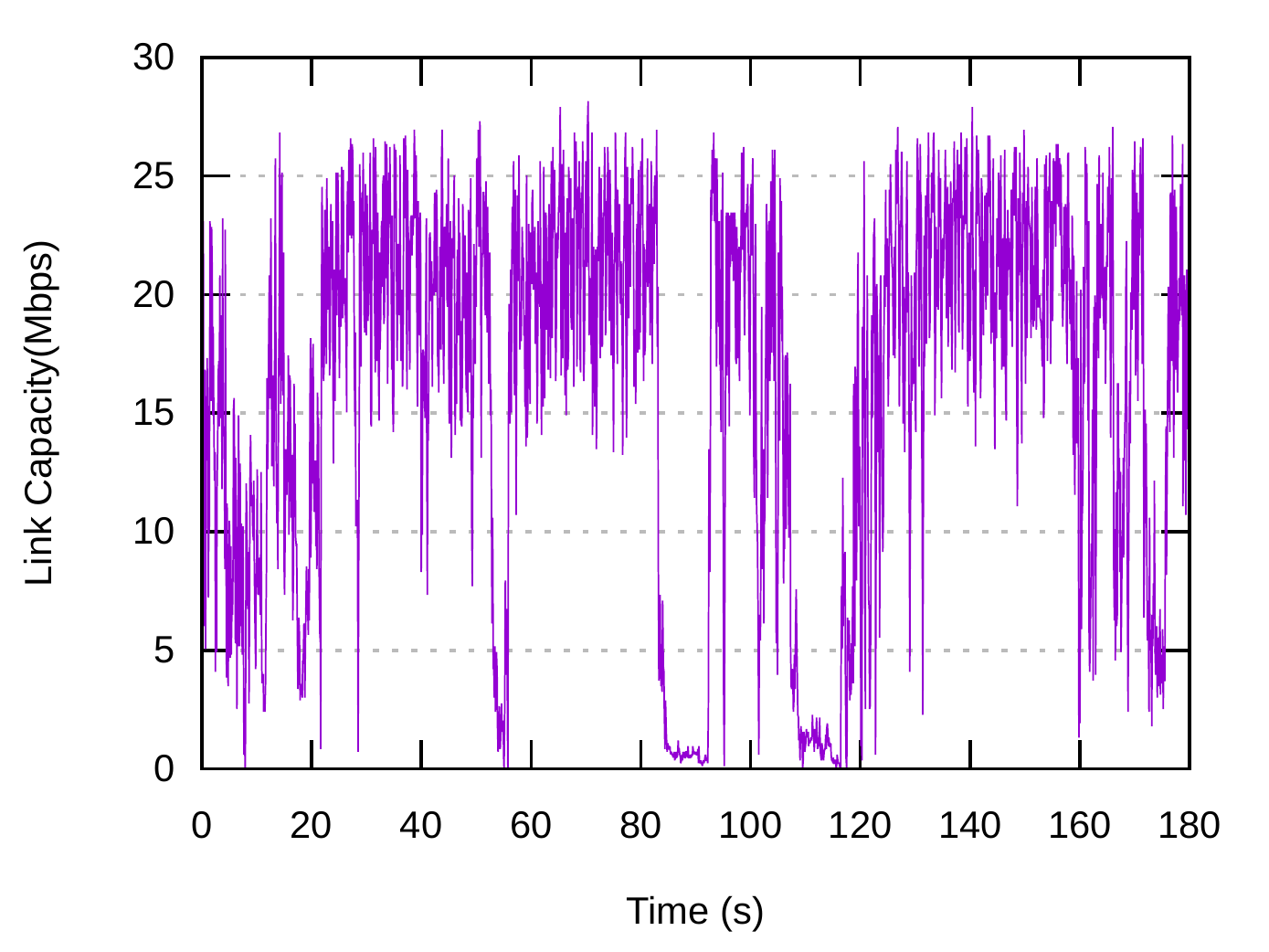}
        \subcaption{Trace \# (8) }
    \end{minipage}        
     \begin{minipage}[b]{0.32\textwidth}
        \centering
        \includegraphics[width=0.95\linewidth,height=1in]{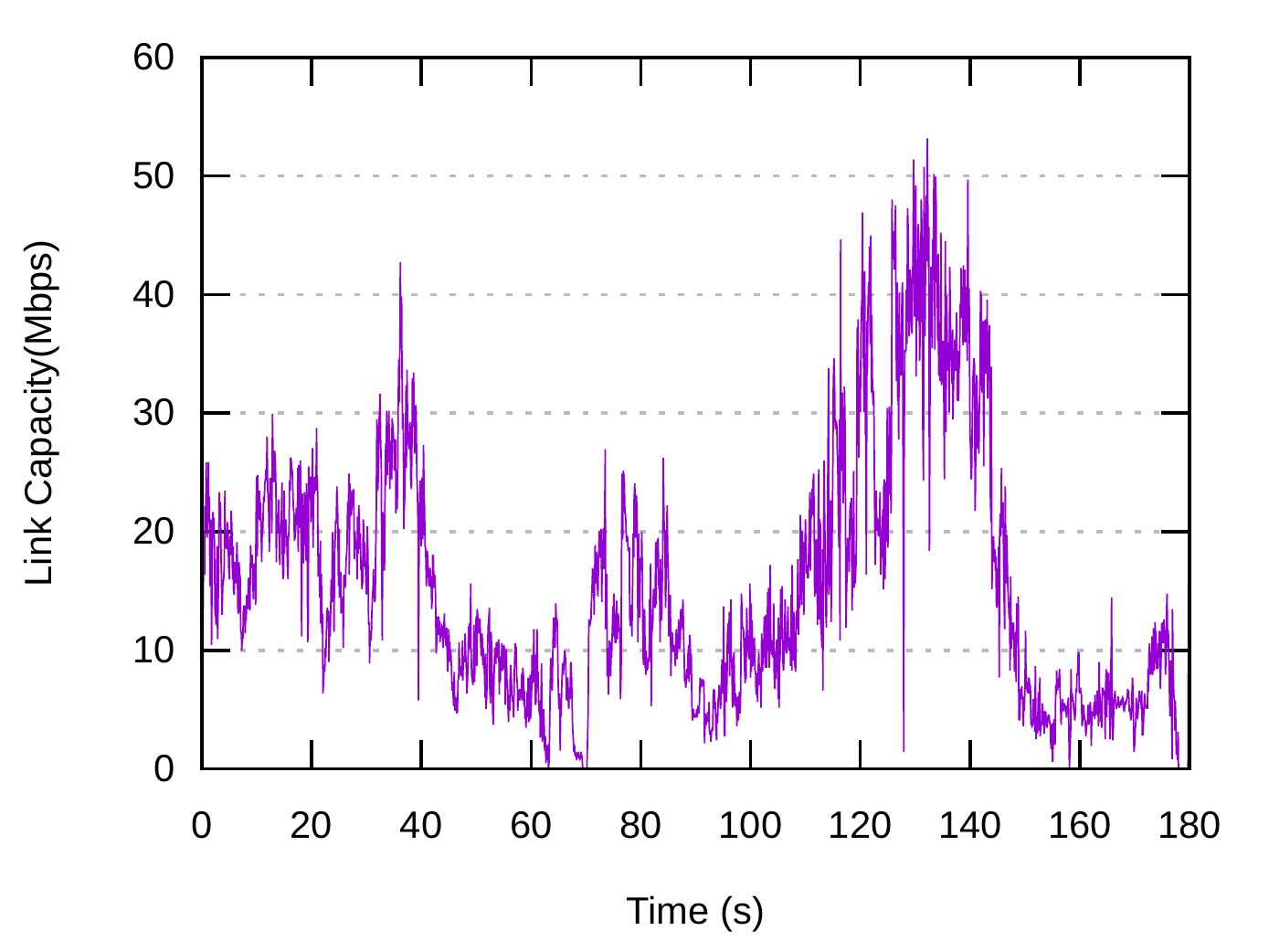}
        \subcaption{Trace \# (9) }
    \end{minipage}        
    \caption{Link capacity variations for the cropped few first minutes of the traces reported in Table~\ref{table_trace_sample}}
    \label{fig_trace_sample}
\end{figure*}

\section{Sample of Cellular Traces Used During Evaluation}
\label{sec_trace}
To have a qualitative view of the collected traces, in Fig.~\ref{fig_trace_sample}, we show the sample of link capacity variations for the cropped first few minutes of the traces reported in Table~\ref{table_trace_sample}. 

\section{Background}
\label{sec_back}
\subsection{Deep Neural Networks}

\begin{figure}[!b]
\centering
    \includegraphics[width=0.7\linewidth,height=1.3in]{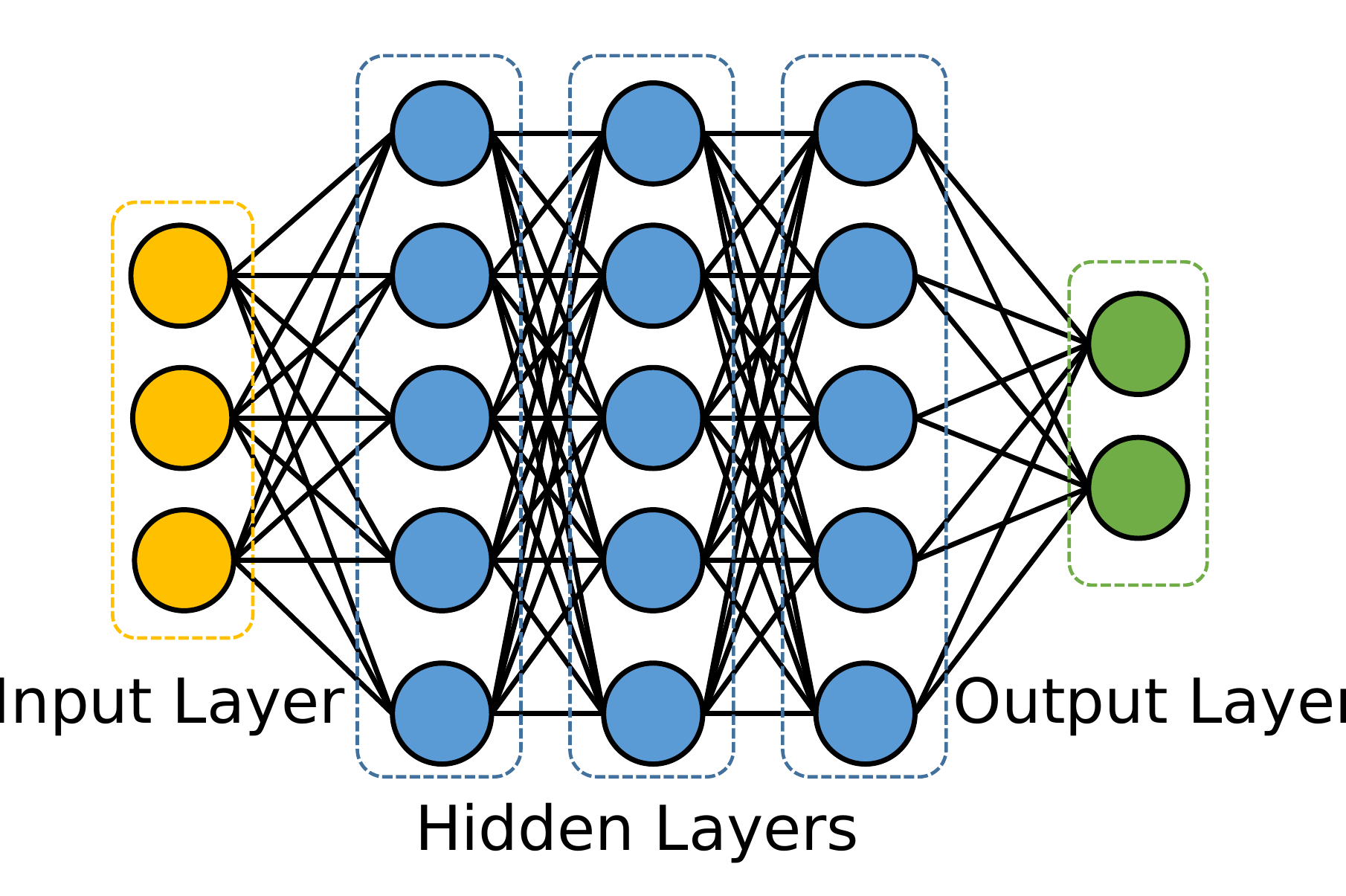}
\caption{The structure of a general DNN}
\label{fig_dnn}
\end{figure}

DNNs enable the automatic feature extraction from raw sensory input. This property reduces the efforts of handcraft feature engineering that requires different domain knowledge. In essence, DNN can be described as a non-linear function transformation. The goal of DNN is to find the function approximation $F^*$ that models the relationship between input vector $x$ and output vector $y$. Fig.~\ref{fig_dnn} declares a generic DNN. DNN consists of multiple layers each consisting of multiple neurons. Each neuron is a computational unit that computes the weighted sum of input values and applies a non-linear activation function to the weighted sum. The output of a neuron is passed to the connected neurons in the next layer. The computation for layer $l$ can be written as:
\begin{equation}
z^{(l)}_j= g(a^{(l)}_j), \;\;\;\;\;\;  a^{(l)}_j = \sum_{i} w^{(l)}_{i,j} \cdot z^{(l-1)}_i,
\end{equation}
where $g$ is an activation function, $w$ is the connection weights, and $z^{(0)}$ corresponds to the input $x$. A DNN has many stages of layers: an input layer which takes the input vector $x$, more than one hidden layers, and an output layer which produces the final output. Each layer processes its input information to a higher-level of representations. The DNN allows the model to learn multiple-level abstractions of data.

The training process of a DNN is done through adjusting the value of the weights $w$ to find a set of weights, $w$, that maximizes the objective or minimizes the cost. 

\subsection{Reinforcement Learning}
\begin{figure}[!t]
\centering
    \includegraphics[width=0.45\linewidth,height=.7in]{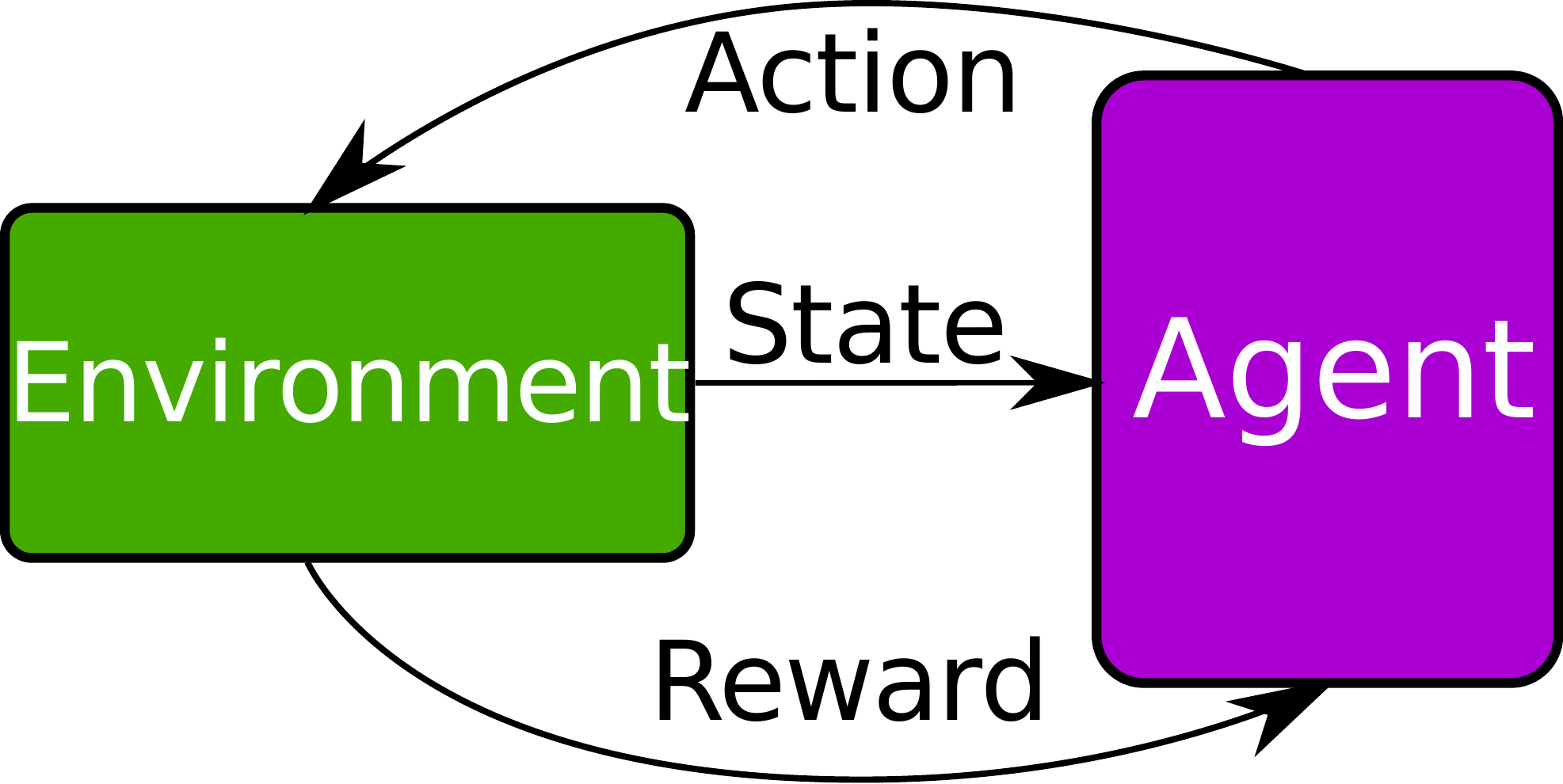}
\caption{Big picture of an RL system}
\label{fig_rl}
\end{figure}

Reinforcement learning is learning how to act from interaction, by mapping a situation to action, to achieve a goal. As Fig~\ref{fig_rl} illustrates, an RL system consists of an agent (the decision-maker block) and the environment (where agent interacts with and draws observations).
The \textit{reward} is a feedback signal to the agent showing whether the goal is achieved. The agent selects an action and the environment responds to the action and presents a new situation, and the interactions are repeated. The agent's objective is to select a sequence of actions that maximize cumulative future rewards. 


\subsection{RL problem setup}

The foundation of solving the general RL problem is to model the task as a Markov Decision Process (MDP)~\cite{sutton}. An MDP is a stochastic decision-making model, defined as a tuple $\mathcal{(S,A,R,T,\gamma)}$, where $\mathcal{S}$ denotes the state space and $\mathcal{A}$ denotes the set of actions. At each time step $t$, the environment reveals the current state $s_t \in \mathcal{S}$, and the agent chooses an actions $a_t \in \mathcal{A}$. The environment then reveals the reward $r_t \sim \mathcal{R}(s_t,a_t)$ and the next state $s_{t+1}$ as a consequence of the action. 
The probability function of state transition satisfies Markov property, which means the state from the environment at time $t+1$ only depends on state and action at time $t$, but not the history of them. The environment dynamics can be written as:
\begin{equation}
\mathcal{T}(s'|s,a) = \mathbb{P}(s_{t+1}=s'|s_t=s,a_t=a),   
\label{statedis}
\end{equation}

A policy $\pi: \mathcal{S} \rightarrow \mathcal{A}$, specifies a mapping from any given state to an action.
The learning goal of RL is to find the policy, that maximize the cumulative reward
\begin{equation}
\mathbb{E}_{s\sim \mathcal{T}, a\sim \pi} [ \sum_{t=0}^{T} \gamma^{t} r_t ],
\label{cul-reward}
\end{equation}
where $\gamma \in [0,1]$ is the discount factor. 
The q-function is defined as the expected return taking the action $a$ in state $s$, and thereafter following policy $\pi$:
\begin{equation}
Q^\pi(s,a) = \mathbb{E} [\sum_{i=t}^{T} \gamma^{i-t} r_i  |s_t = s, a_t=a ],
\label{q-function}
\end{equation}

The optimal policy $\pi^*$ maximizes the optimal q-function $Q^*$. The optimal q-function corresponding to the optimal policy can be unrolled by Bellman optimal equation~\cite{Bellman}:
\begin{equation}
Q^*(s,a) =  \mathbb{E}_{s'} [r_t+ \gamma \max_{a'}Q^*(s',a') |s_t = s, a_t=a ]
\end{equation}
The optimal policy can be written as:
\begin{equation}
\pi^* = \arg \max_{a} Q^*(s,a)
\label{eq_optimal}
\end{equation}

\subsection{Deep Reinforcement Learning}
The challenge of using RL in a real-world task is how to represent the policy and the q-function effectively. The function approximations for computing $\pi$ and $Q(s,a)$ are indispensable, and the state representation needs to be tackled. 
 
Deep Reinforcement Learning (DRL) is the integration of RL with DNN, where the DNN is used as function approximations in the RL framework. DRL learns the essential features directly from the raw inputs, reducing the need for specialized handcraft features used to be carefully designed previously for successful RL application. DRL enables the training in an end-to-end fashion, reaching or surpassing human-level performance.

\end{document}